\documentclass[lettersize,journal]{IEEEtran}
\usepackage{cite}
\usepackage{amsmath,amssymb,amsfonts}
\usepackage{graphicx}
\usepackage{textcomp}

\usepackage{cite}
\usepackage{amsmath,amssymb,amsfonts}
\usepackage{graphicx,color}

\usepackage{algorithm}
\usepackage{array}
\usepackage[caption=false,font=normalsize,labelfont=sf,textfont=sf]{subfig}
\usepackage{textcomp}
\usepackage{stfloats}
\usepackage{url}
\usepackage{verbatim}

\usepackage{amsthm}
\newcommand{\bDiamond}{\mathbin{\Diamond}}
\newcommand{\quotes}[1]{``#1''}

\newtheoremstyle{ieeestyle}%
  {3pt}
  {3pt}
  {\itshape}
  {1em}
  {\normalfont}
  {:}
  { }
  {\thmname{#1}\thmnumber{ #2}\thmnote{ (#3)}}
\theoremstyle{ieeestyle}

\renewcommand\qedsymbol{$\blacksquare$} 

\definecolor{cadmiumgreen}{rgb}{0.0, 0.42, 0.24}
\definecolor{dandelion}{rgb}{0.94, 0.88, 0.19}
\definecolor{emerald}{rgb}{0.31, 0.78, 0.47}
\usepackage{fancyhdr}
\usepackage{amssymb}
\usepackage{epsfig}
\usepackage{xcolor}
\usepackage{pifont}
\usepackage{tipa}
\usepackage{graphics}
\usepackage{algpseudocode}
\usepackage{hyperref}
\usepackage{multirow}
\usepackage{etoolbox}
\usepackage{array,multirow,graphicx}
\usepackage{float}
\usepackage{comment}
\usepackage{longtable}

\newcommand{\cmark}{\ding{51}}%
\newcommand{\xmark}{\ding{55}}%

\begin{document}

\title{Negative Log-Shifted Gamma: A Generalized Pointing Error Model for Directional Wideband Systems}

\author{Evla S. Ahrazoglu,
        Ahmet C. Gul,
        Mehmet N. Akinci,
        Ibrahim Altunbas,
        and~Eylem Erdogan\vspace{-0.75cm}
\thanks{Evla S. Ahrazoglu, Ahmet C. Gul, Mehmet N. Akinci, and Ibrahim Altunbas are with the Department of Electronics and Communication Engineering, Istanbul Technical University, 34469 Istanbul, Turkey (e-mail: ahrazoglu16@itu.edu.tr; gula22@itu.edu.tr; akincime@itu.edu.tr; ibraltunbas@itu.edu.tr).}
\thanks{Eylem Erdogan is with the Department of Electrical and Electronics Engineering, Izmir Institute of Technology, 35433 Izmir, Turkey (e-mail: eylemerdogan@iyte.edu.tr).}
}

\markboth{Journal of \LaTeX\ Class Files,~Vol.~14, No.~8, August~2021}%
{Shell \MakeLowercase{\textit{et al.}}: A Sample Article Using IEEEtran.cls for IEEE Journals}


\maketitle

\begin{abstract}
Pointing error is critical in highly directional communication systems, where even slight misalignment leads to severe received power reduction. Existing pointing error models fail to cover diverse practical scenarios due to restrictive distributional and design limitations. This paper proposes a novel pointing error model incorporating jitters, angle-of-departure/arrival estimation errors, and comprehensive design impacts, including phase shifter quantization, antenna patterns, array design, beamforming architectures, steering capabilities, and bandwidth. To accomplish this, antenna patterns are modeled for both high- and low-gain configurations. Then, the array factor’s main lobe is approximated as a Gaussian beam for subarrays and superarrays, accommodating analog and digital beamforming stages in wideband systems under beam squint presence or compensation. Jitters and estimation errors are incorporated via 3-dimensional rotations, while phase shift quantization is captured through the phase excitations of the arrays. It is shown that the resulting pointing error follows a novel distribution: \textit{the negative log-shifted Gamma (NLSG)}. Fully characterized by only shape, scale, and location parameters, the NLSG features closed-form statistical expressions. This formulation ensures tractable mathematical manipulation, making it a practical, scalable tool for system design and analysis. Validation via normalized moment error, Jensen-Shannon divergence, and Kolmogorov-Smirnov statistics confirms exceptional accuracy, yielding values primarily below 1\%, 0.02, and 0.05, respectively. Furthermore, capacity analysis has demonstrated that specific design choices are paramount for achieving optimal capacity. Finally, important guidelines for designing highly directional wideband communication systems are provided to mitigate the detrimental effects of pointing errors, thereby paving the way for robust next-generation networks.
\end{abstract}


\section{Introduction}

With the widespread deployment of fifth-generation (5G) communication systems, research and development efforts have increasingly shifted toward sixth-generation (6G) and beyond networks. These future communication systems are expected to deliver unprecedented data rates, ultra-low latency, ubiquitous coverage, hyper-reliability, and superior energy efficiency \cite{10816608}. To achieve these ambitious goals, several technologies stand out as promising solutions, including millimeter-wave (mmWave) and terahertz (THz) communications \cite{9318751,11095286,10918779}, non-terrestrial networks (NTNs) \cite{11505754,11003502,RAGO2024110725}, and integrated sensing and communications (ISAC) \cite{11404239,liu2025advances,yang2025physical}.

Fundamentally, most of these enabling technologies rely on highly directional wideband radiation. In high-frequency communications (mmWave and THz bands), pencil-sharp beams are required to overcome the severe attenuation caused by free-space propagation, molecular absorption, and weather conditions \cite{gao2025terahertz,10776962}. Similarly, in NTNs and long-range backhaul links, high-gain antenna arrays are essential to convey information across vast distances, such as in inter-satellite, high-altitude platform station (HAPS)-to-ground, or \textcolor{black}{uncrewed aerial vehicle} (UAV)-to-ground communications \cite{11317441}. Furthermore, highly directional radiation is critical in ISAC systems, where narrow beams significantly enhance the spatial resolution of sensing operations, enabling high-fidelity 3-dimensional mapping, precise target tracking, and gesture recognition \cite{11127106}. Despite exploiting high directivity to enhance performance, these systems are susceptible to imperfect alignment of antennas or arrays since the received power is significantly influenced by the slight orientational fluctuations at either transmitter or receiver.

Misalignment in highly directional communications can be attributed to several distinct phenomena \cite{kaushal2017acquisition}. One of them is the mechanical vibrations (jitters) at transmitter or receiver \cite{11510319}. In the presence of jitters, the boresight direction of the antenna array rotates towards an undesired direction \cite{10528867,salih2024enhancing}. In this case, the expected gains cannot be achieved. Another factor that contributes to the misalignment is the angle-of-departure (AoD) and angle-of-arrival (AoA) estimation errors \cite{11510319}. This is caused by the movement of the communicating nodes. In scenarios where the transmitter and/or the receiver are not stationary, precise tracking is required to maintain high directivity gains \cite{10812951}. Otherwise, the desired array gains cannot be observed, deteriorating the system performance. Additionally, phase shift quantization might play a critical role in alignment in highly directional communications \cite{11287101}. Since the generated beams at transmitter and receiver are steered towards each other via independent phase excitations, quantized phase shift values in analog beamforming may hinder the accuracy of the steering \cite{8269275}. In this case, slight quantization errors may lead to non-negligible impact \cite{10177699}. Furthermore, in wideband systems, analog beamforming architectures also suffer from the beam squint effect which arises from the frequency-dependent array behaviour and frequency-independent phase shifts \cite{naser2026beam}. This mismatch leads different sub-bands to be steered towards slightly different angles, making them out-of-use at high steering angles. On top of these, the severity of the pointing errors is strongly related to the selected antenna type and physical dimensions of it as they determine the width of the generated beam \cite{10841410}. All of these factors must be taken into consideration in the system design to be able to exploit the advantages of highly directional wideband communication links. Thus, accurate and inclusive modeling of pointing errors is vital to reduce the impact of misalignment.

In the current literature, several pointing error models can be found which are valid in certain conditions. The most widely used pointing error model is proposed in \cite{4267802}. However, this model is valid in free-space optical communication scenarios, and it only accounts for the jitters at {the} transmitter. Additionally, radio frequency (RF)/mmWave/THz transceivers have very different structures and characteristics due to massive arrays, which cannot be captured by this model. Simple analytical pointing error models are considered in \cite{9541155,10968382,10891202}. Yet, these models are not directly related to either hardware or software (such as the accuracy of the tracking algorithms). Therefore, they provide very limited insight into the relationships between the system design and the pointing error statistics. A pointing error model for mmWave to THz transmission is proposed in \cite{9887787}. {In this model}, the Gaussian approximation is used {for the} radiated beam from a square antenna array. In \cite{9931325}, the model is extended to the cases where the jitters in different directions are not identical. However, both models are limited to standard uniform arrays of isotropic antennas. A recent model in \cite{11389924} incorporates the platform vibrations (or jitters) in an inter-satellite communication and ranging scenario with similar limitations about isotropic radiation and array configuration. In \cite{10841410}, the proposed pointing error model accounts for the impacts of antenna selection and its physical dimensions in square antenna array case in the presence of symmetrical Gaussian jitters. However, none of these studies consider the inaccurate beam steering in analog/digital/hybrid beamforming architectures, the impact of phase shift quantization, the beam squint effect, and the influence of different tracking modes.

\begin{figure*}[!t]
\centering
\resizebox{0.95\linewidth}{!}{\includegraphics{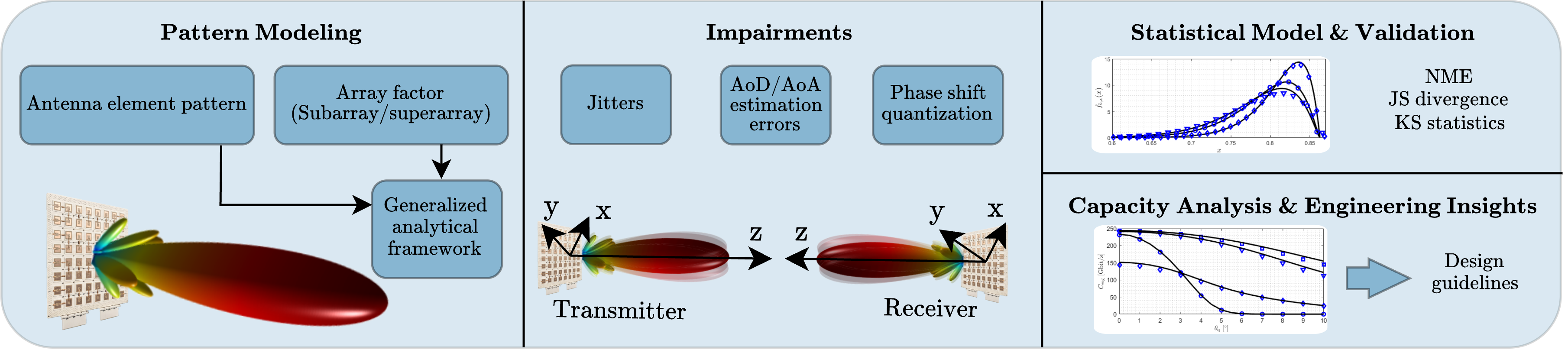}}
\vspace{-5pt}
\caption{Overview of the proposed NLSG pointing error model. \hspace{1cm}}
\label{NLSG-Fig00}
\end{figure*}

Motivated by the lack of holistic models that connect physical transceiver design (both including hardware and software considerations) and the system performance in highly directional wideband systems, a general pointing error model is proposed in this paper, namely, \textit{the negative log-shifted Gamma (NLSG)} model. The main contributions of this paper can be listed as follows:
\begin{itemize}
    \item To achieve such a generalized framework, the array pattern is modeled based on the pattern of a single antenna (for both high- and low-gain antenna element characteristics) and the array factor \textcolor{black}{(as illustrated in the left panel of Fig. \ref{NLSG-Fig00})}. Here, the array factor incorporates subarrays and superarrays of arbitrary sizes to accommodate analog and digital beamforming stages, respectively. This approach also allows analyzing the beam squint effect in the presence of frequency-independent phase shifts, which is critically important to analyze wideband systems.
    \item The influence of mechanical jitters is modeled via 3-dimensional rotations of the overall transceiver architecture for an arbitrary jitter distribution. Additionally, the AoD/AoA estimation errors and phase shift quantization errors are taken into account \textcolor{black}{(see the middle panel of Fig. \ref{NLSG-Fig00})}. Here, three tracking modes are considered: mutual tracking for dynamic alignment, unilateral tracking where only one node tracks the other, and non-tracking mode for stationary deployments where no active steering is performed at either node. These tracking modes can be interpreted as different practical tracking cases in different communication scenarios, e.g. inter-satellite communications, satellite-ground communications, or terrestrial communications, respectively. 
    \item Based on the developed array pattern, jitters, angle estimation errors, and phase shift quantization, the statistics of the pointing error are derived. It is shown that the pointing error follows the NLSG distribution. The statistics of the NLSG distribution are defined by using only three distribution parameters, namely, the shape, scale, and location parameter, and it has a very simple, easy-to-use mathematical form.
    \item It is shown that the proposed model can be used in a wide range of scenarios, including various distributional assumptions, diverse antenna element and array configurations (rectangular arrays of arbitrary sizes) with analog/digital/hybrid beamforming strategies. The model also accounts for the cases where the beam squint effect compensated or uncompensated.
    \item The special cases of the proposed model are also provided. It is shown that the proposed model not only covers the scenarios considered in the previous works but also characterizes the pointing error with much simpler mathematical expressions.
    \item \textcolor{black}{To test the accuracy of the proposed model, theoretical results are compared with simulations where the simulated pointing error coefficients are generated by using the antenna patterns and array factors without any approximations.} Comprehensive validation of the proposed model is performed via {normalized moment error (NME)}, {Jensen-Shannon (JS) divergence}, and {Kolmogorov-Smirnov (KS) statistics} \textcolor{black}{(see the top right panel of Fig. \ref{NLSG-Fig00})}. It is demonstrated that the proposed model is exceptionally accurate in the case of different antenna and array configurations, in various beamforming architectures, in the presence of beam squint, and under various distributional assumptions for jitters or AoD/AoA estimation errors.
    \item Furthermore, the ergodic capacity analysis is performed to gain insights about the relationship between the system design and performance through pointing error statistics \textcolor{black}{(as shown in the bottom right panel of Fig. \ref{NLSG-Fig00})}. The most important tradeoffs between the design choices and the system performance are summarized to guide the design of a highly directional wideband communication link.
\end{itemize}

The rest of the paper is organized as follows: The system model is described in Section II. The element pattern, the array factor for the analog and digital beamforming stages, and the overall array pattern are modeled in Section III along with the examination of the beam squint effect. The orientational fluctuations due to jitters are modeled in Section IV, whereas the angle estimation errors and quantization errors are modeled in Section V. The statistical characterization of the pointing error is presented in Section VI. The practical use cases of the proposed pointing error model are listed in Section VII along with its special cases. The accuracy of the proposed model is examined in Section VIII for a wide range of scenarios. In Section IX, the ergodic capacity is analyzed to highlight the most important engineering insights. Finally, Section X concludes the paper.

{
\textit{Notation:} $f_{X}(x)$, $F_{X}(x)$, and $\mathcal{M}_{X} (s)$ are the probability density function (PDF), cumulative distribution function (CDF), and moment generating function (MGF) of the random variable $X$, respectively. $E[\cdot]$ and $\text{Var}[\cdot]$ denote the expectation and variance operators, respectively, and $\text{Cov}(X,Y)$ represents the covariance between the random variables $X$ and $Y$. $\mathcal{CN}(\mu,\sigma^2)$ and $\mathcal{N}(\mu,\sigma^{2})$ show the complex Gaussian distribution and Gaussian distribution, respectively, with mean $\mu$ and variance $\sigma^2$. $\text{Uniform}(a,b)$ denotes the uniform distribution between $a$ and $b$. $\Gamma(\cdot)$, $\gamma(\cdot,\cdot)$, and $\Gamma(\cdot,\cdot)$ represent the Gamma function \cite[Eq. (8.310.1)]{gradshteyn2014table}, lower incomplete Gamma function \cite[Eq. (8.350.1)]{gradshteyn2014table}, and upper incomplete Gamma function \cite[Eq. (8.350.2)]{gradshteyn2014table}, respectively.}

\section{System Model}

In the considered wideband link, 
the transmitter (${\rm{t}}$) and receiver (${\rm{r}}$) are equipped with arrays of subarrays with subarray size $M_{\rm{qx}} \times M_{\rm{qy}}$ and superarray size $\frac{N_{\rm{qx}}}{M_{\rm{qx}}} \times \frac{N_{\rm{qy}}}{M_{\rm{qy}}}$, forming a {uniform planar array (UPA)} of size $N_{\rm{qx}} \times N_{\rm{qy}}$ with $N_{\rm{q}} = N_{\rm{qx}} N_{\rm{qy}}$ number of antenna elements, where ${\rm{q}} \in \{{\rm{t}},{\rm{r}}\}$. 
To exploit high directivity gains, a {line-of-sight (LOS)} link is assumed where the nodes are positioned within each other's main lobes, located at elevation angle $\theta_{\rm{q}}$ and azimuth angle $\phi_{\rm{q}}$. This architecture utilizes a hybrid beamforming approach, employing analog beamforming at the subarrays and digital beamforming at the superarrays, to support three distinct tracking modes: mutual tracking for dynamic alignment, unilateral tracking where only one node tracks the other, and non-tracking mode for stationary deployments where no active steering is performed at either node. Dividing the total bandwidth $B$ in $K$ number of sub-bands, the received signal in the $k$-th sub-band ($1 \leq k \leq K$) is expressed as
\begin{equation}
{\small
\begin{split}
y_{k} = \sqrt{{P_{k} G^{{\rm{t}}k}(\theta_{\rm{t}},\phi_{\rm{t}})G^{{\rm{r}}k}(\theta_{\rm{r}},\phi_{\rm{r}})}/{(L^{\rm{abs}}_{k}L^{\rm{fsl}}_{k}})} h_{{\rm{f}}k} s_{k} + w_{k},
\label{NLSG-Eq01}
\end{split}}
\end{equation}
where $P_{k}$ is the transmit power and $L^{\rm{fsl}}_{k} = \left({4\pi D}/{\lambda_{k}}\right)^{2}$ denotes the free-space loss in the $k$-th sub-band. Moreover, $L^{\rm{abs}}_{k}$ is the atmospheric absorption loss which can be evaluated as shown in \cite{series2019attenuation} for different atmospheric conditions specified in \cite{series2019atmosphere}. In this formulation, $\lambda_{k}$ represents the operating wavelength of the $k$-th sub-band, $D$ is the link distance, $s_{k}$ is the transmitted signal in the $k$-th sub-band of unit power, and $w_{k} \sim \mathcal{CN}(0,\sigma_{w}^{2})$ stands for the additive Gaussian noise, where $\sigma_{w}^{2}$ shows the noise power. Additionally, $h_{{\rm{f}}k}$ is the $\alpha$-$\mu$ distributed fading coefficient \textcolor{black}{with unit power} \cite{4067122}, where $\alpha$ and $\mu$ are the distribution parameters. The CDF of $h_{{\rm{f}}k}$ is given as 
$F_{h_{{\rm{f}}k}}(x) = \frac{1}{\Gamma(\mu)} \gamma\left(\mu,\mu\frac{x^{\alpha}}{\hat{h}_{{\rm{f}}k}^{\alpha}}\right)$. 
Here, $\hat{h}_{{\rm{f}}k} = \sqrt[\alpha]{\text{E}\left[h_{{\rm{f}}k
}^\alpha\right]}$ is the $\alpha$-root mean value of the fading coefficient $h_{{\rm{f}}k}$. Moreover, $G^{{\rm{q}}k}(\theta_{\rm{q}},\phi_{\rm{q}})$ for $\rm q  \in \{\rm t,\rm r\}$ is the gain of the antenna array in the direction of the elevation angle $\theta_{\rm{q}}  \in [0, \pi]$ and the azimuth angle $\phi_{\rm{q}}  \in [0,2\pi)$, and it depends on the gain of the antenna element $G^{{\rm{q}}k}_{\rm{e}}(\theta_{\rm{q}},\phi_{\rm{q}})$ and the normalized array factors of both the subarray $a^{{\rm{q}}k}(\theta_{\rm{q}},\phi_{\rm{q}})$ and superarray $A^{{\rm{q}}k}(\theta_{\rm{q}},\phi_{\rm{q}})$ as follows {\cite[Eq. (2.30)]{mailloux2017phased}}:
\begin{equation}
{\small
\begin{aligned}
G^{{\rm{q}}k}(\theta_{\rm{q}},\phi_{\rm{q}}) = N_{\rm{q}}G^{{\rm{q}}k}_{\rm{e}}(\theta_{\rm{q}},\phi_{\rm{q}}) \left( a^{{\rm{q}}k}(\theta_{\rm{q}},\phi_{\rm{q}})  A^{{\rm{q}}k}(\theta_{\rm{q}},\phi_{\rm{q}}) \right)^{2}.
\end{aligned}}
\label{NLSG-Eq03}
\end{equation}
In the presence of disturbances, such as mechanical jitters, angle estimation errors, and imperfect phase adjustments within the beamforming architecture, the maximum achievable gains may not be observed at transmitter and/or receiver, reducing the received power. This phenomenon is referred to as pointing error. To characterize the pointing errors, $G^{{\rm{q}}k}(\theta_{\rm{q}},\phi_{\rm{q}})$ and the disturbances are modeled, and the pointing error statistics are derived accordingly in the subsequent sections.

\section{Pattern Modeling and the Beam Squint Effect}

{\label{Sec-2}}

In this section, the factors that contribute to the array pattern are formulated. To describe the pattern's inherent shape, it is first defined within the array's local frame of reference \textcolor{black}{(denoted as (${\rm{x}}'$, ${\rm{y}}'$, ${\rm{z}}'$) in cartesian)}, using the elevation angle $\theta'$ and azimuth angle $\phi'$. This formulation describes the pattern relative to the array's own physical axes, providing the necessary foundation for Sections IV and V where the impacts of mechanical jitters, angle estimation errors, and quantization are examined.

\subsection{Antenna Element Pattern}

In highly directional wideband systems, both high-gain and low-gain antenna elements might be preferred based on several design criteria.\footnote{\textcolor{black}{Especially in high-frequency communications, very high array gains are required to combat excessive path loss. Achieving such gains with low-gain antenna elements leads to massive element counts. In such scenarios, high-gain antenna elements can be preferred to reduce hardware complexity, simplify the array architecture, and lower power consumption.}} To model the patterns of such antennas, first, the pattern function is defined as $g^{{\rm{q}}k}(\theta',\phi')$, then the gain pattern is written as \cite[Eq. (8.1.10)]{vaughan2003channels} 
\begin{equation}
G^{{\rm{q}}k}_{\rm{e}}(\theta',\phi') = G^{{\rm{q}}k}_{\rm{e,max}} (g^{{\rm{q}}k}(\theta',\phi'))^{2}.
\label{NLSG-Eq03.5}
\end{equation}
Here, $G^{{\rm{q}}k}_{\rm{e,max}}$ denotes the maximum gain of the antenna element, which can be either calculated as 
$G^{{\rm{q}}k}_{\rm{e,max}} = \left( \frac{1}{4 \pi} \int_{0}^{2 \pi} \int_{0}^{\pi}  (g^{{\rm{q}}k}(\theta',\phi'))^{2} \sin \theta' d \theta' d \phi' \right)^{-1}$ \textcolor{black}{\cite[Eq. (8.1.10)]{vaughan2003channels}}
or extracted from recent publications on antenna design.

For high-gain antennas \textcolor{black}{(with gains larger than about 8 dBi \cite[Sec. 8.1.1.5]{vaughan2003channels})} with a single dominant main lobe, the radiation pattern can be approximated by a Gaussian beam. To model such a Gaussian beam pointed to $\theta' = 0^\circ$, the radiation function is defined as \cite{Carter}
\begin{equation}
g^{{\rm{q}}k}(\theta',\phi') = \exp\left(-\frac{(\theta' \cos\phi')^{2}}{2\sigma_{\rm{ex}}^{2}} - \frac{(\theta' \sin\phi')^{2}}{2\sigma_{\rm{ey}}^{2}}\right),
\label{NLSG-Eq03.9}
\end{equation}
where the Gaussian spread parameters $\sigma_{\rm{ex}}$ and $\sigma_{\rm{ey}}$ are determined by the 3 dB beamwidths of the antenna element $\theta_{\rm{ex,3dB}}$ and $\theta_{\rm{ey,3dB}}$ on the ${\rm{x}}'$-${\rm{z}}'$ plane and ${\rm{y}}'$-${\rm{z}}'$ plane, respectively, as $\sigma_{\rm{ex}} \approx 0.6\theta_{\rm{ex,3dB}}$ and $\sigma_{\rm{ey}} \approx 0.6\theta_{\rm{ey,3dB}}$ \cite[Eq. (8.1.12)]{vaughan2003channels}. Therefore, the gain of the antenna element can be obtained as
\begin{equation}
G^{{\rm{q}}k}_{\rm{e}}(\theta',\phi') = G^{{\rm{q}}k}_{\rm{e,max}} \exp\left(-\frac{(\theta' \cos\phi')^{2}}{\sigma_{\rm{ex}}^{2}} - \frac{(\theta' \sin\phi')^{2}}{\sigma_{\rm{ey}}^{2}}\right).
\label{NLSG-Eq04}
\end{equation}

For low-gain antennas radiating towards $\theta' = 0^\circ$, the pattern function \textcolor{black}{can be modeled with circular functions and} is written as \textcolor{black}{\cite[Sec. 8.1.1.6]{vaughan2003channels}}
\begin{equation}
\begin{split}
g^{{\rm{q}}k}(\theta',\phi') &= (1-\sin^{2}\theta'\cos^{2}\phi')^{n_{\rm{xz}}/2} \\
&\times \left(\frac{1}{2} + \frac{\cos\theta'}{2 \sqrt{1-\sin^{2}\theta'\cos^{2}\phi'}}\right)^{n_{\rm{yz}}/2},
\end{split}
\label{NLSG-Eq03.1}
\end{equation}
where $n_{\rm{xz}} = -({\log_2\left(\sin\left({\theta_{\rm{ex,3dB}}}/{2}\right)\right)})^{-1}$ and $n_{\rm{yz}} = -({\log_2\left(\cos\left({\theta_{\rm{ey,3dB}}}/{4}\right)\right)})^{-1}$ are the beamwidth parameters related to the 3 dB beamwidth of the antenna element $\theta_{\rm{ex,3dB}}$ and $\theta_{\rm{ey,3dB}}$ on the ${\rm{x}}'$-${\rm{z}}'$ plane and ${\rm{y}}'$-${\rm{z}}'$ plane, respectively.\footnote{This pattern function expression is obtained by 3-dimensional rotation of that in \cite[Eq. (8.1.18)]{vaughan2003channels}. Please refer to Appendix A for the detailed derivation of \eqref{NLSG-Eq03.1}.} In this case, the gain of the antenna element can be obtained as
\begin{equation}
\begin{split}
G^{{\rm{q}}k}_{\rm{e}}(\theta',\phi') &= G^{{\rm{q}}k}_{\rm{e,max}} (1-\sin^{2}\theta'\cos^{2}\phi')^{n_{\rm{xz}}} \\
&\times \left(\frac{1}{2} + \frac{\cos\theta'}{2 \sqrt{1-\sin^{2}\theta'\cos^{2}\phi'}}\right)^{n_{\rm{yz}}}.
\end{split}
\label{NLSG-Eq03.2}
\end{equation}

\subsection{Array Factor}

The normalized array factor of the $M_{\rm{qx}} \times M_{\rm{qy}}$ subarray can be formulated as $a^{{\rm{q}}k}(\theta',\phi') = a_{\rm{x}}^{{\rm{q}}k}(\theta',\phi') a_{\rm{y}}^{{\rm{q}}k}(\theta',\phi')$, where $a_{\rm{x}}^{{\rm{q}}k}(\theta',\phi') = \frac{1}{M_{\rm{qx}}}\sum_{p = 1}^{M_{\rm{qx}}}I_{p \rm{x}} e^{j \kappa_{k} {\rm{x}}_p \sin\theta' \cos\phi' + \Tilde{b}_{p \rm{x}}}$ and $a_{\rm{y}}^{{\rm{q}}k}(\theta',\phi') = \frac{1}{M_{\rm{qy}}}\sum_{m = 1}^{M_{\rm{qy}}} I_{m \rm{y}} e^{j \kappa_{k} {\rm{y}}_m \sin\theta' \sin\phi' + \Tilde{b}_{m \rm{y}}}$ are the normalized array factors of the arrays on the ${\rm{x}'}$-axis and ${\rm{y}'}$-axis, respectively. In this formulation, $I_{p \rm{x}}$ and ${{\rm{x}}_p}$ are the amplitude excitation and {$\rm{x}'$-coordinate of the antenna element}, respectively. Here, $\Tilde{b}_{p \rm{x}}$ denotes the quantized phase excitation along the ${\rm{x}'}$-axis, and it is obtained by the uniform quantization of the continuous phase excitation $b_{p \rm{x}}$. The same quantities for the ${\rm{y}'}$-axis are denoted as $I_{m \rm{y}}$, ${{\rm{y}}_m}$, and $\Tilde{b}_{m \rm{y}}$. Moreover, $\kappa_{k} = 2\pi / \lambda_{k}$ is the wave number. For $M_{\rm{qx}} \times M_{\rm{qy}}$ subarray, $a_{\rm{x}}^{{\rm{q}}k}(\theta',\phi')$ and $a_{\rm{y}}^{{\rm{q}}k}(\theta',\phi')$ can be written in closed-form as {\cite[Eq. (6.91)]{balanis2016antenna}}
\begin{subequations}
{\small
\begin{align}
a_{\rm{x}}^{{\rm{q}}k}(\theta',\phi') &= \frac{1}{M_{\rm{qx}}}\frac{\sin\left(\frac{M_{\rm{qx}}}{2} \left(\kappa_{k} d_{\rm{qx}}\sin\theta'\cos\phi' + \Tilde{b}_{\rm{qx}}\right)\right)}{\sin\left(\frac{1}{2} \left(\kappa_{k} d_{\rm{qx}}\sin\theta'\cos\phi' + \Tilde{b}_{\rm{qx}}\right)\right)},\\
a_{\rm{y}}^{{\rm{q}}k}(\theta',\phi') &= \frac{1}{M_{\rm{qy}}}\frac{\sin\left(\frac{M_{\rm{qy}}}{2} \left(\kappa_{k} d_{\rm{qy}}\sin\theta'\sin\phi' + \Tilde{b}_{\rm{qy}}\right)\right)}{\sin\left(\frac{1}{2} \left(\kappa_{k} d_{\rm{qy}}\sin\theta'\sin\phi' + \Tilde{b}_{\rm{qy}}\right)\right)},
\end{align}}
\label{NLSG-Eq05}
\end{subequations}

\vspace{-6.2pt}
\noindent where $d_{\rm{qx}}$ and $d_{\rm{qy}}$ are the spacings between antenna elements in the subarray. \textcolor{black}{Notice here that the array factor expressions in the above-equation follows the Dirichlet kernel form, i.e. $\mathcal{D}_{N}(\psi) = \textcolor{black}{\frac{1}{N}} \frac{\sin(\frac{N}{2}\psi)}{\sin(\frac{1}{2}\psi)}$.} Also, $\Tilde{b}_{\rm{qx}}$ and $\Tilde{b}_{\rm{qy}}$ show the quantized phase excitation values which are adjusted to steer the radiation direction at both transmitter and receiver.\footnote{The framework developed in this paper accounts for the tracking modes in which the phase excitations are set to zero to incorporate the non-steering nodes. However, the phase excitation terms are preserved with the aim of generalization of the model.}

For the array factor of the superarray $A^{{\rm{q}}k}(\theta',\phi')$, the same mathematical form in \eqref{NLSG-Eq05} can be used by replacing the subarray sizes ($M_{\rm{qx}}$ and $M_{\rm{qy}}$), element spacings ($d_{\rm{qx}}$ and $d_{\rm{qy}}$), and quantized phase excitations ($\Tilde{b}_{\rm{qx}}$ and $\Tilde{b}_{\rm{qy}}$) by the superarray sizes $\left(\frac{N_{\rm{qx}}}{M_{\rm{qx}}}\text{ and }\frac{N_{\rm{qy}}}{M_{\rm{qy}}}\right)$, subarray spacings ($D_{\rm{qx}} = M_{\rm{qx}} d_{\rm{qx}}$ and $D_{\rm{qy}} = M_{\rm{qy}} d_{\rm{qy}}$), and digital phase excitations ($B_{\rm{qx}}$ and $B_{\rm{qy}}$), respectively.\footnote{Here, quantization is not considered for the digital beamforming stage since steering can be performed almost perfectly in digital baseband.}

In \eqref{NLSG-Eq05}, the array response terms in the analog beamforming stage, i.e. $\kappa_{k} d_{\rm{qx}}\sin\theta'\cos\phi'$ and $\kappa_{k} d_{\rm{qy}}\sin\theta'\sin\phi'$, are frequency-dependent as they depend on the wave number of the $k$-th sub-band $\kappa_{k}$.\footnote{For the digital beamforming stage, frequency-dependent phase excitations can be adjusted in digital baseband. Hence, in this work, the beam squint effect is considered only for the subarrays.} In the case of frequency-independent phase excitation, the maximum array gain in different sub-bands can be observed in different radiation directions. This phenomenon is referred to as the beam squint effect which is analyzed in the next subsection.

\subsection{The Beam Squint Effect}

In wideband systems utilizing UPAs, the array is designed according to the center frequency $f_{\rm{c}}$ (or wavelength $\lambda_{\rm{c}}$). To steer the main lobe at $f_{\rm{c}}$ to a desired direction $\theta_{\rm{q}}$ and $\phi_{\rm{q}}$, the phase excitations can be adjusted as
\begin{equation}
\begin{split}
b_{\rm{qx}} &= -\kappa_{\rm{c}} d_{\rm{qx}} \sin\theta_{\rm{q}} \cos\phi_{\rm{q}},\\
b_{\rm{qy}} &= -\kappa_{\rm{c}} d_{\rm{qy}} \sin\theta_{\rm{q}} \sin\phi_{\rm{q}},
\end{split}
\label{NLSG-Eq06}
\end{equation}
where $\kappa_{\rm{c}}$ is the wave number at $f_{\rm{c}}$. However, the array factor depends on the sub-band frequency $f_{k}$ (or wavelength $\lambda_{k}$) through the wave number $\kappa_{k}$, making the array response frequency-dependent. Since the phase excitations are adjusted according to $f_{\rm{c}}$, the array responses and the phase excitations are not matched, which causes the main lobe at $f_{k}$ to be steered to an undesired direction. Assuming no quantization and perfect angle estimation for the sake of analysis of the beam squint effect, the radiation direction in the $k$-th sub-band, i.e. $(\theta_{{\rm{q}}k},\phi_{{\rm{q}}k})$, can be found by jointly solving the set of equations
\begin{equation}
\begin{split}
\kappa_{k} d_{\rm{qx}} \sin\theta_{{\rm{q}}k} \cos\phi_{{\rm{q}}k} -\kappa_{\rm{c}} d_{\rm{qx}} \sin\theta_{\rm{q}} \cos\phi_{\rm{q}} &= 0\\
\kappa_{k} d_{\rm{qy}} \sin\theta_{{\rm{q}}k} \sin\phi_{{\rm{q}}k} -\kappa_{\rm{c}} d_{\rm{qy}} \sin\theta_{\rm{q}} \sin\phi_{\rm{q}} &= 0
\end{split}
\label{NLSG-Eq07}
\end{equation}
as 
$\phi_{{\rm{q}}k} = \phi_{\rm{q}}$ and 
$\theta_{{\rm{q}}k} = \sin^{-1}\left(\frac{\kappa_{\rm{c}}}{\kappa_{k}}\sin\theta_{\rm{q}}\right)$,
which indicates that the beam squint effect is observed only in the elevation direction. When the beam squint is compensated by baseband signal processing or by using {true time delay (TTD)} elements, $\kappa_{\rm{c}}$ can be replaced by $\kappa_{k}$ in \eqref{NLSG-Eq06}, which indicates that the phase shift values are determined depending on the sub-band frequency, eliminating the beam squint effect.

\begin{figure}[!t]
\centering
 	\resizebox*{0.95\linewidth}{!}{\includegraphics{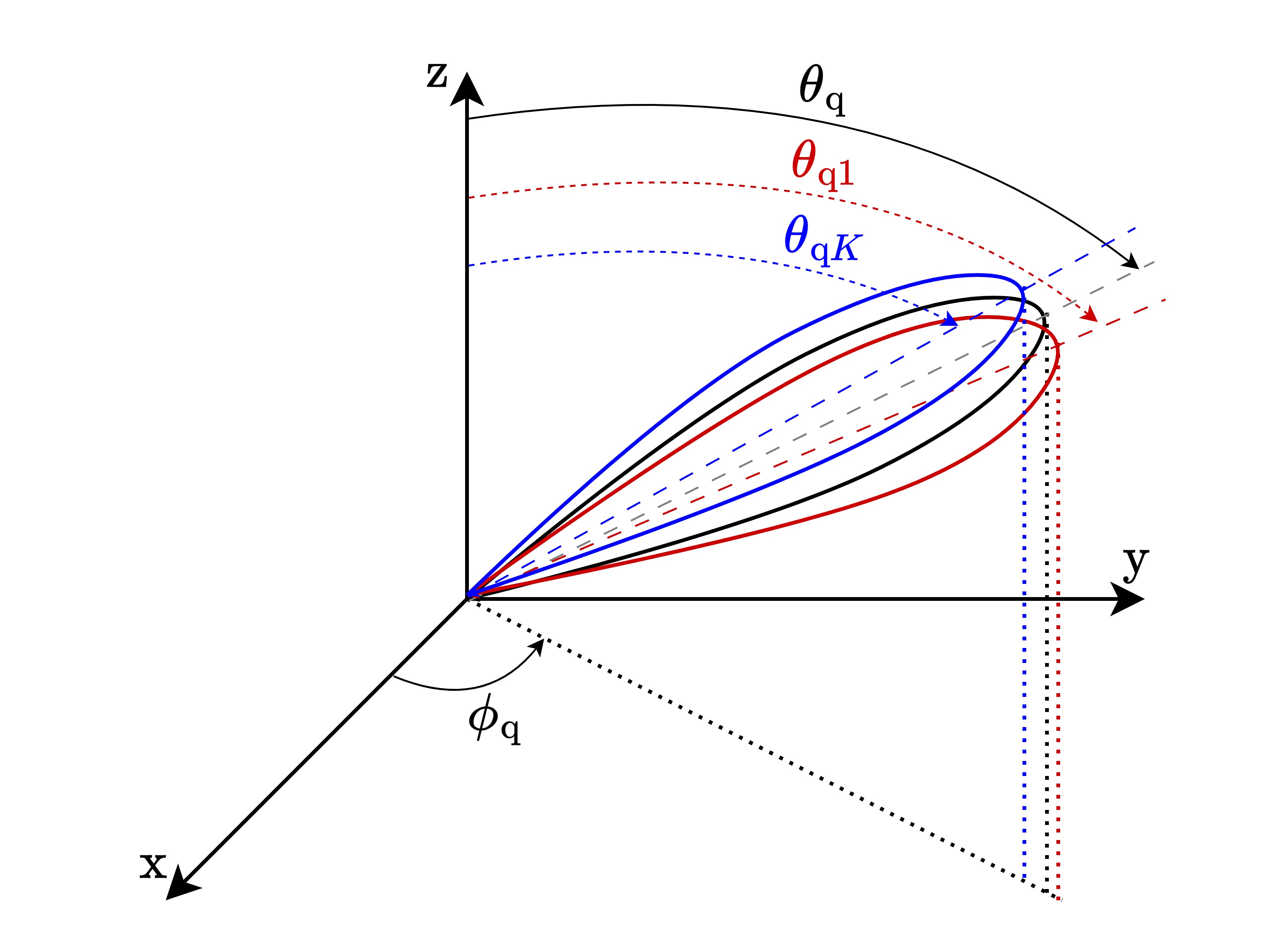}}
    \vspace{-10pt}
\caption{Illustration of the beam squint effect with the center frequency (black), lowest sub-band (red), and highest sub-band (blue).}
\label{NLSG-Fig01}
\end{figure}

In Fig. \ref{NLSG-Fig01}, the beam squint effect is illustrated for the desired radiation direction $(\theta_{\rm{q}},\phi_{\rm{q}})$. The lowest ($k = 1$) and the highest $(k = K)$ sub-bands are radiating towards $(\theta_{{\rm{q}}k},\phi_{\rm{q}})$ where $\theta_{{\rm{q}}k} \neq \theta_{\rm{q}}$. In such cases, not only the receiver observes a reduced array gain but also the system may not fully exploit the corresponding sub-band if the squint effect is exacerbated. This leads to an additional bandwidth limitation on the system, inevitably reducing the achievable capacity. When the frequency of the $k$-th sub-band $f_{k}$ significantly deviates from the center frequency $f_{\rm{c}}$ and the steering angles are high, the transmitter (receiver) cannot cover the receiver (transmitter) with its main lobe \cite{9735144,10008487}. This condition can be characterized by the following theorem.

\textit{Theorem 1:} For the considered subarray configuration, the beam generated at the transmitter (receiver) cannot cover the receiver (transmitter) with its main lobe if the radiation direction $\theta_{{\rm{q}}k}$ in the $k$-th sub-band satisfies
\begin{equation}
\begin{split}
\left|\frac{f_{k}}{f_{\rm{c}}}\sin\theta_{{\rm{q}}k}\cos\phi_{{\rm{q}}k} - \sin\theta_{{\rm{q}}}\cos\phi_{{\rm{q}}}\right| & \geq \frac{2}{{l_{\rm{qx}}} M_{\rm{qx}}}, \\
\left|\frac{f_{k}}{f_{\rm{c}}}\sin\theta_{{\rm{q}}k}\sin\phi_{{\rm{q}}k} - \sin\theta_{{\rm{q}}}\sin\phi_{{\rm{q}}}\right| & \geq \frac{2}{{l_{\rm{qy}}} M_{\rm{qy}}},
\end{split}
\label{NLSG-Eq09}
\end{equation}
where ${l_{\rm{qx}}} = \frac{2d_{\rm{qx}}}{\lambda_{\rm{c}}}$ and ${l_{\rm{qy}}} = \frac{2d_{\rm{qy}}}{\lambda_{\rm{c}}}$ denote the ratio of $d_{\rm{qx}}$ and $d_{\rm{qy}}$ to $\frac{\lambda_{\rm{c}}}{2}$, respectively, to account for element spacings different than $\frac{\lambda_{\rm{c}}}{2}$. 


\textit{Proof: }By using the power-focusing property of the Dirichlet kernel $\mathcal{D}_{N}(\psi) = \textcolor{black}{\frac{1}{N}} \frac{\sin(\frac{N}{2}\psi)}{\sin(\frac{1}{2}\psi)}$, i.e. $\psi$ locates out of the main lobe of $|\mathcal{D}_{N}(\psi)|$ when $|\psi| \geq \frac{2\pi}{N}$ \cite{9735144,10008487},\footnote{Please note here that unlike the conventional notation used in \cite{9735144,10008487}, the definition of the Dirichlet kernel in this paper is given with its argument scaled by $\pi$. This notational change does not affect the obtained results. For the special case $d_{\rm{qx}} = d_{\rm{qy}} = \frac{\lambda_{\rm{c}}}{2}$, \eqref{NLSG-Eq09} coincide with \cite[Eq. (5)]{10008487}.} the transmitter (receiver) cannot cover the receiver (transmitter) with its main lobe when
\begin{equation}
\begin{split}
\left|\kappa_{k} d_{\rm{qx}} \sin \theta_{{\rm{q}}k} \cos \phi_{{\rm{q}}k} - \kappa_{\rm{c}} d_{\rm{qx}} \sin\theta_{{\rm{q}}}\cos\phi_{{\rm{q}}}\right| & \geq \frac{2 \pi}{M_{\rm{qx}}}, \\
\left|\kappa_{k} d_{\rm{qy}} \sin \theta_{{\rm{q}}k} \sin \phi_{{\rm{q}}k} - \kappa_{\rm{c}} d_{\rm{qy}} \sin\theta_{{\rm{q}}}\sin\phi_{{\rm{q}}}\right| & \geq \frac{2 \pi}{M_{\rm{qy}}}.
\end{split}
\label{NLSG-Eq09-t1}
\end{equation}
Here, by substituting $\kappa_{k} = \frac{2 \pi}{\lambda_{k}}$, $\kappa_{\rm{c}} = \frac{2 \pi}{\lambda_{\rm{c}}}$, $d_{\rm{qx}} = \frac{\lambda_{\rm{c}}}{2} l_{\rm{qx}}$, and $d_{\rm{qy}} = \frac{\lambda_{\rm{c}}}{2} l_{\rm{qy}}$, the above-inequalities can be rewritten as in \eqref{NLSG-Eq09}. \hfill $\blacksquare$

\textit{Remark 1:} The inequalities in \eqref{NLSG-Eq09} define the critical limit at which the beam squint effect becomes so severe that the transmitter (receiver) main lobe completely misses the receiver (transmitter). For all practical wideband communication scenarios where the $k$-th sub-band is intended for use, the system parameters (bandwidth, center frequency, array size, and steering angle) must be designed such that these inequalities are not satisfied. By doing this, the system maintains its full intended bandwidth usability. 
Specifically, in this work, where LOS link is assumed, the considered scenarios are those in which the transmitter and receiver can communicate via their main lobes in the $k$-th sub-band.

\subsection{Array Pattern}

Since the aim of this study is to model the pointing error in a LOS communication scenario, the main lobe of the array pattern is sufficient to characterize the impact of misalignment. With this objective, the main lobe of the array factors in \eqref{NLSG-Eq05} can be approximated by a Gaussian beam by the following theorem.

\textit{Theorem 2:} The Dirichlet kernel $D_{N}(\psi) = \frac{1}{N}\frac{\sin\left(\frac{N}{2}\psi\right)}{\sin\left(\frac{1}{2}\psi\right)}$ can be approximated for small $\psi$ as 
\begin{equation}
\begin{split}
D_{N}(\psi) \approx \exp\left(-\frac{N^2 - 1}{24} \psi^2\right).
\end{split}
\label{NLSG-Eq10}
\end{equation}

\textit{Proof:} By using the first two terms of the Taylor expansion for the sine function, i.e. $\sin x \approx x - \frac{1}{3!}x^{3}$, the Dirichlet kernel can be approximated as
\begin{equation}
{\small
\begin{split}
D_{N}(\psi) &\approx \frac{1}{N} \frac{\frac{N}{2}\psi -\frac{1}{3!}\left(\frac{N}{2}\psi\right)^{3}}{\frac{1}{2}\psi -\frac{1}{3!}\left(\frac{1}{2}\psi\right)^{3}} = \left(1 - \frac{N^{2}}{24}\psi^{2}\right) \left(1 - \frac{1}{24}\psi^{2}\right)^{-1} \hspace{-3pt}.
\end{split}}
\label{NLSG-EqA-01}
\end{equation}
With the aid of \cite[Eq. (1.112.1)]{gradshteyn2014table}, the above-equation can be rewritten as
\begin{equation}
\begin{split}
D_{N}(\psi) 
&\approx 1 - \frac{N^{2} - 1}{24}\psi^{2} - \textcolor{black}{\frac{N^{2} - 1}{576}}\psi^{4} \approx 1 - \frac{N^{2} - 1}{24}\psi^{2}.
\end{split}
\label{NLSG-EqA-02dd}
\end{equation}
The last approximation follows from the fact that for small values of $\psi$, the term $\left(- \textcolor{black}{\frac{N^{2} - 1}{576}}\psi^{4}\right)$ becomes negligible. By matching the two terms in \eqref{NLSG-EqA-02dd} with the first two terms of the Taylor expansion of the exponential function, i.e. $e^{x} \approx 1 + x$, the Dirichlet kernel $D_{N}(\psi)$ can be approximated as in \eqref{NLSG-Eq10}. \hfill $\blacksquare$

By using \eqref{NLSG-Eq03}, \eqref{NLSG-Eq04}, \eqref{NLSG-Eq05}, and Theorem 2, the main lobe of the array pattern can be written as
\begin{equation}
{\small
\begin{split}
G^{{\rm{q}}k}(\theta',\phi') &= N_{\rm{q}}  G^{{\rm{q}}k}_{\rm{e}}(\theta',\phi')\\
&\times \exp\Bigg(- \frac{M_{\rm{qx}}^{2} - 1}{12}(\kappa_{k}d_{\rm{qx}}\sin \theta' \cos \phi' + \Tilde{b}_{\rm{qx}})^{2} \\
&- \frac{M_{\rm{qy}}^{2} - 1}{12}(\kappa_{k}d_{\rm{qy}}\sin \theta' \sin \phi' + \Tilde{b}_{\rm{qy}})^{2} \\
&- \frac{\frac{N_{\rm{qx}}^{2}}{M_{\rm{qx}}^{2}} - 1}{12}(\kappa_{k}D_{\rm{qx}}\sin \theta' \cos \phi' + B_{\rm{qx}})^{2} \\
&- \frac{\frac{N_{\rm{qy}}^{2}}{M_{\rm{qy}}^{2}} - 1}{12}(\kappa_{k}D_{\rm{qy}}\sin \theta' \sin \phi' + B_{\rm{qy}})^{2}\Bigg).
\end{split}}
\label{NLSG-Eq11}
\end{equation}

By utilizing the array pattern expression above, the impacts of mechanical jitters, angle estimation errors, and imperfect phase adjustments due to quantization can be modeled accurately.

\section{Impact of Mechanical Jitters}
One of the major sources of the pointing error is the mechanical jitters. Vibrations observed at the transmitter or receiver may impact the alignment between them, significantly reducing the achieved gain and received signal power. To model mechanical jitters in the x-z plane and y-z plane, two random variables are introduced as $\vartheta_{\rm{jqx}}$ and $\vartheta_{\rm{jqy}}$, respectively. It can be considered that the jitter in the x-z plane rotates the array around the y-axis physically. Similarly, the jitter in the y-z plane rotates the array around the x-axis. Hence, the array pattern can be written in the array's local frame of reference $(\theta',\phi')$, or $({\rm{x}}',{\rm{y}}',{\rm{z}}')$ in cartesian, as in \eqref{NLSG-Eq11},\footnote{Recall that the pattern expression in \eqref{NLSG-Eq11} is written in the array's local frame of reference by using $(\theta',\phi')$.} and it must be expressed in the fixed reference frame $(\theta,\phi)$. Here, these two reference frames are related through the following matrix operation: 
\begin{equation}
\begin{split}
\begin{bmatrix}
\sin \theta \cos \phi \\ \sin \theta \sin \phi \\ \cos \theta
\end{bmatrix}
= \mathcal{R}_{\rm{y}}(\vartheta_{\rm{jqx}}) \mathcal{R}_{\rm{x}}(\vartheta_{\rm{jqy}})
\begin{bmatrix}
\sin \theta' \cos \phi' \\ \sin \theta' \sin \phi' \\ \cos \theta'
\end{bmatrix},
\end{split}
\label{NLSG-EqB-01}
\end{equation}
where the rotation matrices are given as
\begin{equation}
\begin{split}
\mathcal{R}_{\rm{x}}(\vartheta_{\rm{jqy}}) &= 
\begin{bmatrix}
1 & 0 & 0 \\
0 & \cos \vartheta_{\rm{jqy}} & - \sin \vartheta_{\rm{jqy}} \\
0 & \sin \vartheta_{\rm{jqy}} & \cos \vartheta_{\rm{jqy}}
\end{bmatrix},\\
\mathcal{R}_{\rm{y}}(\vartheta_{\rm{jqx}}) &= 
\begin{bmatrix}
\cos \vartheta_{\rm{jqx}} & 0 & \sin \vartheta_{\rm{jqx}} \\
0 & 1 & 0 \\
- \sin \vartheta_{\rm{jqx}} & 0 & \cos \vartheta_{\rm{jqx}}
\end{bmatrix}.
\end{split}
\label{NLSG-EqB-02}
\end{equation}
According to the characteristics of rotation matrices, the inverse of the total rotation matrix can be written as 
\begin{equation}
\begin{split}
(\mathcal{R}_{\rm{y}}(\vartheta_{\rm{jqx}}) \mathcal{R}_{\rm{x}}(\vartheta_{\rm{jqy}}))^{-1} &= \mathcal{R}_{\rm{x}}^{-1}(\vartheta_{\rm{jqy}}) \mathcal{R}_{\rm{y}}^{-1}(\vartheta_{\rm{jqx}}) \\
&= \mathcal{R}_{\rm{x}}^{T}(\vartheta_{\rm{jqy}}) \mathcal{R}_{\rm{y}}^{T}(\vartheta_{\rm{jqx}}).
\end{split}
\label{NLSG-EqB-03}
\end{equation}
By using \eqref{NLSG-EqB-01} and \eqref{NLSG-EqB-03}, the following substitutions can be made for small $\vartheta_{\rm{jqx}}$ and $\vartheta_{\rm{jqy}}$, i.e. $\sin \vartheta_{\rm{jqx}} \approx \vartheta_{\rm{jqx}}$, $\cos \vartheta_{\rm{jqx}} \approx 1$, $\sin \vartheta_{\rm{jqy}} \approx \vartheta_{\rm{jqy}}$, $\cos \vartheta_{\rm{jqy}} \approx 1$, and $\vartheta_{\rm{jqx}} \vartheta_{\rm{jqy}} \approx 0$ :
\begin{equation}
\begin{split}
\sin \theta' \cos \phi' &= \sin\theta \cos\phi - \vartheta_{\rm{jqx}}\cos\theta,\\
\sin \theta' \sin \phi' &= \sin\theta \sin\phi + \vartheta_{\rm{jqy}}\cos\theta.
\end{split}
\label{NLSG-Eq12}
\end{equation}
Hence, the array pattern can be written in the fixed reference frame as\footnote{Note here that the impact of mechanical jitters via the element pattern is ignored. This is because the array pattern is dominated by the array factor since the main lobe of the array factor is narrower than that of the antenna element, making the impact of angular fluctuations due to jitters via the antenna element insignificant. Hence, $G^{{\rm{q}}k}_{\rm{e}}(\theta',\phi')$ in \eqref{NLSG-Eq13} is replaced by $G^{{\rm{q}}k}_{\rm{e}}(\theta,\phi)$.}
\begin{equation}
{\small
\begin{aligned}
G^{{\rm{q}}k}(\theta,\phi) &= N_{\rm{q}} G^{{\rm{q}}k}_{\rm{e}}(\theta,\phi) \\
& \hspace{-30pt} \times \exp\left(-\frac{M_{\rm{qx}}^{2} - 1}{12}(\kappa_{k}d_{\rm{qx}}(\sin\theta \cos\phi - \vartheta_{\rm{jqx}}\cos\theta) + \Tilde{b}_{\rm{qx}})^{2} \right) \\
& \hspace{-30pt} \times \exp\left(- \frac{M_{\rm{qy}}^{2} - 1}{12}(\kappa_{k}d_{\rm{qy}}(\sin\theta\sin\phi + \vartheta_{\rm{jqy}}\cos\theta) + \Tilde{b}_{\rm{qy}})^{2} \right) \\
& \hspace{-30pt} \times \exp\left(- \frac{\frac{N_{\rm{qx}}^{2}}{M_{\rm{qx}}^{2}} - 1}{12}(\kappa_{k}D_{\rm{qx}}(\sin\theta \cos\phi - \vartheta_{\rm{jqx}}\cos\theta) + B_{\rm{qx}})^{2}\right) \\
& \hspace{-30pt} \times \exp\left(- \frac{\frac{N_{\rm{qy}}^{2}}{M_{\rm{qy}}^{2}} - 1}{12}(\kappa_{k}D_{\rm{qy}}(\sin\theta\sin\phi + \vartheta_{\rm{jqy}}\cos\theta) + B_{\rm{qy}})^{2} \right) \hspace{-3pt}.\\
\end{aligned}}
\label{NLSG-Eq13}
\end{equation}
Notice here that for the jitter-free case, i.e. $\vartheta_{\rm{jqx}} = 0^\circ$ and $\vartheta_{\rm{jqy}} = 0^\circ$, \eqref{NLSG-Eq13} reduces to \eqref{NLSG-Eq11}, which means that the fixed and local reference frames, $(\theta,\phi)$ and $(\theta',\phi')$, respectively, coincide exactly.

\section{Impact of Angle Estimation Errors and Phase Shift Quantization}

\subsection{Angle Estimation Errors}
\begin{figure}[!t]
\centering
 	\resizebox*{0.90\linewidth}{!}{\includegraphics{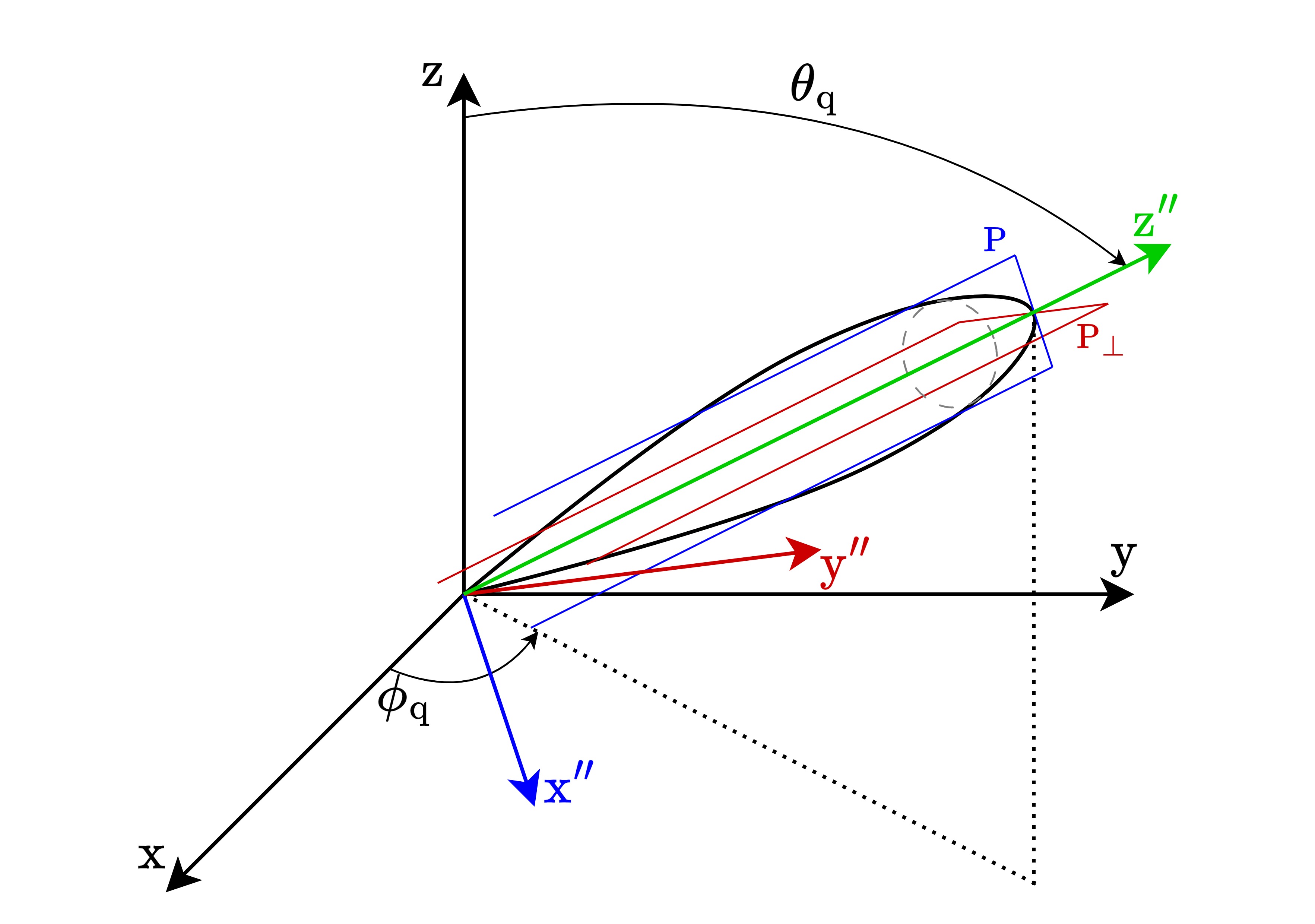}}
    \vspace{-13pt}
\caption{Illustration of the main lobe at the center frequency radiating towards $(\theta_{\rm{q}},\phi_{\rm{q}})$, the elevation plane ${\rm{P}}$, the plane that is perpendicular to the elevation plane ${\rm{P}}_{\perp}$, the global frame of reference $\rm{(x,y,z)}$, and the local frame of reference $\rm{(x'',y'',z'')}$ for angle estimation error modeling.}
\label{NLSG-Fig02}
\end{figure}

In scenarios where the transmitter and/or the receiver is not stationary, the mobility introduced by the relative movements of the communicating nodes necessitates frequent AoD and AoA estimations. Due to rapid movements or insufficient algorithms, these estimations can be hindered, resulting in erroneous angle estimates. In the case of phase adjustments based on the erroneous angle estimations, the radiation patterns are steered inaccurately, leading to pointing errors. To model this impact, two planes are considered, namely, the elevation plane and the plane that is perpendicular to the elevation plane which are shown in Fig. \ref{NLSG-Fig02} as ${\rm{P}}$ and ${\rm{P}}_\perp$, respectively. 
These planes define a local frame of reference ($\rm{x'',y'',z''}$) with ${\rm{P}}$ and ${\rm{P}}_\perp$ corresponding to $\rm{x}''$-$\rm{z}''$ and $\rm{y}''$-$\rm{z}''$ planes, respectively. The new local reference frame can be viewed as the rotated version of the fixed reference frame by $\theta_{\rm{q}}$ around y-axis and $\phi_{\rm{q}}$ around z-axis of the fixed reference frame. The angle estimation errors are defined on ${\rm{P}}$ and ${\rm{P}}_\perp$ and denoted as two independent random variables as $\vartheta_{\rm{qP}}$ and $\vartheta_{\rm{qP}_\perp}$, respectively. Therefore, the erroneous estimates can be written in the local frame of reference ($\rm{x'',y'',z''}$) as $\theta_{\rm{qe}}'' = \tan^{-1}\left( \sqrt{\tan^{2}\vartheta_{\rm{qP}} + \tan^{2}\vartheta_{{\rm{qP}}_\perp}} \right)$ and $\phi_{\rm{qe}}'' = \tan^{-1}\left(\frac{\tan\vartheta_{{\rm{qP}}_\perp}}{\tan\vartheta_{\rm{qP}}}\right)$, 
and they can be expressed in the global frame of reference by using the following relation between the local ($\rm{x'',y'',z''}$) and the global ($\rm{x,y,z}$) frames of references:
\begin{equation}
\begin{split}
\begin{bmatrix}
\sin \theta_{\rm{qe}} \cos \phi_{\rm{qe}} \\ \sin \theta_{\rm{qe}} \sin \phi_{\rm{qe}} \\ \cos \theta_{\rm{qe}}
\end{bmatrix}
= \mathcal{R}_{\rm{z}}(\phi_{\rm{q}}) \mathcal{R}_{\rm{y}}(\theta_{\rm{q}})
\begin{bmatrix}
\sin \theta_{\rm{qe}}'' \cos \phi_{\rm{qe}}'' \\ \sin \theta_{\rm{qe}}'' \sin \phi_{\rm{qe}}'' \\ \cos \theta_{\rm{qe}}''
\end{bmatrix}.
\end{split}
\label{NLSG-Eq-17}
\end{equation}
Here, $\mathcal{R}_{\rm{y}}(\theta_{\rm{q}})$ is written similarly as in \eqref{NLSG-EqB-02}, and $\mathcal{R}_{\rm{z}}(\phi_{\rm{q}})$ is given as
\begin{equation}
\begin{split}
\mathcal{R}_{\rm{z}}(\phi_{\rm{q}}) &= 
\begin{bmatrix}
\cos \phi_{\rm{q}} & - \sin \phi_{\rm{q}} & 0 \\
\sin \phi_{\rm{q}} & \cos \phi_{\rm{q}} & 0 \\
0 & 0 & 1
\end{bmatrix}.
\end{split}
\label{NLSG-Eq-18}
\end{equation}
By performing the matrix operation in \eqref{NLSG-Eq-17}, the following set of equations can be obtained:
\begin{equation}
{\small
\begin{split}
\sin \theta_{\rm{qe}} \cos \phi_{\rm{qe}} &= \cos \theta_{\rm{q}} \cos \phi_{\rm{q}} \sin \theta_{\rm{qe}}'' \cos \phi_{\rm{qe}}'' \\ 
&- \sin \phi_{\rm{q}} \sin \theta_{\rm{qe}}'' \sin \phi_{\rm{qe}}'' + \sin \theta_{\rm{q}} \cos \phi_{\rm{q}} \cos \theta_{\rm{qe}}'' \\
\sin \theta_{\rm{qe}} \sin \phi_{\rm{qe}} &= \cos \theta_{\rm{q}} \sin \phi_{\rm{q}} \sin \theta_{\rm{qe}}'' \cos \phi_{\rm{qe}}'' \\ 
&+ \cos \phi_{\rm{q}} \sin \theta_{\rm{qe}}'' \sin \phi_{\rm{qe}}'' + \sin \theta_{\rm{q}} \sin \phi_{\rm{q}} \cos \theta_{\rm{qe}}''.
\end{split}}
\label{NLSG-Eq-19}
\end{equation}
In both subarray and superarray, the phase shifts are adjusted according to the erroneous angle estimates. Thus, the phase excitation values can be obtained by using geometric relations ($\vartheta_{\rm{qP}} = \sin \theta_{\rm{qe}}'' \cos \phi_{\rm{qe}}''$ and $\vartheta_{\rm{qP}_\perp} = \sin \theta_{\rm{qe}}'' \sin \phi_{\rm{qe}}''$) as
\begin{equation}
{\small
\begin{split}
b_{\rm{qx}} &= - \kappa_{\rm{c}} d_{\rm{qx}} (\vartheta_{\rm{qP}} \cos \theta_{\rm{q}} \cos \phi_{\rm{q}} - \vartheta_{{\rm{qP}}_\perp} \sin \phi_{\rm{q}} + \sin \theta_{\rm{q}} \cos \phi_{\rm{q}}), \\
b_{\rm{qy}} &= - \kappa_{\rm{c}} d_{\rm{qy}} (\vartheta_{\rm{qP}} \cos \theta_{\rm{q}} \sin \phi_{\rm{q}} + \vartheta_{{\rm{qP}}_\perp} \cos \phi_{\rm{q}} + \sin \theta_{\rm{q}} \sin \phi_{\rm{q}}), \\
B_{\rm{qx}} &= - \kappa_{k} D_{\rm{qx}} (\vartheta_{\rm{qP}} \cos \theta_{\rm{q}} \cos \phi_{\rm{q}} - \vartheta_{{\rm{qP}}_\perp} \sin \phi_{\rm{q}} + \sin \theta_{\rm{q}} \cos \phi_{\rm{q}}), \\
B_{\rm{qy}} &= - \kappa_{k} D_{\rm{qy}} (\vartheta_{\rm{qP}} \cos \theta_{\rm{q}} \sin \phi_{\rm{q}} + \vartheta_{{\rm{qP}}_\perp} \cos \phi_{\rm{q}} + \sin \theta_{\rm{q}} \sin \phi_{\rm{q}}).
\end{split}}
\label{NLSG-Eq-20}
\end{equation}
Notice here that $B_{\rm{qx}}$ and $B_{\rm{qy}}$ are the scaled versions of $b_{\rm{qx}}$ and $b_{\rm{qy}}$ by $\frac{\kappa_{k}}{\kappa_{\rm{c}}} M_{\rm{qx}}$ and $\frac{\kappa_{k}}{\kappa_{\rm{c}}} M_{\rm{qy}}$, respectively.

{\textit{Remark 2:} The statistical framework is generalized to accommodate the three considered tracking modes. For nodes that are not actively steering their beams, the phase excitations are not adjusted to track the target at $(\theta_{\rm{q}}, \phi_{\rm{q}})$. Instead, the intended phase excitations are set to deterministic values of $b_{\rm{qx}} = b_{\rm{qy}} = B_{\rm{qx}} = B_{\rm{qy}} = 0$, representing boresight radiation. 
}

In the subarrays, the $b_{\rm{qx}}$ and $b_{\rm{qy}}$ are quantized, yielding $\Tilde{b}_{\rm{qx}}$ and $\Tilde{b}_{\rm{qy}}$, respectively. The impact of quantization is modeled in the next subsection.

\subsection{Quantization}

In practice, phase shifters in analog beamforming have finite resolution due to hardware limitations. Instead of adjusting the phase shifts to radiate towards a desired direction, a set of predetermined phase shift levels are used, which is achieved by the quantization of the continuous phase shift values. This leads to a slight misalignment because the main lobe is not steered to the exact desired direction but around it. In this study, it is assumed that the array is designed to radiate $\theta_{\rm{q}} \in [0, \theta_{\rm{q,max}}]$ and $\phi_{\rm{q}} \in [0,2 \pi)$, where $\theta_{\rm{q,max}}$ denotes the maximum steering range.\footnote{Here, $\theta_{\rm{q,max}}$ determines the angular range that can be supported by the beamforming architecture. If the communicating nodes are highly mobile, i.e. transmitter and receiver must steer their main lobes within a wide range of values, $\theta_{\rm{q,max}}$ must be set to a larger value, e.g. $50^\circ$. However, in the case of quasi-stationary communication link, lower steering angles might be sufficient.} The maximum and minimum continuous phase shifts can be given as
\begin{equation}
\begin{split}
b_{\rm{qx, max}} &= \kappa_{\rm{c}} d_{\rm{qx}} \sin \theta_{\rm{q,max}}, \\  
b_{\rm{qy, max}} &= \kappa_{\rm{c}} d_{\rm{qy}} \sin \theta_{\rm{q,max}}, \\  
b_{\rm{qx, min}} &= - \kappa_{\rm{c}} d_{\rm{qx}} \sin \theta_{\rm{q,max}},\\
b_{\rm{qy, min}} &= - \kappa_{\rm{c}} d_{\rm{qy}} \sin \theta_{\rm{q,max}}.\\
\end{split}
\label{NLSG-Eq-23}
\end{equation}
Assuming $\ell_{\rm{q}}$-bit phase shifters in the analog beamforming stage where $\ell_{\rm{q}}$ denotes the number of quantization bits, the total quantization interval $(b_{\rm{qx, max}} - b_{\rm{qx, min}})$ along ${\rm{x}}$-axis is divided into $2^{\ell_{\rm{q}}}$ number of intervals of widths $\Delta_{\rm{qx}} = \frac{b_{\rm{qx, max}} - b_{\rm{qx, min}}}{2^{\ell_{\rm{q}}}}$. Similarly, the total quantization interval and widths of each quantization interval along ${\rm{y}}$-axis are represented as $(b_{\rm{qy, max}} - b_{\rm{qy, min}})$ and $\Delta_{\rm{qy}} = \frac{b_{\rm{qy, max}} - b_{\rm{qy, min}}}{2^{\ell_{\rm{q}}}}$. 

In \cite{492748}, it is stated that if the Quantization Theorem I and II hold, the quantization error follows the uniform distribution, i.e. $\varepsilon_{\rm{qx}} \sim \text{Uniform}\left(-{\Delta_{\rm{qx}}}/{2}, {\Delta_{\rm{qx}}}/{2}\right)$ and $\varepsilon_{\rm{qy}} \sim \text{Uniform}\left(-{\Delta_{\rm{qy}}}/{2}, {\Delta_{\rm{qy}}}/{2}\right)$, where $\varepsilon_{\rm{qx}} = b_{\rm{qx}} - \Tilde{b}_{\rm{qx}}$ and $\varepsilon_{\rm{qy}} = b_{\rm{qy}} - \Tilde{b}_{\rm{qy}}$, and the moments of $\Tilde{b}_{\rm{qx}}$ ($\Tilde{b}_{\rm{qy}}$) are equal to the moments of the sum of $b_{\rm{qx}}$ and $\varepsilon_{\rm{qx}}$ ($b_{\rm{qy}}$ and $\varepsilon_{\rm{qy}}$). 

{\textit{Remark 3:} For non-steering nodes, the quantization does not introduce uncertainty since the boresight direction corresponds to a zero-phase state that is perfectly representable in the quantization codebook. Consequently, the quantization errors are nullified by setting quantization intervals $\Delta_{\rm{qx}} = 0$ and $\Delta_{\rm{qy}} = 0$. However, these terms are retained to maintain a unified model applicable to mutual tracking, unilateral tracking, and non-tracking modes.}

To incorporate the impact of phase shift quantization, $\Tilde{b}_{\rm{qx}}$ and $\Tilde{b}_{\rm{qy}}$ in \eqref{NLSG-Eq13} can be replaced by $\Tilde{b}_{\rm{qx}} = b_{\rm{qx}} - \varepsilon_{\rm{qx}}$ and $\Tilde{b}_{\rm{qy}} = b_{\rm{qy}} - \varepsilon_{\rm{qy}}$, respectively, and the array pattern can be rewritten as
\begin{equation}
{\footnotesize
\begin{split}
G^{{\rm{q}}k}(\theta,\phi) &= N_{\rm{q}} G^{{\rm{q}}k}_{\rm{e}}(\theta,\phi) \\
& \hspace{-30pt} \times \exp\left(-\frac{M_{\rm{qx}}^{2} - 1}{12}(\kappa_{k}d_{\rm{qx}}(\sin\theta \cos\phi - \vartheta_{\rm{jqx}}\cos\theta) + b_{\rm{qx}} - \varepsilon_{\rm{qx}})^{2} \right) \\
& \hspace{-30pt} \times \exp\left(- \frac{M_{\rm{qy}}^{2} - 1}{12}(\kappa_{k}d_{\rm{qy}}(\sin\theta\sin\phi + \vartheta_{\rm{jqy}}\cos\theta) + b_{\rm{qy}} - \varepsilon_{\rm{qy}})^{2} \right) \\
& \hspace{-30pt} \times \exp\left(- \frac{\frac{N_{\rm{qx}}^{2}}{M_{\rm{qx}}^{2}} - 1}{12}(\kappa_{k}D_{\rm{qx}}(\sin\theta \cos\phi - \vartheta_{\rm{jqx}}\cos\theta) + B_{\rm{qx}})^{2}\right) \\
& \hspace{-30pt} \times \exp\left(- \frac{\frac{N_{\rm{qy}}^{2}}{M_{\rm{qy}}^{2}} - 1}{12}(\kappa_{k}D_{\rm{qy}}(\sin\theta\sin\phi + \vartheta_{\rm{jqy}}\cos\theta) + B_{\rm{qy}})^{2} \right) \hspace{-3pt}.\\
\end{split}}
\label{NLSG-Eq24}
\end{equation}

By using \eqref{NLSG-Eq24}, the joint effects of mechanical jitters, AoD/AoA estimation errors, and phase‐shift quantization can be investigated for both narrowband ($\kappa_{k} \approx \kappa_{\rm{c}}$) and wideband ($\kappa_{k} \neq \kappa_{\rm{c}}$) systems, considering analog, digital, or hybrid beamforming architectures with either square ($N_{\rm{qx}}=N_{\rm{qy}}$) or non-square ($N_{\rm{qx}}\neq N_{\rm{qy}}$) UPAs, operating in steering or non-steering configurations and supporting symmetrical or asymmetrical radiation patterns at both the transmitter and receiver. Moreover, the array pattern formulation remains applicable for both uncompensated and compensated beam squint conditions in wideband systems, where the compensated case is obtained by replacing $\kappa_{\rm{c}}$ by $\kappa_{k}$ in \eqref{NLSG-Eq-20}. This formulation provides a unified analytical basis for studying and designing practical wideband directional beamforming systems across diverse application scenarios. The statistical characterization of the pointing error in the aforementioned scenarios is presented in the next section.

\section{Pointing Error Modeling}

\label{Sec-4}

\subsection{System Model Revisited}

By defining a distinct variable $h_{{\rm{p}}k}$ to isolate the impact of pointing errors, the received signal expression in the $k$-th sub-band in \eqref{NLSG-Eq01} rewritten as
$y_{k} = \sqrt{{P_{k}}/{L_{k}}} h_{{\rm{f}}k} h_{{\rm{p}}k} s_{k} + w_{k}$.
Here, $L_{k} = L^{\rm{abs}}_{k}L^{\rm{fsl}}_{k}/(G^{{\rm{t}}k}_{\rm{max}} G^{{\rm{r}}k}_{\rm{max}})$ is the total loss factor, with $G^{{\rm{q}}k}_{\rm{max}} = G^{{\rm{q}}k}_{\rm{e,max}} N_{\rm{q}}$ representing the maximum array gain at ${\rm{q}}$ under ideal boresight alignment. The pointing error coefficient $h_{{\rm{p}}k}$ is defined as the square root of the ratio between the observed total gain and the maximum array gain under ideal boresight alignment. Mathematically, this is expressed as
\begin{equation}
\begin{split}
h_{{\rm{p}}k} = \sqrt{\frac{G^{{\rm{t}}k}(\theta_{\rm{t}},\phi_{\rm{t}}) G^{{\rm{r}}k}(\theta_{\rm{r}},\phi_{\rm{r}})}{G^{{\rm{t}}k}_{\rm{max}} G^{{\rm{r}}k}_{\rm{max}}}}.
\end{split}
\label{NLSG-Eq25.5}
\end{equation}
By using \eqref{NLSG-Eq03.5} and the array pattern expression in \eqref{NLSG-Eq24}, the pointing error coefficient can be expressed as
\begin{equation}
\begin{split}
h_{{\rm{p}}k} = e^{-\mathcal{T}_{k}},
\end{split}
\label{NLSG-Eq26}
\end{equation}
where $\mathcal{T}_{k} = Z_{{\rm{t}}k} + Z_{{\rm{r}}k} - \ln(g^{{\rm{t}}k}(\theta_{\rm{t}},\phi_{\rm{t}})) - \ln(g^{{\rm{r}}k}(\theta_{\rm{r}},\phi_{\rm{r}}))$. In this formulation $Z_{{\rm{q}}k}$ is written for ${\rm{q}} \in \{{\rm{t}}, {\rm{r}}\}$ as
\begin{equation}
{\small
\begin{split}
Z_{{\rm{q}}k} &= \frac{M_{\rm{qx}}^{2} - 1}{24}(\kappa_{k}d_{\rm{qx}}(\sin\theta \cos\phi - \vartheta_{\rm{jqx}}\cos\theta) + b_{\rm{qx}} - \varepsilon_{\rm{qx}})^{2}  \\
& + \frac{M_{\rm{qy}}^{2} - 1}{24}(\kappa_{k}d_{\rm{qy}}(\sin\theta\sin\phi + \vartheta_{\rm{jqy}}\cos\theta) + b_{\rm{qy}} - \varepsilon_{\rm{qy}})^{2}  \\
& + \frac{\frac{N_{\rm{qx}}^{2}}{M_{\rm{qx}}^{2}} - 1}{24}(\kappa_{k}D_{\rm{qx}}(\sin\theta \cos\phi - \vartheta_{\rm{jqx}}\cos\theta) + B_{\rm{qx}})^{2} \\
& + \frac{\frac{N_{\rm{qy}}^{2}}{M_{\rm{qy}}^{2}} - 1}{24}(\kappa_{k}D_{\rm{qy}}(\sin\theta\sin\phi + \vartheta_{\rm{jqy}}\cos\theta) + B_{\rm{qy}})^{2}.
\end{split}}
\label{NLSG-Eq27}
\end{equation}
With the aid of $D_{\rm{qx}} = M_{\rm{qx}} d_{\rm{qx}}$, $D_{\rm{qy}} = M_{\rm{qy}} d_{\rm{qy}}$, $B_{\rm{qx}} = \frac{\kappa_{k}}{\kappa_{\rm{c}}} M_{\rm{qx}} b_{\rm{qx}}$, and $B_{\rm{qy}} = \frac{\kappa_{k}}{\kappa_{\rm{c}}} M_{\rm{qy}} b_{\rm{qy}}$, \eqref{NLSG-Eq27} can be rewritten after some manipulations as 
\begin{equation}
{\small
\begin{split}
Z_{{\rm{q}}k} &= T_{\rm q1} + T_{\rm q2} \vartheta_{\rm{jqx}}^{2} + T_{\rm q3} \vartheta_{\rm{jqx}} + T_{\rm q4} b_{\rm{qx}}^{2} + T_{\rm q5} b_{\rm{qx}} + T_{\rm q6} \vartheta_{\rm{jqx}} b_{\rm{qx}} \\
&\hspace{-10pt} + T_{\rm q7} + T_{\rm q8} \vartheta_{\rm{jqy}}^{2} + T_{\rm q9} \vartheta_{\rm{jqy}} + T_{\rm q10} b_{\rm{qy}}^{2} + T_{\rm q11} b_{\rm{qy}} + T_{\rm q12} \vartheta_{\rm{jqy}} b_{\rm{qy}} \\
&\hspace{-10pt} + T_{\rm q13} \varepsilon_{\rm{qx}}^{2} + T_{\rm q14} b_{\rm{qx}} \varepsilon_{\rm{qx}} + T_{\rm q15} \varepsilon_{\rm{qx}} + T_{\rm q16} \vartheta_{\rm{jqx}} \varepsilon_{\rm{qx}} \\
&\hspace{-10pt} + T_{\rm q17} \varepsilon_{\rm{qy}}^{2} + T_{\rm q18} b_{\rm{qy}} \varepsilon_{\rm{qy}} + T_{\rm q19} \varepsilon_{\rm{qy}} + T_{\rm q20} \vartheta_{\rm{jqy}} \varepsilon_{\rm{qy}},
\end{split}}
\label{NLSG-Eq28}
\end{equation}
\begin{sloppypar}
\noindent where 
$T_{\rm q1} = \mathcal{A}_{\rm{qx}} \sin^{2}\theta_{\rm{q}} \cos^{2}\phi_{\rm{q}}$,  
$T_{\rm q2} = \mathcal{A}_{\rm{qx}} \cos^{2}\theta_{\rm{q}}$, 
$T_{\rm q3} = - 2 \mathcal{A}_{\rm{qx}} \sin\theta_{\rm{q}} \cos\phi_{\rm{q}} \cos\theta_{\rm{q}}$, 
$T_{\rm q4} = \mathcal{B}_{\rm{qx}}$, 
$T_{\rm q5} = \mathcal{C}_{\rm{qx}} \sin\theta_{\rm{q}} \cos\phi_{\rm{q}}$, 
$T_{\rm q6} = - \mathcal{C}_{\rm{qx}} \cos\theta_{\rm{q}}$, 
$T_{\rm q7} = \mathcal{A}_{\rm{qy}} \sin^{2}\theta_{\rm{q}} \sin^{2}\phi_{\rm{q}}$, 
$T_{\rm q8} = \mathcal{A}_{\rm{qy}} \cos^{2}\theta_{\rm{q}}$, 
$T_{\rm q9} = 2 \mathcal{A}_{\rm{qy}} \sin\theta_{\rm{q}} \sin\phi_{\rm{q}} \cos\theta_{\rm{q}}$, 
$T_{\rm q10} = \mathcal{B}_{\rm{qy}}$, 
$T_{\rm q11} = \mathcal{C}_{\rm{qy}} \sin\theta_{\rm{q}} \sin\phi_{\rm{q}}$, 
$T_{\rm q12} = \mathcal{C}_{\rm{qy}} \cos\theta_{\rm{q}}$, 
$T_{\rm q13} = \frac{M_{\rm{qx}}^{2} - 1}{24}$, 
$T_{\rm q14} = - \frac{M_{\rm{qx}}^{2} - 1}{\textcolor{black}{12}}$, 
$T_{\rm q15} = - \mathcal{D}_{\rm{qx}} \sin\theta_{\rm{q}} \cos\phi_{\rm{q}}$, 
$T_{\rm q16} = \mathcal{D}_{\rm{qx}} \cos\theta_{\rm{q}}$, 
$T_{\rm q17} = \frac{M_{\rm{qy}}^{2} - 1}{24}$, 
$T_{\rm q18} = - \frac{M_{\rm{qy}}^{2} - 1}{\textcolor{black}{12}}$, 
$T_{\rm q19} = - \mathcal{D}_{\rm{qy}} \sin\theta_{\rm{q}} \sin\phi_{\rm{q}}$, 
$T_{\rm q20} = - \mathcal{D}_{\rm{qy}} \cos\theta_{\rm{q}}$, 
$\mathcal{A}_{\rm{qx}} = \left(\frac{N_{\rm{qx}}^{2} - 1}{24}\right) \kappa_{k}^{2} d_{\rm{qx}}^{2}$, 
$\mathcal{B}_{\rm{qx}} = \left({M_{\rm{qx}}^{2} - 1 + (N_{\rm{qx}}^{2} - M_{\rm{qx}}^{2})\frac{\kappa_{k}^{2}}{\kappa_{\rm{c}}^{2}}}\right)\big/{24}$, 
$\mathcal{C}_{\rm{qx}} = \left({M_{\rm{qx}}^{2} - 1 + (N_{\rm{qx}}^{2} - M_{\rm{qx}}^{2})\frac{\kappa_{k}}{\kappa_{\rm{c}}}}\right) \textcolor{black}{\kappa_{k} d_{\rm{qx}}} \big/{12}$, and
$\mathcal{D}_{\rm{qx}} = ({M_{\rm{qx}}^{2} - 1}) \kappa_{k} d_{\rm{qx}}/{12}$. Here,
$\mathcal{A}_{\rm{qy}}$, 
$\mathcal{B}_{\rm{qy}}$, 
$\mathcal{C}_{\rm{qy}}$, and
$\mathcal{D}_{\rm{qy}}$ are defined similar to 
$\mathcal{A}_{\rm{qx}}$, 
$\mathcal{B}_{\rm{qx}}$, 
$\mathcal{C}_{\rm{qx}}$, and
$\mathcal{D}_{\rm{qx}}$, respectively.
\end{sloppypar}

By using the statistics of the mechanical jitters ($\vartheta_{\rm{jqx}}$ and $\vartheta_{\rm{jqy}}$), angle estimation errors through phase adjustments ($b_{\rm{qx}}$, $b_{\rm{qy}}$, $B_{\rm{qx}}$, and $B_{\rm{qy}}$), and quantization errors ($\varepsilon_{\rm{qx}}$ and $\varepsilon_{\rm{qy}}$), the statistics of the pointing error $h_{{\rm{p}}k}$ can be derived.

\subsection{Statistical Characterization of the Pointing Error}

The pointing error in a wideband directional beamforming system can be characterized by obtaining the statistics of $\mathcal{T}_{k}' = Z_{{\rm{t}}k} + Z_{{\rm{r}}k}$, $\mathcal{T}_{k} = \mathcal{T}_{k}' - \ln(g^{{\rm{t}}k}(\theta_{\rm{t}},\phi_{\rm{t}})) - \ln(g^{{\rm{r}}k}(\theta_{\rm{r}},\phi_{\rm{r}}))$, and then $h_{{\rm{p}}k}$. 

First, it can be observed from \eqref{NLSG-Eq27} that all the summation terms in $Z_{{\rm{q}}k}$ are squared, making $Z_{{\rm{q}}k}$ and $\mathcal{T}_{k}'$ positive. Hence, \textcolor{black}{utilizing the moment-matching principle of the Welch-Satterthwaite approach \cite{satterthwaite1946approximate,welch1947generalization}}, $\mathcal{T}_{k}'$ can be modeled as a Gamma distributed random variable, and its CDF can be expressed as
\begin{equation}
\begin{split}
F_{\mathcal{T}_{k}'}(x) &= \frac{1}{\Gamma(\varsigma_{k})}\gamma\left(\varsigma_{k},\frac{x}{\nu_{k}}\right),
\label{NLSG-Eq28.5}
\end{split}
\end{equation}
with the shape parameter
\begin{equation}
\begin{split}
\varsigma_{k} = \frac{E^{2}[\mathcal{T}_{k}']}{\text{Var}(\mathcal{T}_{k}')} \overset{({\rm a})}{=} \frac{\left(E[Z_{{\rm{t}}k}] + E[Z_{{\rm{r}}k}]\right)^{2}}{\text{Var}(Z_{{\rm{t}}k}) + \text{Var}(Z_{{\rm{r}}k})}
\label{NLSG-Eq28.6}
\end{split}
\end{equation}
and the scale parameter
\begin{equation}
\begin{split}
\nu_{k} = \frac{\text{Var}(\mathcal{T}_{k}')}{E[\mathcal{T}_{k}']} \overset{({\rm b})}{=} \frac{\text{Var}(Z_{{\rm{t}}k}) + \text{Var}(Z_{{\rm{r}}k})}{E[Z_{{\rm{t}}k}] + E[Z_{{\rm{r}}k}]}.
\label{NLSG-Eq28.7}
\end{split}
\end{equation}
In this formulation, equations (a) and (b) in \eqref{NLSG-Eq28.6} and \eqref{NLSG-Eq28.7}, respectively, hold since $Z_{{\rm{t}}k}$ and $Z_{{\rm{r}}k}$ are statistically independent as they are related to the transmitter and receiver, respectively.\footnote{Here, $E[Z_{{\rm{q}}k}]$ and $\text{Var}(Z_{{\rm{q}}k})$ must be evaluated to obtain the shape parameter $\varsigma_{k}$ and scale parameter $\nu_{k}$, which is performed in the next subsection.}

From the statistics of $\mathcal{T}_{k}'$, it can be readily said that $\mathcal{T}_{k} = \mathcal{T}_{k}' - \ln(g^{{\rm{t}}k}(\theta_{\rm{t}},\phi_{\rm{t}})) - \ln(g^{{\rm{r}}k}(\theta_{\rm{r}},\phi_{\rm{r}}))$ is shifted Gamma distributed with the following CDF:
\begin{equation}
\begin{split}
F_{\mathcal{T}_{k}}(x) &= \frac{1}{\Gamma(\varsigma_{k})}\gamma\left(\varsigma_{k},\frac{x-\rho_{k}}{\nu_{k}}\right),
\label{NLSG-Eq46}
\end{split}
\end{equation}
where $\rho_{k} = - \ln(g^{{\rm{t}}k}(\theta_{\rm{t}},\phi_{\rm{t}})) - \ln(g^{{\rm{r}}k}(\theta_{\rm{r}},\phi_{\rm{r}}))$ is the location parameter and is related to the scanning loss due to the impact of antenna element in beamforming systems.

Finally, the pointing error coefficient $h_{{\rm{p}}k} = e^{-\mathcal{T}_{k}}$ follows the NLSG distribution with the shape parameter $\varsigma_{k}$, the scale parameter $\nu_{k}$, and the location parameter $\rho_{k}$, and its CDF can be derived by using \eqref{NLSG-Eq46} as 
\begin{equation}
\begin{split}
F_{h_{{\rm{p}}k}}(x) &= 
\Pr[e^{-\mathcal{T}_{k}} \leq x] = 1 - F_{\mathcal{T}_{k}}(- \ln x) \\
&= 1 - \frac{1}{\Gamma(\varsigma_{k})}\gamma\left(\varsigma_{k},\frac{- \ln x-\rho_{k}}{\nu_{k}}\right),
\label{NLSG-Eq47}
\end{split}
\end{equation}
\textcolor{black}{for $x \in [0, e^{-\rho_{k}}]$}. By using the relationship between the upper and lower incomplete Gamma functions, i.e. $\Gamma(z) = \gamma(z,x) + \Gamma(z,x)$ \cite[Eq. 8.356.3]{gradshteyn2014table}, the CDF of $h_{{\rm{p}}k}$ can be rewritten as 
\begin{equation}
\begin{split}
F_{h_{{\rm{p}}k}}(x) &= \frac{1}{\Gamma(\varsigma_{k})}\Gamma\left(\varsigma_{k},\frac{- \ln x-\rho_{k}}{\nu_{k}}\right),
\label{NLSG-Eq48}
\end{split}
\end{equation}
\textcolor{black}{for $x \in [0, e^{-\rho_{k}}]$}. By taking the first derivative of the above-equation with the aid of \cite[Eq. 45.10.1]{oldham2009atlas}, the PDF of $h_{{\rm{p}}k}$ can be obtained as
\begin{equation}
\begin{split}
f_{h_{{\rm{p}}k}}(x) &= \frac{1}{\Gamma(\varsigma_{k}) \nu_{k}^{\varsigma_{k}}} (- \ln x-\rho_{k})^{\varsigma_{k} - 1} e^{\rho_{k}/\nu_{k}} x^{\frac{1}{\nu_{k}}-1}.
\label{NLSG-Eq49}
\end{split}
\end{equation}
Moreover, the $r$-th moment of $h_{{\rm{p}}k}$ can be given as
\begin{equation}
\begin{split}
E[h_{{\rm{p}}k}^{r}] &= E[e^{- \mathcal{T}_{k} r}] = E[e^{- \mathcal{T}_{k}' r}] e^{-\rho_{k}r},
\label{NLSG-Eq50}
\end{split}
\end{equation}
where $E[e^{- \mathcal{T}_{k}' r}]$ is simply the MGF of the Gamma distributed random variable $\mathcal{T}_{k}'$, i.e. $\mathcal{M}_{\mathcal{T}_{k}'}(\varrho) = E[e^{- \mathcal{T}_{k}' \varrho}] = (1 + \nu_{k}\varrho)^{-\varsigma_{k}}$ for $\textcolor{black}{\varrho > -\frac{1}{\nu_{k}}}$, with its argument $\varrho = r$. Hence, the $r$-th moment of $h_{{\rm{p}}k}$ can be found as 
\begin{equation}
\begin{split}
E[h_{{\rm{p}}k}^{r}] &= (1 + \nu_{k} r)^{-\varsigma_{k}} e^{-\rho_{k}r}.
\label{NLSG-Eq52}
\end{split}
\end{equation}

\begin{figure}[!t]
    \centering
    \includegraphics[width=1\linewidth]{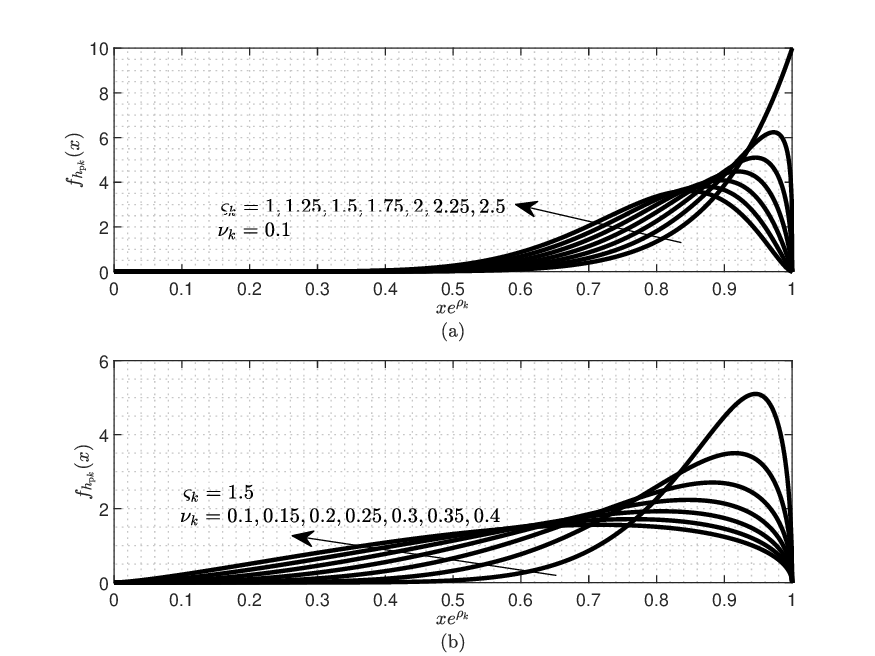}
    \vspace{-20pt}
    \caption{The PDF of NLSG distribution for different values of (a) the shape parameter and (b) the scale parameter.}
    \label{fig000}
\end{figure}

{\textit{Remark 4:} It is crucial to note that for all scenarios that can be covered by the NLSG model, the pointing error can be fully characterized by only the three distribution parameters, i.e. the shape, scale, and location parameters. Considering the PDF expression in \eqref{NLSG-Eq49} which involves elementary functions, such as the logarithm and the power function, the mathematical interpretation and manipulation of the NLSG model is exceptionally convenient, rendering it a highly practical and scalable tool for the system design and performance analysis in diverse communication scenarios. Additionally, in Fig. \ref{fig000}, the PDF of the NLSG distribution is shown for different values of the shape and scale parameters. It can be seen that as either of the parameters increase, the PDF leans towards left, meaning that the pointing error coefficient is more likely to have lower values. This reflects a clear insight about the relationship between the distribution parameters and the pointing error characteristics: the higher the values of the distribution parameters, the poorer the pointing accuracy.\footnote{Note that the shape and scale parameters are strictly positive ($\varsigma_{k} > 0$ and $\nu_{k} > 0$), whereas the location parameter is non-negative ($\rho_{k} \geq 0$). However, for large values of the product $\varsigma_{k}\nu_{k}$, the PDF of the NLSG distribution exhibits a pronounced leftward skew, concentrating near zero. This condition corresponds to the \quotes{beam alignment failure} rather than beam misalignment. Since the focus of this study is to characterize the pointing error coefficient specifically for the beam misalignment case, large values of the product $\varsigma_{k}\nu_{k}$ are excluded from Fig. \ref{fig000}.} }

The closed-form expressions for the CDF, PDF, and moments of $h_{{\rm{p}}k}$ derived above constitute the core of the proposed pointing error model. Unlike currently existing models that often rely on restrictive assumptions, such as boresight radiation or symmetric Gaussian jitter, the proposed NLSG model captures the cumulative impact of hardware impairments, beamforming architectures, and steering dynamics within a unified statistical characterization.

\subsection{Derivation of the Distribution Parameters under Distributional Assumptions}

Although the proposed pointing error model is valid for arbitrary distributions, obtaining tractable analytical expressions for the distribution parameters through $E[Z_{{\rm{q}}k}]$ and $\text{Var}(Z_{{\rm{q}}k})$ by using \eqref{NLSG-Eq28.6} and \eqref{NLSG-Eq28.7} requires specific distributional assumptions. Therefore, following common practice in the literature, statistically independent Gaussian distributed mechanical jitters and angle estimation errors are assumed, i.e. $\vartheta_{\rm{jqx}} \sim \mathcal{N}(0,\sigma_{\rm{jqx}}^{2})$, $\vartheta_{\rm{jqy}} \sim \mathcal{N}(0,\sigma_{\rm{jqy}}^{2})$, $\vartheta_{\rm{qP}} \sim \mathcal{N}(0,\sigma_{\rm{qP}}^{2})$, and $\vartheta_{{\rm{qP}}_\perp} \sim \mathcal{N}(0,\sigma_{{\rm{qP}}_\perp}^{2})$. 
Here, $\sigma_{\rm{jqx}}^{2}$ and $\sigma_{\rm{jqy}}^{2}$ denote the jitter variances on the x-z plane and y-z plane for ${\rm{q}} \in \{ {\rm{t}}, {\rm{r}} \}$, respectively. Moreover, $\sigma_{\rm{qP}}^{2}$ and $\sigma_{\rm{qP}_\perp}^{2}$ show the angle estimation error variances on $\rm{P}$ plane and $\rm{P}_\perp$ plane for ${\rm{q}} \in \{ {\rm{t}}, {\rm{r}} \}$, respectively. As can be deduced from \eqref{NLSG-Eq-20}, the phase excitations follow Gaussian distribution, i.e. 
$b_{\rm{qx}} \sim \mathcal{N}(\mu_{b_{\rm{qx}}},\sigma^{2}_{b_{\rm{qx}}})$, 
$b_{\rm{qy}} \sim \mathcal{N}(\mu_{b_{\rm{qy}}},\sigma^{2}_{b_{\rm{qy}}})$, 
$B_{\rm{qx}} \sim \mathcal{N}(\mu_{B_{\rm{qx}}},\sigma^{2}_{B_{\rm{qx}}})$, and 
$B_{\rm{qy}} \sim \mathcal{N}(\mu_{B_{\rm{qy}}},\sigma^{2}_{B_{\rm{qy}}})$, with means
\begin{equation}
\begin{split}
\mu_{b_{\rm{qx}}} &= - \kappa_{\rm{c}} d_{\rm{qx}} \sin \theta_{\rm{q}} \cos \phi_{\rm{q}}, \\
\mu_{b_{\rm{qy}}} &= - \kappa_{\rm{c}} d_{\rm{qy}} \sin \theta_{\rm{q}} \sin \phi_{\rm{q}}, \\
\mu_{B_{\rm{qx}}} &= - \kappa_{k} D_{\rm{qx}} \sin \theta_{\rm{q}} \cos \phi_{\rm{q}}, \\
\mu_{B_{\rm{qy}}} &= - \kappa_{k} D_{\rm{qy}} \sin \theta_{\rm{q}} \sin \phi_{\rm{q}},
\end{split}
\label{NLSG-Eq-21}
\end{equation}
and variances
\begin{equation}
\begin{split}
\sigma^{2}_{b_{\rm{qx}}} &= \kappa_{\rm{c}}^{2} d_{\rm{qx}}^{2} (\sigma^{2}_{\rm{qP}} \cos^{2} \theta_{\rm{q}} \cos^{2} \phi_{\rm{q}} + \sigma^{2}_{{\rm{qP}}_\perp} \sin^{2} \phi_{\rm{q}}), \\
\sigma^{2}_{b_{\rm{qy}}} &= \kappa_{\rm{c}}^{2} d_{\rm{qy}}^{2} (\sigma^{2}_{\rm{qP}} \cos^{2} \theta_{\rm{q}} \sin^{2} \phi_{\rm{q}} + \sigma^{2}_{{\rm{qP}}_\perp} \cos^{2} \phi_{\rm{q}}), \\
\sigma^{2}_{B_{\rm{qx}}} &= \kappa_{k}^{2} D_{\rm{qx}}^{2} (\sigma^{2}_{\rm{qP}} \cos^{2} \theta_{\rm{q}} \cos^{2} \phi_{\rm{q}} + \sigma^{2}_{{\rm{qP}}_\perp} \sin^{2} \phi_{\rm{q}}), \\
\sigma^{2}_{B_{\rm{qy}}} &= \kappa_{k}^{2} D_{\rm{qy}}^{2} (\sigma^{2}_{\rm{qP}} \cos^{2} \theta_{\rm{q}} \sin^{2} \phi_{\rm{q}} + \sigma^{2}_{{\rm{qP}}_\perp} \cos^{2} \phi_{\rm{q}}).
\end{split}
\label{NLSG-Eq-22}
\end{equation}
Note here that the continuous phase shift values $b_{\rm{qx}}$, $b_{\rm{qy}}$, $B_{\rm{qx}}$, and $B_{\rm{qy}}$ are statistically dependent. Recalling that $\varepsilon_{\rm{qx}} \sim \text{Uniform}\left(-{\Delta_{\rm{qx}}}/{2}, {\Delta_{\rm{qx}}}/{2}\right)$ and $\varepsilon_{\rm{qy}} \sim \text{Uniform}\left(-{\Delta_{\rm{qy}}}/{2}, {\Delta_{\rm{qy}}}/{2}\right)$, the expected value of $Z_{{\rm{q}}k}$ for ${\rm{q}} \in \{{\rm{t}}, {\rm{r}}\}$ can be easily obtained by the linearity property of the expectation operator as
\begin{equation}
\begin{split}
E[Z_{{\rm{q}}k}] &= T_{\rm q1} + T_{\rm q2} \sigma_{\rm{jqx}}^{2} + T_{\rm q4} (\mu_{b_{\rm{qx}}}^{2} + \sigma_{b_{\rm{qx}}}^{2}) + T_{\rm q5} \mu_{b_{\rm{qx}}} \\
&+ T_{\rm q7} + T_{\rm q8} \sigma_{\rm{jqy}}^{2} + T_{\rm q10} (\mu_{b_{\rm{qy}}}^{2} + \sigma_{b_{\rm{qy}}}^{2}) + T_{\rm q11} \mu_{b_{\rm{qy}}} \\
&+ T_{\rm q13} \frac{\Delta_{\rm{qx}}^{2}}{12} + T_{\rm q17} \frac{\Delta_{\rm{qy}}^{2}}{12}.
\end{split}
\label{NLSG-Eq29}
\end{equation}
To evaluate the variance of $Z_{{\rm{q}}k}$, the law of total variance can be used as
\begin{equation}
\begin{split}
\text{Var}(Z_{{\rm{q}}k}) = \text{Var}(E[Z_{{\rm{q}}k} | b_{\rm{qx}}, b_{\rm{qy}}]) + E[\text{Var}(Z_{{\rm{q}}k} | b_{\rm{qx}}, b_{\rm{qy}})].
\end{split}
\label{NLSG-Eq30}
\end{equation}
Here, $E[Z_{{\rm{q}}k} | b_{\rm{qx}}, b_{\rm{qy}}]$ can be written by using the linearity of the expectation operator and the moments of $\vartheta_{\rm{jqx}}$ and $\vartheta_{\rm{jqy}}$ as
\begin{equation}
\begin{split}
E[Z_{{\rm{q}}k} | b_{\rm{qx}}, b_{\rm{qy}}] &= T_{\rm q1} + T_{\rm q2} \sigma_{\rm{jqx}}^{2} + T_{\rm q4} b_{\rm{qx}}^{2} + T_{\rm q5} b_{\rm{qx}} \\
&+ T_{\rm q7} + T_{\rm q8} \sigma_{\rm{jqy}}^{2} + T_{\rm q10} b_{\rm{qy}}^{2} + T_{\rm q11} b_{\rm{qy}} \\
&+ T_{\rm q13} \frac{\Delta_{\rm{qx}}^{2}}{12} + T_{\rm q17} \frac{\Delta_{\rm{qy}}^{2}}{12}.
\end{split}
\label{NLSG-Eq32}
\end{equation}
Hence, $\text{Var}(E[Z_{{\rm{q}}k} | b_{\rm{qx}}, b_{\rm{qy}}])$ yields
\begin{equation}
{\footnotesize
\begin{split}
\text{Var}(E[Z_{{\rm{q}}k} | b_{\rm{qx}}, b_{\rm{qy}}]) &= T_{\rm q4}^{2} \text{Var}(b_{\rm{qx}}^{2}) + T_{\rm q5}^{2} \text{Var}(b_{\rm{qx}}) + T_{\rm q10}^{2} \text{Var}(b_{\rm{qy}}^{2}) \\
&\hspace{-20pt} + T_{\rm q11}^{2} \text{Var}(b_{\rm{qy}}) + 2 T_{\rm q4} T_{\rm q5}  \text{Cov}(b_{\rm{qx}}^{2},b_{\rm{qx}}) \\
&\hspace{-20pt} + 2 T_{\rm q4} T_{\rm q10}  \text{Cov}(b_{\rm{qx}}^{2},b_{\rm{qy}}^{2}) + 2 T_{\rm q4} T_{\rm q11} \text{Cov}(b_{\rm{qx}}^{2},b_{\rm{qy}}) \\
&\hspace{-20pt} + 2 T_{\rm q5} T_{\rm q10}  \text{Cov}(b_{\rm{qx}},b_{\rm{qy}}^{2}) + 2 T_{\rm q5} T_{\rm q11} \text{Cov}(b_{\rm{qx}},b_{\rm{qy}})     \\
&\hspace{-20pt} + 2 T_{\rm q10} T_{\rm q11} \text{Cov}(b_{\rm{qy}}^{2},b_{\rm{qy}}),
\end{split}}
\label{NLSG-Eq33}
\end{equation}
where $\text{Var}(b_{\rm{qx}})$ and $\text{Var}(b_{\rm{qy}})$ can be found in \eqref{NLSG-Eq-22}, and the following lemma can be used to evaluate the other variance and covariance terms in the above-equation.

\textit{Lemma 1:} The continuous phase adjustments $b_{\rm{qx}} = \zeta_{\rm{qx}}\vartheta_{\rm{qP}} + \eta_{\rm{qx}}\vartheta_{\rm{qP}_\perp} + \mu_{b_{\rm{qx}}}$ and $b_{\rm{qy}} = \zeta_{\rm{qy}}\vartheta_{\rm{qP}} + \eta_{\rm{qy}}\vartheta_{\rm{qP}_\perp} + \mu_{b_{\rm{qy}}}$ are statistically dependent with the following statistics:
\begin{subequations}
{\small
\begin{align}
\text{Var}(b_{\rm{qx}}^{2}) &= 4 \mu_{b_{\rm{qx}}}^{2}\sigma_{b_{\rm{qx}}}^{2} + 2\sigma_{b_{\rm{qx}}}^{4}, \label{NLSG-Eq34a}\\
\text{Cov}(b_{\rm{qx}}^{2},b_{\rm{qx}}) &= 2\mu_{b_{\rm{qx}}}\sigma_{b_{\rm{qx}}}^{2}, \label{NLSG-Eq34b}\\
\text{Cov}(b_{\rm{qx}},b_{\rm{qy}}) &= \zeta_{\rm{qx}} \zeta_{\rm{qy}} \sigma_{\rm{qP}}^{2} + \eta_{\rm{qx}} \eta_{\rm{qy}} \sigma_{\rm{qP}_\perp}^{2}, \label{NLSG-Eq34c}\\
\text{Cov}(b_{\rm{qx}}^{2},b_{\rm{qy}}) &= 2 \zeta_{\rm{qx}} \mu_{b_{\rm{qx}}} \zeta_{\rm{qy}} \sigma_{\rm{qP}}^{2} + 2 \eta_{\rm{qx}} \mu_{b_{\rm{qx}}} \eta_{\rm{qy}} \sigma_{\rm{qP}_\perp}^{2}, \label{NLSG-Eq34d}\\
\begin{split}
\text{Cov}(b_{\rm{qx}}^{2},b_{\rm{qy}}^{2}) &= 2 \zeta_{\rm{qx}}^{2} \zeta_{\rm{qy}}^{2} \sigma_{\rm{qP}}^{4} + 2 \eta_{\rm{qx}}^{2} \eta_{\rm{qy}}^{2} \sigma_{\rm{qP}_\perp}^{4} \\
&+ 4 \zeta_{\rm{qx}} \mu_{b_{\rm{qx}}} \zeta_{\rm{qy}} \mu_{b_{\rm{qy}}} \sigma_{\rm{qP}}^{2} + 4 \eta_{\rm{qx}} \mu_{b_{\rm{qx}}} \eta_{\rm{qy}} \mu_{b_{\rm{qy}}} \sigma_{\rm{qP}_\perp}^{2}\\
&+ \textcolor{black}{4 \zeta_{\rm{qx}} \eta_{\rm{qx}} \zeta_{\rm{qy}} \eta_{\rm{qy}} \sigma_{\rm{qP}}^{2} \sigma_{\rm{qP}_\perp}^{2}}, \label{NLSG-Eq34e}
\end{split}
\end{align}}
\label{NLSG-Eq34}
\end{subequations}

\vspace{-10pt}
\noindent where 
$\zeta_{\rm{qx}} = - \kappa_{\rm{c}} d_{\rm{qx}} \cos \theta_{\rm{q}} \cos \phi_{\rm{q}}$, 
$\eta_{\rm{qx}} = \kappa_{\rm{c}} d_{\rm{qx}} \sin \phi_{\rm{q}}$, 
$\mu_{b_{\rm{qx}}} = - \kappa_{\rm{c}} d_{\rm{qx}} \sin \theta_{\rm{q}} \cos \phi_{\rm{q}}$, 
$\zeta_{\rm{qy}} = - \kappa_{\rm{c}} d_{\rm{qy}} \cos \theta_{\rm{q}} \sin \phi_{\rm{q}}$, 
$\eta_{\rm{qy}} = - \kappa_{\rm{c}} d_{\rm{qy}} \cos \phi_{\rm{q}}$, and
$\mu_{b_{\rm{qy}}} = - \kappa_{\rm{c}} d_{\rm{qy}} \sin \theta_{\rm{q}} \sin \phi_{\rm{q}}$ for brevity. Moreover, $\text{Var}(b_{\rm{qy}}^{2})$, $\text{Cov}(b_{\rm{qy}}^{2},b_{\rm{qy}})$, and $\text{Cov}(b_{\rm{qx}},b_{\rm{qy}}^{2})$ can be found similar to
\eqref{NLSG-Eq34a}, \eqref{NLSG-Eq34b}, and \eqref{NLSG-Eq34d}, respectively. 

\textit{Proof:} Please see Appendix B. \hfill $\blacksquare$

\begin{figure*}[!b]
\vspace*{0pt} 
\noindent\rule{\textwidth}{0.4pt} 
\begin{equation}
{\small
\begin{split}
\text{Var}(E[Z_{{\rm{q}}k} | b_{\rm{qx}} , b_{\rm{qy}}]) &=  
{2 T_{\rm q4}^{2} \kappa_{\rm{c}}^{2} d_{\rm{qx}}^{2} \left[ 2 \mu_{b_{\rm{qx}}}^{2} (\cos^{2}\theta_{\rm{q}}\cos^{2}\phi_{\rm{q}} \sigma_{\rm{qP}}^{2} + \sin^{2}\phi_{\rm{q}} \sigma_{\rm{qP}_\perp}^{2}) + \kappa_{\rm{c}}^{2} d_{\rm{qx}}^{2}(\cos^{2}\theta_{\rm{q}} \cos^{2}\phi_{\rm{q}} \sigma_{\rm{qP}}^{2} + \sin^{2}\phi_{\rm{q}} \sigma_{\rm{qP}_\perp}^{2})^{2} \right]} \\ 
&\hspace{-55pt} + T_{\rm q5}^{2} \kappa_{\rm{c}}^{2} d_{\rm{qx}}^{2} \left(\cos^{2}\theta_{\rm{q}}\cos^{2}\phi_{\rm{q}} \sigma_{\rm{qP}}^{2} + \sin^{2}\phi_{\rm{q}} \sigma_{\rm{qP}_\perp}^{2}\right) + T_{\rm q11}^{2} \kappa_{\rm{c}}^{2} d_{\rm{qy}}^{2} \left(\cos^{2}\theta_{\rm{q}} \sin^{2}\phi_{\rm{q}} \sigma_{\rm{qP}}^{2} + \cos^{2}\phi_{\rm{q}} \sigma_{\rm{qP}_\perp}^{2}\right) \\ 
&\hspace{-55pt} + {2 T_{\rm q10}^{2} \kappa_{\rm{c}}^{2} d_{\rm{qy}}^{2} \left[ 2 \mu_{b_{\rm{qy}}}^{2} (\cos^{2}\theta_{\rm{q}} \sin^{2}\phi_{\rm{q}} \sigma_{\rm{qP}}^{2} + \cos^{2}\phi_{\rm{q}} \sigma_{\rm{qP}_\perp}^{2}) + \kappa_{\rm{c}}^{2} d_{\rm{qy}}^{2} (\cos^{2}\theta_{\rm{q}} \sin^{2}\phi_{\rm{q}} \sigma_{\rm{qP}}^{2} + \cos^{2}\phi_{\rm{q}} \sigma_{\rm{qP}_\perp}^{2})^{2} \right]} \\ 
&\hspace{-55pt} + {4 {T_{\rm q4} T_{\rm q5}} \kappa_{\rm{c}}^{2} d_{\rm{qx}}^{2} \mu_{b_{\rm{qx}}} \left(\cos^{2}\theta_{\rm{q}}\cos^{2}\phi_{\rm{q}} \sigma_{\rm{qP}}^{2} + \sin^{2}\phi_{\rm{q}} \sigma_{\rm{qP}_\perp}^{2}\right)}   
+ {4 T_{\rm q10} T_{\rm q11} \kappa_{\rm{c}}^{2} d_{\rm{qy}}^{2}  \mu_{b_{\rm{qy}}}  \left(\cos^{2}\theta_{\rm{q}} \sin^{2}\phi_{\rm{q}}\sigma_{\rm{qP}}^{2} + \cos^{2}\phi_{\rm{q}} \sigma_{\rm{qP}_\perp}^{2}\right)} \\ 
&\hspace{-55pt} + 2 T_{\rm q5} T_{\rm q11} \kappa_{\rm{c}}^{2} d_{\rm{qx}} d_{\rm{qy}} \cos\phi_{\rm{q}} \sin\phi_{\rm{q}} (\cos^{2}\theta_{\rm{q}} \sigma_{\rm{qP}}^{2} - \sigma_{\rm{qP}_\perp}^{2} )   
+ {4 T_{\rm q4} T_{\rm q11} \kappa_{\rm{c}}^{2} d_{\rm{qx}} d_{\rm{qy}} \mu_{b_{\rm{qx}}} \cos\phi_{\rm{q}} \sin\phi_{\rm{q}} (\cos^{2}\theta_{\rm{q}} \sigma_{\rm{qP}}^{2} - \sigma_{\rm{qP}_\perp}^{2})} \\ 
&\hspace{-55pt} + {4 T_{\rm q5} T_{\rm q10} \kappa_{\rm{c}}^{2} d_{\rm{qx}} d_{\rm{qy}} \mu_{b_{\rm{qy}}} \cos\phi_{\rm{q}} \sin\phi_{\rm{q}} (\cos^{2}\theta_{\rm{q}}  \sigma_{\rm{qP}}^{2} - \sigma_{\rm{qP}_\perp}^{2})}   
 + 8 T_{\rm q4} T_{\rm q10} \kappa_{\rm{c}}^{2} d_{\rm{qx}} d_{\rm{qy}} \mu_{b_{\rm{qx}}} \mu_{b_{\rm{qy}}} \cos\phi_{\rm{q}} \sin\phi_{\rm{q}} \left( {\cos^{2}\theta_{\rm{q}} \sigma_{\rm{qP}}^{2} - \sigma_{\rm{qP}_\perp}^{2}} \right) \\ 
&\hspace{-55pt} + 4 T_{\rm q4} T_{\rm q10} \kappa_{\rm{c}}^{4} d_{\rm{qx}}^{2} d_{\rm{qy}}^{2} \cos^{2}\phi_{\rm{q}} \sin^{2}\phi_{\rm{q}} \left( \cos^{4}\theta_{\rm{q}} \sigma_{\rm{qP}}^{4} + \sigma_{\rm{qP}_\perp}^{4} - 2 \cos^{2}\theta_{\rm{q}}  \sigma_{\rm{qP}}^{2}\sigma_{\rm{qP}_\perp}^{2} \right).
\end{split}}
\label{NLSG-Eq38}
\end{equation}
\end{figure*}

By using \eqref{NLSG-Eq-21}, \eqref{NLSG-Eq-22}, \eqref{NLSG-Eq33}, and Lemma 1, $\text{Var}(E[Z_{{\rm{q}}k} | b_{\rm{qx}} , b_{\rm{qy}}])$ can be found as in \eqref{NLSG-Eq38} at \textcolor{black}{the bottom of the this page}.

For the second term in \eqref{NLSG-Eq30}, $\text{Var}(Z_{{\rm{q}}k} | b_{\rm{qx}}, b_{\rm{qy}})$ can be written as 
\begin{equation}
{\footnotesize
\begin{split}
\text{Var}(Z_{{\rm{q}}k} | b_{\rm{qx}}, b_{\rm{qy}}) &= \sum_{i = 1}^{20} C_{i}^{2} \text{Var}(V_{i}) + 2 \sum_{1 \leq i < j \leq 20} C_{i} C_{j}\text{Cov}(V_{i},V_{j}), 
\end{split}}
\label{NLSG-Eq39}
\end{equation}
where 
$V_{1} = V_{4} = V_{5} = V_{7} = V_{10} = V_{11} = 1$, 
$V_{2} = \vartheta_{\rm{jqx}}^{2}$, 
$V_{3} = V_{6} = \vartheta_{\rm{jqx}}$, 
$V_{8} = \vartheta_{\rm{jqy}}^{2}$, 
$V_{9} = V_{12} = \vartheta_{\rm{jqy}}$, 
$V_{13} = \varepsilon_{\rm{qx}}^{2}$, 
$V_{14} = V_{15} = \varepsilon_{\rm{qx}}$, 
$V_{16} = \vartheta_{\rm{jqx}} \varepsilon_{\rm{qx}}$, 
$V_{17} = \varepsilon_{\rm{qy}}^{2}$, 
$V_{18} = V_{19} = \varepsilon_{\rm{qy}}$, 
$V_{20} = \vartheta_{\rm{jqy}} \varepsilon_{\rm{qy}}$, 
and 
$C_{i} = T_{{\rm{q}}i}$ for all $i$ except for $C_{4} = T_{\rm q4} b_{\rm{qx}}^{2}$, $C_{5} = T_{\rm q5} b_{\rm{qx}}$, $C_{6} = T_{\rm q6} b_{\rm{qx}}$, $C_{10} = T_{\rm q10} b_{\rm{qy}}^{2}$, $C_{11} = T_{\rm q11} b_{\rm{qy}}$, $C_{12} = T_{\rm q12} b_{\rm{qy}}$, $C_{14} = T_{\rm q14} b_{\rm{qx}}$, and $C_{18} = T_{\rm q18} b_{\rm{qy}}$ for notational brevity. Note here that the terms $V_{i} = 1$ for $i \in \{1,4,5,7,10,11\}$ are deterministic, resulting in $\text{Var}(V_{i}) = 0$ and $\text{Cov}(V_{i},V_{j}) = 0$ for $i \neq j$. Furthermore, $\text{Cov}(V_{i},V_{j})$ is non-zero only when the variables $V_{i}$ and $V_{j}$ share common underlying variables, i.e. $\vartheta_{\rm{jqx}}$, $\vartheta_{\rm{jqy}}$, $\varepsilon_{\rm{qx}}$, or $\varepsilon_{\rm{qy}}$. For the non-zero variance and covariance terms, the following lemma can be used.

\textit{Lemma 2:} Let $\mathcal{W}_{1} = \mathcal{X}^{\tau_{1}} \epsilon^{\varpi_{1}}$ and $\mathcal{W}_{2} = \mathcal{X}^{\tau_{2}} \epsilon^{\varpi_{2}}$ denote two random variables, where $\mathcal{X} \sim \mathcal{N}(0,\sigma_{\mathcal{X}}^{2})$ is a zero-mean Gaussian random variable with variance $\sigma_{\mathcal{X}}^{2}$, and $\epsilon \sim \text{Uniform}(-{\Delta}/{2},{\Delta}/{2})$ is uniformly distributed in the interval $[-{\Delta}/{2},{\Delta}/{2}]$. The variance $\text{Var}(\mathcal{W}_{1})$ and the covariance $\text{Cov}(\mathcal{W}_{1},\mathcal{W}_{2})$ are given by \eqref{NLSG-Eq40} and \eqref{NLSG-Eq41}, respectively, at \textcolor{black}{the top of the next page}, while $\text{Var}(\mathcal{W}_{2})$ can be written similar to $\eqref{NLSG-Eq40}$.
\begin{figure*}[!t]
\begin{equation}
{
\begin{split}
\text{Var}(\mathcal{W}_{1}) &= \sigma_{\mathcal{X}}^{2\tau_{1}} (2\tau_{1}-1)!! \frac{(\Delta/2)^{2\varpi_{1}}}{2\varpi_{1} + 1} - 
\begin{cases} 
\sigma_{\mathcal{X}}^{2\tau_{1}} ((\tau_{1}-1)!!)^{2} \frac{(\Delta/2)^{2\varpi_{1}}}{(\varpi_{1} + 1)^{2}}, & \text{both $\tau_{1}$ and $\varpi_{1}$ are even} \\        
0, & \text{otherwise} \\ 
\end{cases}.
\end{split}}
\label{NLSG-Eq40}
\end{equation}
\begin{equation}
{
\begin{split}
\text{Cov}(\mathcal{W}_{1},\mathcal{W}_{2}) = 
\begin{cases} 
      \sigma_{\mathcal{X}}^{\tau_{1} + \tau_{2}} (\tau_{1} + \tau_{2} - 1)!! \frac{(\Delta/2)^{\varpi_{1} + \varpi_{2}}}{\varpi_{1} + \varpi_{2} + 1} , & \text{$\tau_{1}$, $\tau_{2}$, $\varpi_{1}$, and $\varpi_{2}$ are odd} \\
      \sigma_{\mathcal{X}}^{\tau_{1} + \tau_{2}}(\Delta/2)^{\varpi_{1} + \varpi_{2}} \left(\frac{(\tau_{1} + \tau_{2} - 1)!!}{(\varpi_{1} + \varpi_{2} + 1)} - \frac{(\tau_{1} - 1)!!(\tau_{2} - 1)!!}{(\varpi_{1} + 1)(\varpi_{2} + 1)}\right), & \text{$\tau_{1}$, $\tau_{2}$, $\varpi_{1}$, and $\varpi_{2}$ are even} \\
      0, & \text{otherwise}
   \end{cases}.
\end{split}}
\label{NLSG-Eq41}
\end{equation}
\end{figure*}

\textit{Proof:} The variance of $\mathcal{W}_{1} = \mathcal{X}^{\tau_{1}}\epsilon^{\varpi_{1}}$ is expressed as
\begin{equation}
\begin{split}
\text{Var}(\mathcal{W}_{1}) &= E[\mathcal{X}^{2\tau_{1}}]E[\epsilon^{2\varpi_{1}}] - E^{2}[\mathcal{X}^{\tau_{1}}]E^{2}[\epsilon^{\varpi_{1}}].
\end{split}
\label{NLSG-Eq42}
\end{equation}
Since the distributions of $\mathcal{X}$ and $\epsilon$ are symmetric around zero, their odd moments vanish. Consequently, non-zero contributions occur only for even moments, i.e. $E[\mathcal{X}^{u}] = \sigma_{\mathcal{X}}^{u}(u-1)!!$ and $E[\epsilon^{u}] = \frac{(\Delta/2)^{u}}{u+1}$ for even $u$. Thus, $\text{Var}(\mathcal{W}_{1})$ results in \eqref{NLSG-Eq40}, and $\text{Var}(\mathcal{W}_{2})$ can be written similarly.

The covariance between $\mathcal{W}_{1} = \mathcal{X}^{\tau_{1}}\epsilon^{\varpi_{1}}$ and $\mathcal{W}_{2} = \mathcal{X}^{\tau_{2}}\epsilon^{\varpi_{2}}$ is evaluated as
\begin{equation}
{
\begin{split}
\text{Cov}(\mathcal{W}_{1},\mathcal{W}_{2}) &= E[\mathcal{X}^{\tau_{1}+\tau_{2}}]E[\epsilon^{\varpi_{1}+\varpi_{2}}] \\
&- E[\mathcal{X}^{\tau_{1}}]E[\epsilon^{\varpi_{1}}]E[\mathcal{X}^{\tau_{2}}]E[\epsilon^{\varpi_{2}}].
\end{split}}
\label{NLSG-Eq43}
\end{equation}
Similar to the previous discussion, considering the symmetric distribution of $\mathcal{X}$ and $\epsilon$ around zero, the first term is non-zero if $(\tau_{1}+\tau_{2})$ and $(\varpi_{1}+\varpi_{2})$ are even, which occurs when $\tau_{1}, \tau_{2}, \varpi_{1}, \varpi_{2}$ are either all odd or all even. However, the second term contributes only if $\tau_{1}, \tau_{2}, \varpi_{1}, \varpi_{2}$ are all even. In all other cases, the covariance is zero. This leads to the expression in \eqref{NLSG-Eq41}. \hfill $\blacksquare$

By using Lemma 2, \eqref{NLSG-Eq39} can be rewritten as
\begin{equation}
{\footnotesize
\begin{split}
\text{Var}(Z_{{\rm{q}}k} | b_{\rm{qx}}, b_{\rm{qy}}) 
&= T_{\rm q2}^{2} 2 \sigma_{\rm{jqx}}^{4} 
+ (T_{\rm q3}^{2} + T_{\rm q6}^{2} b_{\rm{qx}}^{2}) \sigma_{\rm{jqx}}^{2} 
+ T_{\rm q8}^{2} 2 \sigma_{\rm{jqy}}^{4} \\
&\hspace{-40pt} + (T_{\rm q9}^{2} + T_{\rm q12}^{2} b_{\rm{qy}}^{2}) \sigma_{\rm{jqy}}^{2} 
+ T_{\rm q13}^{2} \frac{\Delta_{\rm{qx}}^{4}}{180} + (T_{\rm q14}^{2} b_{\rm{qx}}^{2} 
+ T_{\rm q15}^{2}) \frac{\Delta_{\rm{qx}}^{2}}{12}\\
&\hspace{-40pt}  
+ T_{\rm q16}^{2} \sigma_{\rm{jqx}}^{2} \frac{\Delta_{\rm{qx}}^{2}}{12} 
+ T_{\rm q17}^{2} \frac{\Delta_{\rm{qy}}^{4}}{180} 
+ (T_{\rm q18}^{2} b_{\rm{qy}}^{2} + T_{\rm q19}^{2}) \frac{\Delta_{\rm{qy}}^{2}}{12}\\
&\hspace{-40pt} + T_{\rm q20}^{2} \sigma_{\rm{jqy}}^{2} \frac{\Delta_{\rm{qy}}^{2}}{12} 
+ 2 T_{\rm q3} T_{\rm q6} b_{\rm{qx}} \sigma_{\rm{jqx}}^{2} 
+ 2 T_{\rm q9} T_{\rm q12} b_{\rm{qy}} \sigma_{\rm{jqy}}^{2} \\
&\hspace{-40pt} + 2 T_{\rm q14} T_{\rm q15} b_{\rm{qx}} \frac{\Delta_{\rm{qx}}^{2}}{12} 
+ 2 T_{\rm q18} T_{\rm q19} b_{\rm{qy}} \frac{\Delta_{\rm{qy}}^{2}}{12}.
\end{split}}
\label{NLSG-Eq44}
\end{equation}
By taking the expectation of $\text{Var}(Z_{{\rm{q}}k} | b_{\rm{qx}}, b_{\rm{qy}})$ over $b_{\rm{qx}}$ and $b_{\rm{qy}}$ with the aid of \eqref{NLSG-Eq-21} and \eqref{NLSG-Eq-22}, $E[\text{Var}(Z_{{\rm{q}}k} | b_{\rm{qx}} , b_{\rm{qy}})]$ can be found as in \eqref{NLSG-Eq45} at \textcolor{black}{the top of the next page}. 
\begin{figure*}[!t]
\begin{equation}
{\small
\begin{split}
E[\text{Var}(Z_{{\rm{q}}k} | b_{\rm{qx}} , b_{\rm{qy}})] 
&= T_{\rm q2}^{2} 2 \sigma_{\rm{jqx}}^{4} 
+ {[T_{\rm q3}^{2} + T_{\rm q6}^{2} (\mu_{b_{\rm{qx}}}^{2} +  \sigma^{2}_{\rm{qP}} \kappa_{\rm{c}}^{2} d_{\rm{qx}}^{2} \cos^{2} \theta_{\rm{q}} \cos^{2} \phi_{\rm{q}} + \sigma^{2}_{{\rm{qP}}_\perp} \kappa_{\rm{c}}^{2} d_{\rm{qx}}^{2} \sin^{2} \phi_{\rm{q}})] \sigma_{\rm{jqx}}^{2}} \\
&+ T_{\rm q8}^{2} 2 \sigma_{\rm{jqy}}^{4} 
+ {[T_{\rm q9}^{2} + T_{\rm q12}^{2} (\mu_{b_{\rm{qy}}}^{2} + \sigma^{2}_{\rm{qP}} \kappa_{\rm{c}}^{2} d_{\rm{qy}}^{2} \cos^{2} \theta_{\rm{q}} \sin^{2} \phi_{\rm{q}} + \sigma^{2}_{{\rm{qP}}_\perp} \kappa_{\rm{c}}^{2} d_{\rm{qy}}^{2} \cos^{2} \phi_{\rm{q}})] \sigma_{\rm{jqy}}^{2}} 
+ T_{\rm q13}^{2} \frac{\Delta_{\rm{qx}}^{4}}{180} \\
&+ {[T_{\rm q14}^{2} (\mu_{b_{\rm{qx}}}^{2} +  \sigma^{2}_{\rm{qP}} \kappa_{\rm{c}}^{2} d_{\rm{qx}}^{2} \cos^{2} \theta_{\rm{q}} \cos^{2} \phi_{\rm{q}} + \sigma^{2}_{{\rm{qP}}_\perp} \kappa_{\rm{c}}^{2} d_{\rm{qx}}^{2} \sin^{2} \phi_{\rm{q}}) + T_{\rm q15}^{2}] \frac{\Delta_{\rm{qx}}^{2}}{12}}  
+ T_{\rm q16}^{2} \sigma_{\rm{jqx}}^{2} \frac{\Delta_{\rm{qx}}^{2}}{12} \\
&+ T_{\rm q17}^{2} \frac{\Delta_{\rm{qy}}^{4}}{180} 
+ {[T_{\rm q18}^{2} (\mu_{b_{\rm{qy}}}^{2} + \sigma^{2}_{\rm{qP}} \kappa_{\rm{c}}^{2} d_{\rm{qy}}^{2} \cos^{2} \theta_{\rm{q}} \sin^{2} \phi_{\rm{q}} + \sigma^{2}_{{\rm{qP}}_\perp} \kappa_{\rm{c}}^{2} d_{\rm{qy}}^{2} \cos^{2} \phi_{\rm{q}}) + T_{\rm q19}^{2}] \frac{\Delta_{\rm{qy}}^{2}}{12}} \\
&+ T_{\rm q20}^{2} \sigma_{\rm{jqy}}^{2} \frac{\Delta_{\rm{qy}}^{2}}{12} 
+ {2 T_{\rm q3}  T_{\rm q6} \mu_{b_{\rm{qx}}} \sigma_{\rm{jqx}}^{2}} 
+ {2 T_{\rm q9} T_{\rm q12} \mu_{b_{\rm{qy}}} \sigma_{\rm{jqy}}^{2}} 
+ {2 T_{\rm q14} T_{\rm q15} \mu_{b_{\rm{qx}}} \frac{\Delta_{\rm{qx}}^{2}}{12}} 
+ {2 T_{\rm q18} T_{\rm q19} \mu_{b_{\rm{qy}}} \frac{\Delta_{\rm{qy}}^{2}}{12}}.  
\end{split}}
\label{NLSG-Eq45}
\end{equation}
\vspace*{0pt} 
\noindent\rule{\textwidth}{0.4pt} 
\end{figure*}
By using \eqref{NLSG-Eq30}, \eqref{NLSG-Eq38}, and \eqref{NLSG-Eq45}, the variance of $Z_{{\rm{q}}k}$ can be obtained. 
Afterwards, the shape parameter $\varsigma_{k}$ and the scale parameter $\nu_{k}$ of the Gamma distributed random variable $\mathcal{T}_{k}'$ can be obtained by using \eqref{NLSG-Eq28.6} and \eqref{NLSG-Eq28.7}, respectively. Lastly, the location parameter is already found as $\rho_{k} = - \ln(g^{{\rm{t}}k}(\theta_{\rm{t}},\phi_{\rm{t}})) - \ln(g^{{\rm{r}}k}(\theta_{\rm{r}},\phi_{\rm{r}}))$ in the previous subsection.

While the preceding derivations establish a comprehensive statistical characterization of the pointing error, the proposed model is not strictly limited to these specific jitter and angle estimation error profiles. The core analytical framework is developed to remain highly adaptable. Building upon this mathematical foundation, the practical implications of the model are explored through various use cases, explicitly illustrating its functionality both within and outside the confines of the aforementioned distributional assumptions.

\section{Practical Use Cases of NLSG Distribution and Model Reductions}

\label{Sec-5}

\subsection{Practical Use Cases of NLSG Model}
To highlight the versatility of the proposed framework, the various system configurations and practical impairments that the model accommodates are summarized. By leveraging the CDF in \eqref{NLSG-Eq48}, the PDF in \eqref{NLSG-Eq49}, and the moments in \eqref{NLSG-Eq52}, the impact of pointing errors on highly directional beamforming systems can be analyzed across a wide range of scenarios. Specifically, it can guide system design by accounting for:
\begin{itemize}
\item \textbf{Hardware and estimation impairments:} The model incorporates the impacts of mechanical jitter, AoD/AoA estimation errors, and phase shifter quantization errors. It flexibly accounts for both symmetric and asymmetric error distributions across different spatial axes. Furthermore, the framework accommodates a diverse range of distributional assumptions, remaining robust under both Gaussian and non-Gaussian distributions for mechanical jitter and angular estimation errors.
\item \textbf{Element pattern awareness:} Patterns of individual antenna elements for either high-gain or low-gain antennas are incorporated, which significantly influence model parameters during beam steering. More specifically, through $\rho_{k} = - \ln(g^{{\rm{t}}k}(\theta_{\rm{t}},\phi_{\rm{t}})) - \ln(g^{{\rm{r}}k}(\theta_{\rm{r}},\phi_{\rm{r}}))$, the reduction in the observed gain due to the element pattern $g^{{\rm{q}}k}(\theta,\phi)$, i.e. scanning loss, can be captured.
\item \textbf{Diverse array configurations:} Symmetric and asymmetric radiation patterns can be modeled by considering square ($N_{\rm{qx}} = N_{\rm{qy}}$) or rectangular ($N_{\rm{qx}} \neq N_{\rm{qy}}$) antenna arrays at the transmitter and/or receiver. Moreover, practical antenna arrays beyond standard UPAs can be supported, including configurations with non-half-wavelength element spacing ($d_{\rm{qx}} \neq \lambda_{\rm{c}}/2$ and $d_{\rm{qy}} \neq \lambda_{\rm{c}}/2$). In this way, the physical dimensions of the antenna element can be incorporated in the pointing error statistics.
\item \textbf{Various beamforming architectures:} Fully analog ($M_{\rm{qx}} = N_{\rm{qx}}$, $M_{\rm{qy}} = N_{\rm{qy}}$), fully digital ($M_{\rm{qx}} = 1$, $M_{\rm{qy}} = 1$), or hybrid beamforming architectures can be modeled. Owing to this feature, practical design tradeoffs, such as the impact of phase shift quantization and required number of RF chains can be accounted for.
\item \textbf{Steered or non-steered main lobes:} Both steered beam and non-steered beam configurations can be analyzed. Thanks to the characterization of the main lobes of the patterns for an arbitrary radiation direction, the influence of beam steering on the pointing error statistics can be examined. Additionally, the proposed framework accounts for the impact of the maximum steering angle $\theta_{\rm{q,max}}$ illustrating the critical tradeoff where reducing the number of quantization bits diminishes the angular resolution over the available steering range.
\item \textbf{Bandwidth considerations and beam squint effect:} The proposed model is applicable to both wideband ($\kappa_{k} \neq \kappa_{\rm{c}}$) and narrowband ($\kappa_{k} \approx \kappa_{\rm{c}}$) communication systems. Furthermore, it accounts for the beam squint effect inherent to wideband arrays, accommodating scenarios where this impairment is either present or mitigated through baseband signal processing or TTD units.
\end{itemize}

\subsection{Special Cases of NLSG Model}

In the following, the NLSG model is progressively simplified to isolate the effects of specific parameters. Each subsequent case builds upon the assumptions of the previous one, sequentially narrowing the scope from a general architecture to the cases covered by the currently existing models in the literature.

\subsubsection{Special Case 1 (Negligible Quantization Error)}

Considering the practical deployments, the impact of quantization bits can be easily reduced to a negligible level. As the quantization interval exponentially shrinks with the increasing number of quantization bits ($\Delta_{\rm{qx}},\Delta_{\rm{qy}} \propto {1}/{2^{\ell_{\rm{q}}}}$), the terms related to the quantization error in \eqref{NLSG-Eq29} and \eqref{NLSG-Eq45} become negligible. Hence, \eqref{NLSG-Eq29} and \eqref{NLSG-Eq45} can be reduced to
\begin{equation}
\begin{split}
E[Z_{{\rm{q}}k}] &= T_{\rm q1} + T_{\rm q2} \sigma_{\rm{jqx}}^{2} + T_{\rm q4} (\mu_{b_{\rm{qx}}}^{2} + \sigma_{b_{\rm{qx}}}^{2}) + T_{\rm q5} \mu_{b_{\rm{qx}}} \\
&+ T_{\rm q7} + T_{\rm q8} \sigma_{\rm{jqy}}^{2} + T_{\rm q10} (\mu_{b_{\rm{qy}}}^{2} + \sigma_{b_{\rm{qy}}}^{2}) + T_{\rm q11} \mu_{b_{\rm{qy}}},
\end{split}
\label{NLSG-Eq29-SC-1}
\end{equation}
and 
\begin{equation}
{
\begin{aligned}
E[\text{Var}(Z_{{\rm{q}}k} | b_{\rm{qx}} , b_{\rm{qy}})] 
&= T_{\rm q2}^{2} 2 \sigma_{\rm{jqx}}^{4} 
+ T_{\rm q8}^{2} 2 \sigma_{\rm{jqy}}^{4}  \\
&\hspace{-70pt} +  [T_{\rm q3}^{2}+ T_{\rm q6}^{2} (\mu_{b_{\rm{qx}}}^{2} +  \sigma^{2}_{\rm{qP}} \kappa_{\rm{c}}^{2} d_{\rm{qx}}^{2} \cos^{2} \theta_{\rm{q}} \cos^{2} \phi_{\rm{q}} \\
&\hspace{-70pt}+ \sigma^{2}_{{\rm{qP}}_\perp} \kappa_{\rm{c}}^{2} d_{\rm{qx}}^{2} \sin^{2} \phi_{\rm{q}})] \sigma_{\rm{jqx}}^{2} + [T_{\rm q9}^{2} + T_{\rm q12}^{2} (\mu_{b_{\rm{qy}}}^{2} \\
&\hspace{-70pt}+ \sigma^{2}_{\rm{qP}} \kappa_{\rm{c}}^{2} d_{\rm{qy}}^{2} \cos^{2} \theta_{\rm{q}} \sin^{2} \phi_{\rm{q}} + \sigma^{2}_{{\rm{qP}}_\perp} \kappa_{\rm{c}}^{2} d_{\rm{qy}}^{2} \cos^{2} \phi_{\rm{q}})] \sigma_{\rm{jqy}}^{2} \\
&\hspace{-70pt}+ {2 T_{\rm q3}  T_{\rm q6} \mu_{b_{\rm{qx}}} \sigma_{\rm{jqx}}^{2}} 
+ {2 T_{\rm q9} T_{\rm q12} \mu_{b_{\rm{qy}}} \sigma_{\rm{jqy}}^{2}} ,
\end{aligned}}
\label{NLSG-Eq45-SC-1}
\end{equation}
respectively. These expressions provide less involved computation of the distribution parameters in practical negligible quantization error scenarios.

\subsubsection{Special Case 2 (The Jitter-Dominated Regime)}
Building upon the previous case, angle estimation errors can be significantly reduced by state-of-the-art tracking algorithms. In such cases, the impact of both quantization and angle estimation errors can be neglected, leaving the mechanical jitters as the only dominant source of the misalignment. For negligible quantization errors and angle estimation errors, $\text{Var}(E[Z_{{\rm{q}}k} | b_{\rm{qx}} , b_{\rm{qy}}])$ is entirely nullified as $\sigma_{\rm{qP}} \approx 0$ and $\sigma_{\rm{qP}_\perp} \approx 0$, thus, \eqref{NLSG-Eq29} and \eqref{NLSG-Eq30} can be significantly reduced as 
\begin{equation}
\begin{split}
E[Z_{{\rm{q}}k}] &= T_{\rm q1} + T_{\rm q2} \sigma_{\rm{jqx}}^{2} + T_{\rm q4} \mu_{b_{\rm{qx}}}^{2} + T_{\rm q5} \mu_{b_{\rm{qx}}} \\
&+ T_{\rm q7} + T_{\rm q8} \sigma_{\rm{jqy}}^{2} + T_{\rm q10} \mu_{b_{\rm{qy}}}^{2} + T_{\rm q11} \mu_{b_{\rm{qy}}},
\end{split}
\label{NLSG-Eq29-SC-2}
\end{equation}
and
\begin{equation}
{
\begin{aligned}
\text{Var}(Z_{{\rm{q}}k}) 
&= T_{\rm q2}^{2} 2 \sigma_{\rm{jqx}}^{4} 
+ T_{\rm q8}^{2} 2 \sigma_{\rm{jqy}}^{4}  \\
& +  (T_{\rm q3}^{2}+ T_{\rm q6}^{2} \mu_{b_{\rm{qx}}}^{2} + 2 T_{\rm q3}  T_{\rm q6} \mu_{b_{\rm{qx}}}) \sigma_{\rm{jqx}}^{2} \\
&+ (T_{\rm q9}^{2} + T_{\rm q12}^{2} \mu_{b_{\rm{qy}}}^{2} + 2 T_{\rm q9} T_{\rm q12} \mu_{b_{\rm{qy}}}) \sigma_{\rm{jqy}}^{2},
\end{aligned}}
\label{NLSG-Eq45-SC-2}
\end{equation}
respectively. Owing to this reduction, the statistics of the pointing error can be easily found and analyzed for jitter-dominant cases.

\subsubsection{Special Case 3 (Boresight Deployment in Jitter-Dominated Regime)}

In the case where the transmitter and receiver are located in each other's boresight direction, active steering is not required, which is the considered case in most of the existing pointing error modeling studies in the literature \cite{9887787,9931325,10841410,11389924}. This can be achieved with the NLSG model by setting the elevation and azimuth angles to zero at both sides as $\theta_{\rm{q}} = 0^\circ$ and $\phi_{\rm{q}} = 0^\circ$. In boresight deployment in jitter-dominated regime, \eqref{NLSG-Eq29} and \eqref{NLSG-Eq30} reduce to
\begin{equation}
\begin{split}
E[Z_{{\rm{q}}k}] &=  \mathcal{A}_{\rm{qx}}  \sigma_{\rm{jqx}}^{2} + \mathcal{A}_{\rm{qy}} \sigma_{\rm{jqy}}^{2},
\end{split}
\label{NLSG-Eq29-SC-4}
\end{equation}
and
\begin{equation}
{
\begin{aligned}
\text{Var}(Z_{{\rm{q}}k}) 
&= \mathcal{A}_{\rm{qx}}^{2} 2 \sigma_{\rm{jqx}}^{4} 
+ \mathcal{A}_{\rm{qy}}^{2} 2 \sigma_{\rm{jqy}}^{4},
\end{aligned}}
\label{NLSG-Eq45-SC-4}
\end{equation}
respectively, where $\mathcal{A}_{\rm{qx}}$ and $\mathcal{A}_{\rm{qy}}$ depend on the total number of antennas ($N_{\rm{qx}}$ and $N_{\rm{qy}}$), sub-band wave number $\kappa_{k}$, and element spacing $d_{\rm{qx}}$ and $d_{\rm{qy}}$. The shape parameter $\varsigma_{k}$ and scale parameter $\nu_{k}$ can be easily obtained by substituting the above-equations in \eqref{NLSG-Eq28.6} and \eqref{NLSG-Eq28.7}, respectively. Furthermore, since the elevation and azimuth angles are set to zero ($\theta_{\rm{q}} = 0^\circ$ and $\phi_{\rm{q}} = 0^\circ$) in boresight deployment, the location parameter can be found as $\rho_{k} = - \ln(g^{{\rm{t}}k}(0,0)) - \ln(g^{{\rm{r}}k}(0,0)) = 0$. This results in a special case of the NLSG distribution as the negative log-Gamma (NLG) with shape parameter $\varsigma_{k}$ and scale parameter $\nu_{k}$ whose CDF and PDF can be obtained directly from \eqref{NLSG-Eq48} and \eqref{NLSG-Eq49} as
\begin{equation}
\begin{split}
F_{h_{{\rm{p}}k}}(x) &= \frac{1}{\Gamma(\varsigma_{k})}\Gamma\left(\varsigma_{k},\frac{- \ln x}{\nu_{k}}\right).
\label{NLSG-Eq48-SC-3}
\end{split}
\end{equation}
and
\begin{equation}
\begin{split}
f_{h_{{\rm{p}}k}}(x) &= \frac{1}{\Gamma(\varsigma_{k}) \nu_{k}^{\varsigma_{k}}} (- \ln x)^{\varsigma_{k} - 1}  x^{\frac{1}{\nu_{k}}-1},
\label{NLSG-Eq49-SC-3}
\end{split}
\end{equation}
respectively.

\subsubsection{Special Case 4 (Narrowband Standard UPAs and ULAs of Isotropic Antennas)}

For narrowband ($\kappa_{k} \approx \kappa_{\rm{c}}$) boresight deployment with standard UPAs of isotropic antennas (half-wavelength element spacing $d_{\rm{qx}} = \frac{\lambda_{\rm{c}}}{2}$) in the presence of jitters as the only disturbance, \eqref{NLSG-Eq29-SC-4} and \eqref{NLSG-Eq45-SC-4} can be further simplified as
\begin{equation}
\begin{split}
E[Z_{{\rm{q}}k}] &=  \left(\frac{N_{\rm{qx}}^{2} - 1}{24}\right) \pi^{2}  \sigma_{\rm{jqx}}^{2} + \left(\frac{N_{\rm{qy}}^{2} - 1}{24}\right) \pi^{2} \sigma_{\rm{jqy}}^{2},
\end{split}
\label{NLSG-Eq29-SC-4.5}
\end{equation}
and
\begin{equation}
{
\begin{aligned}
\text{Var}(Z_{{\rm{q}}k}) 
&= \frac{\left(N_{\rm{qx}}^{2} - 1\right)^{2}}{288} \pi^{4} \sigma_{\rm{jqx}}^{4} 
+ \frac{\left(N_{\rm{qy}}^{2} - 1\right)^{2}}{288} \pi^{4} \sigma_{\rm{jqy}}^{4},
\end{aligned}}
\label{NLSG-Eq45-SC-4.5}
\end{equation}
respectively, depending only on the number of antennas and jitter variances along x- and y-axis. In the case of square UPAs ($N_{\rm{qx}} = N_{\rm{qy}}$ and $N_{\rm{q}} = N_{\rm{qx}}^{2} = N_{\rm{qy}}^{2}$), \eqref{NLSG-Eq29-SC-4.5} and \eqref{NLSG-Eq45-SC-4.5} reduce to
\begin{equation}
\begin{split}
E[Z_{{\rm{q}}k}] &=  \left(\frac{N_{\rm{q}} - 1}{24}\right) \pi^{2}  \left(\sigma_{\rm{jqx}}^{2} + \sigma_{\rm{jqy}}^{2}\right),
\end{split}
\label{NLSG-Eq29-SC-4.6}
\end{equation}
and
\begin{equation}
{
\begin{aligned}
\text{Var}(Z_{{\rm{q}}k}) 
&= \frac{\left(N_{\rm{q}} - 1\right)^{2}}{288} \pi^{4} \left(\sigma_{\rm{jqx}}^{4} + \sigma_{\rm{jqy}}^{4}\right),
\end{aligned}}
\label{NLSG-Eq45-SC-4.6}
\end{equation}
respectively. Similarly, for ULAs ($N_{\rm{qy}} = 1$ and $N_{\rm{q}} = N_{\rm{qx}}$), \eqref{NLSG-Eq29-SC-4.5} and \eqref{NLSG-Eq45-SC-4.5} can be simplified as
\begin{equation}
\begin{split}
E[Z_{{\rm{q}}k}] &=  \left(\frac{N_{\rm{q}}^{2} - 1}{24}\right) \pi^{2}  \sigma_{\rm{jqx}}^{2},
\end{split}
\label{NLSG-Eq29-SC-4.7}
\end{equation}
and
\begin{equation}
{
\begin{aligned}
\text{Var}(Z_{{\rm{q}}k}) 
&= \frac{\left(N_{\rm{q}}^{2} - 1\right)^{2}}{288} \pi^{4} \sigma_{\rm{jqx}}^{4},
\end{aligned}}
\label{NLSG-Eq45-SC-4.7}
\end{equation}
respectively. The shape parameter $\varsigma_{k}$ and scale parameter $\nu_{k}$ for these scenarios can be easily obtained by using \eqref{NLSG-Eq28.6} and \eqref{NLSG-Eq28.7}, respectively, together with the above-equations corresponding to the scenario of interest. Here, the location parameter is reduced to $\rho_{k} = 0$ similar to the Special Case 3. Therefore, the CDF and PDF of the pointing error coefficient can be given as in \eqref{NLSG-Eq48-SC-3} and \eqref{NLSG-Eq49-SC-3}, respectively. 

Here, the square UPA setup and ULA setup described above are the exact same setups as those considered in \cite[Theorem 1]{9931325} and \cite[Theorem 3]{9931325}, respectively. In comparison to the models in \cite{9931325}, the proposed model in this paper has relatively simpler mathematical form both for the statistics (PDF and CDF) and for the parameter derivation.

\subsubsection{Special Case 5 (Symmetrical Gaussian Jitters)}

In the presence of symmetrical jitters along x- and y-axis ($\sigma_{\rm{jqx}} = \sigma_{\rm{jqy}} = \sigma_{\rm{jq}}$) as considered in \cite[Theorem 2]{9931325}, \eqref{NLSG-Eq29-SC-4.5} and \eqref{NLSG-Eq45-SC-4.5} simplify to
\begin{equation}
\begin{split}
E[Z_{{\rm{q}}k}] &=  \frac{N_{\rm{qx}}^{2} + N_{\rm{qy}}^{2} - 2}{24} \pi^{2}  \sigma_{\rm{jq}}^{2},
\end{split}
\label{NLSG-Eq29-SC-5}
\end{equation}
and
\begin{equation}
{
\begin{aligned}
\text{Var}(Z_{{\rm{q}}k}) 
&= \frac{\left(N_{\rm{qx}}^{2} - 1\right)^{2} + \left(N_{\rm{qy}}^{2} - 1\right)^{2}}{288} \pi^{4} \sigma_{\rm{jq}}^{4},
\end{aligned}}
\label{NLSG-Eq45-SC-5}
\end{equation}
respectively, where the shape and scale parameters can be evaluated similarly. Also, the location parameter is $\rho_{k} = 0$ because of boresight deployment. Hence, the statistics of the pointing error can be given as in \eqref{NLSG-Eq48-SC-3} and \eqref{NLSG-Eq49-SC-3}. Furthermore, if identical jitter distributions are present along x- and y-axis at both transmitter and receiver ($\sigma_{\rm{jt}} = \sigma_{\rm{jr}} = \sigma_{\rm{j}}$) with square UPAs of the same size ($N_{\rm{tx}} = N_{\rm{ty}} = N_{\rm{rx}} = N_{\rm{ry}} = N$), the shape and scale parameters can be found as $\varsigma_{k} = 2$ and $\nu_{k} = \frac{(N^{2} - 1)}{\textcolor{black}{12}} \pi^{2} \sigma_{\rm{j}}^{2}$, which provides very easy manipulation of the proposed analytical model.

By using the special cases introduced in this section, the pointing error characteristics can be easily derived for various scenarios from the most general to the most specific cases. In the next section, several goodness-of-fit metrics are introduced, and the accuracy of the NLSG model and its special cases are analyzed.

\section{Accuracy Assessment and Model Verification}

\subsection{Goodness-of-Fit Metrics}

In statistical modeling, various metrics can be used to assess the goodness-of-fit for the proposed model. In this study, three different error metrics are used for the accuracy assessment, namely, NME, JS divergence, and KS statistics.

\subsubsection{Normalized Moment Error}

The NME analyzes the fitness of the model by focusing on the statistical characteristics, such as average value (the first moment) or average power (the second moment) of the quantity that is being modeled. Mathematically, the NME for the $r$-th moment can be formulated as
\begin{equation}
\begin{split}
\text{NME}_{r} \text{ [\%]} = \left|\frac{E[h_{{\rm{p}}k}^{r}] - E[h_{{\rm{p}}k,{\rm{sim}}}^{r}]}{E[h_{{\rm{p}}k,{\rm{sim}}}^{r}]}\right| \times 100.
\label{NLSG-Eq58}
\end{split}
\end{equation}
Here, $h_{{\rm{p}}k,{\rm{sim}}}$ denotes the simulated pointing error coefficient by directly using the element pattern and array factor expressions described in Section \ref{Sec-2}-A and Section \ref{Sec-2}-B, respectively.

\subsubsection{Jensen-Shannon Divergence}
JS divergence measures the overall information overlap or entropy difference between the distributions. In other words, JS divergence focuses on the information gained about the actual (simulated) distribution $\mathcal{P}_{\rm{sim}} = \{p_{1}^{\rm{sim}}, p_{2}^{\rm{sim}}, \ldots, p_{I}^{\rm{sim}}\}$ by knowing the analytical distribution $\mathcal{P}_{\rm{mod}} = \{p_{1}^{\rm{mod}}, p_{2}^{\rm{mod}}, \ldots, p_{I}^{\rm{mod}}\}$, where $I$ denotes the number of bins. Here, $\mathcal{P}_{\rm{mod}}$ can be extracted by the discretisation of the proposed model's PDF in \eqref{NLSG-Eq49}. JS divergence can be calculated as
\begin{equation}
\begin{split}
D_{\rm JS}(\mathcal{P}_{\rm{sim}}||\mathcal{P}_{\rm{mod}}) = \frac{1}{2} D_{\rm KL}(\mathcal{P}_{\rm{sim}}||\mathcal{Q}) + \frac{1}{2} D_{\rm KL}(\mathcal{Q}||\mathcal{P}_{\rm{mod}}),
\label{NLSG-Eq59}
\end{split}
\end{equation}
where $\mathcal{Q} = \frac{1}{2}(\mathcal{P}_{\rm{sim}} + \mathcal{P}_{\rm{mod}})$. Here, $D_{\rm KL}(\mathcal{P}_{\rm{sim}}||\mathcal{Q})$ is the Kullback-Leibler divergence between $\mathcal{P}_{\rm{sim}}$ and $\mathcal{Q}$, and it is defined as
\begin{equation}
\begin{split}
D_{\rm KL}(\mathcal{P}_{\rm{sim}}||\mathcal{Q}) = \sum_{i = 1}^{I} p_{i}^{\rm{sim}} \ln \left(\frac{p_{i}^{\rm{sim}}}{q_{i}}\right),
\label{NLSG-Eq60}
\end{split}
\end{equation}
where $q_{i} = \frac{1}{2}(p_{i}^{\rm{sim}} + p_{i}^{\rm{mod}})$, $\forall i$. \textcolor{black}{Also, $D_{\rm KL}(\mathcal{Q}||\mathcal{P}_{\rm{mod}})$ is defined similarly.}


\subsubsection{Kolmogorov-Smirnov Statistics}
KS statistic is a non-parametric goodness-of-fit test which analyzes the accuracy of the model based on the maximum deviation in the statistics of the simulation data and the analytical model, measuring the worst-case local mismatch between CDFs. The KS statistic can be defined as the supremum difference between the simulated CDF $F_{h_{{\rm{p}}k}}^{\rm{sim}}(x)$ and the proposed CDF $F_{h_{{\rm{p}}k}}(x)$ as
\begin{equation}
\begin{split}
D_{\rm KS} = \sup_{x} |F_{h_{{\rm{p}}k}}^{\rm{sim}}(x) - F_{h_{{\rm{p}}k}}(x)|.
\label{NLSG-Eq61}
\end{split}
\end{equation}

By using these metrics, the accuracy of the proposed NLSG model can be investigated across a wide range of highly directional communication scenarios listed in Section \ref{Sec-5}-A.

\subsection{Validation of the General Pointing Error Model}

{In this subsection, the proposed pointing error model's versatility and accuracy are presented for different array configurations, steering directions, mechanical jitter levels, AoD/AoA estimation errors, number of phase shifter quantization bits, and tracking modes. Additionally, the impact of pointing error is analyzed with different design considerations for narrowband systems and wideband systems under beam squint. Throughout this section, the simulation results (markers) are obtained via computer simulations with $10^{6}$ randomly generated samples by using the array pattern in \eqref{NLSG-Eq03}, element pattern described in \ref{Sec-2}-A, array factors in the Dirichlet kernel form in \eqref{NLSG-Eq05}, and pointing error coefficient in \eqref{NLSG-Eq25.5}, 
whereas the theoretical curves are obtained by using the CDF in \eqref{NLSG-Eq48} and PDF in \eqref{NLSG-Eq49}. 
For the simulations, the antenna element is characterized by $G^{{\rm{q}}k}_{\rm{e,max}} = 8$ dBi and $\theta_{\rm{ex,3dB}} = \theta_{\rm{ey,3dB}} = 65^\circ$. Additionally, the operating frequency is taken as 300 GHz, where the system bandwidth is $B = 1$ GHz for narrowband systems and $B = 20$ GHz with $K = 20$ sub-band.\footnote{Please note here that the framework developed in this study is applicable across RF, mmWave, sub-THz, and THz bands as long as either transmitter or receiver employs highly directional antennas/arrays. However, sub-THz system is considered for numerical results for illustrative purposes.}} Unless otherwise stated, the element spacings along the $\rm{x}$- and $\rm{y}$-axis are assumed as $d_{\rm{qx}} = d_{\rm{qy}} = d_{\rm{e}} = \frac{\lambda_{\rm{c}}}{2}$, and the mechanical jitter and angle estimation errors are characterized by the Gaussian distribution.  

\begin{table}[!t]
\centering
{
\caption{Jitter, Angle Estimation Error, and Number of Quantization Bits at the Transmitter and Receiver for Fig. \ref{fig001}}
\label{table_params_fig_9}
\resizebox{1\linewidth}{!}{
\begin{tabular}{|l|c|c|c|}
\hline
Parameters              & Unstable link & Partially stable link & Stable link \\ \hline
$\sigma_{\rm{jt}x}$      & $1.2^\circ$   & $1.2^\circ$           & $0.6^\circ$ \\ \hline
$\sigma_{\rm{jt}y}$      & $1.3^\circ$   & $1.3^\circ$           & $0.5^\circ$ \\ \hline
$\sigma_{\rm tP}$       & $1.4^\circ$   & $1.4^\circ$           & $0.4^\circ$ \\ \hline
$\sigma_{\rm tP_\perp}$ & $1.6^\circ$   & $1.6^\circ$           & $0.6^\circ$ \\ \hline
$\ell_{\rm{t}}$          & $5$           & $6$                   & $7$         \\ \hline
$\sigma_{\rm{jr}x}$      & $1.0^\circ$   & $0.7^\circ$           & $0.7^\circ$ \\ \hline
$\sigma_{\rm{jr}y}$      & $1.1^\circ$   & $0.4^\circ$           & $0.4^\circ$ \\ \hline
$\sigma_{\rm rP}$       & $1.3^\circ$   & $0.4^\circ$           & $0.4^\circ$ \\ \hline
$\sigma_{\rm rP_\perp}$ & $1.5^\circ$   & $0.5^\circ$           & $0.5^\circ$ \\ \hline
$\ell_{\rm{r}}$          & $5$           & $6$                   & $7$         \\ \hline
\end{tabular}}
}
\end{table}

\begin{figure}[!t]
    \centering
    \includegraphics[width=1\linewidth]{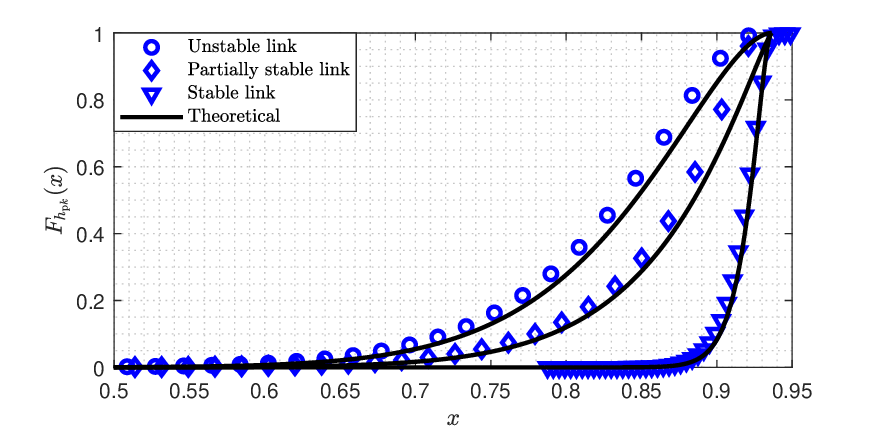}
    \vspace{-20pt}
    \caption{The CDF of the pointing error coefficient $h_{{\rm{p}}k}$ in unstable, partially stable, and stable links.}
    \label{fig001}
\end{figure}

In Fig. \ref{fig001}, the CDF of the pointing error coefficient at the center frequency is illustrated for hybrid beamforming with subarray size $2 \times 3$ and superarray size $4 \times 3$ at both transmitter and receiver for $\theta_{\rm{q,max}} = 60^\circ$. Also, the radiation directions are taken as $\theta_{\rm{q}} = 10^\circ$ and $\phi_{\rm{q}} = 0^\circ$. Here, three distinct stability regimes are considered, namely, unstable link, partially stable link, and stable link. The jitter variances along x- and y-axes, the AoD/AoA estimation error variances on $\rm{P}$ and $\rm{P}_\perp$ planes, and number of quantization bits can be found in Table \ref{table_params_fig_9} for these stability regimes. In the unstable link setup, both transmitter and receiver are subject to high values of jitter and angle estimation error variances. In partially stable link, the receiver experiences lower levels of jitter and AoD/AoA estimation error compared to transmitter. In stable link, both transmitter and receiver are subject to low jitter and AoD/AoA estimation errors. In these setups, the pattern asymmetry introduced by non-square array, and jitter/angle estimation error asymmetry can be captured. It can be inferred from the figure that the proposed model accurately depicts the pointing error statistics in all three complex setups. This reflects the capability of the proposed model in capturing complex characteristics in diverse scenarios.


\begin{figure}[!t]
    \centering
    \vspace{-4.25pt}
    \includegraphics[width=1\linewidth]{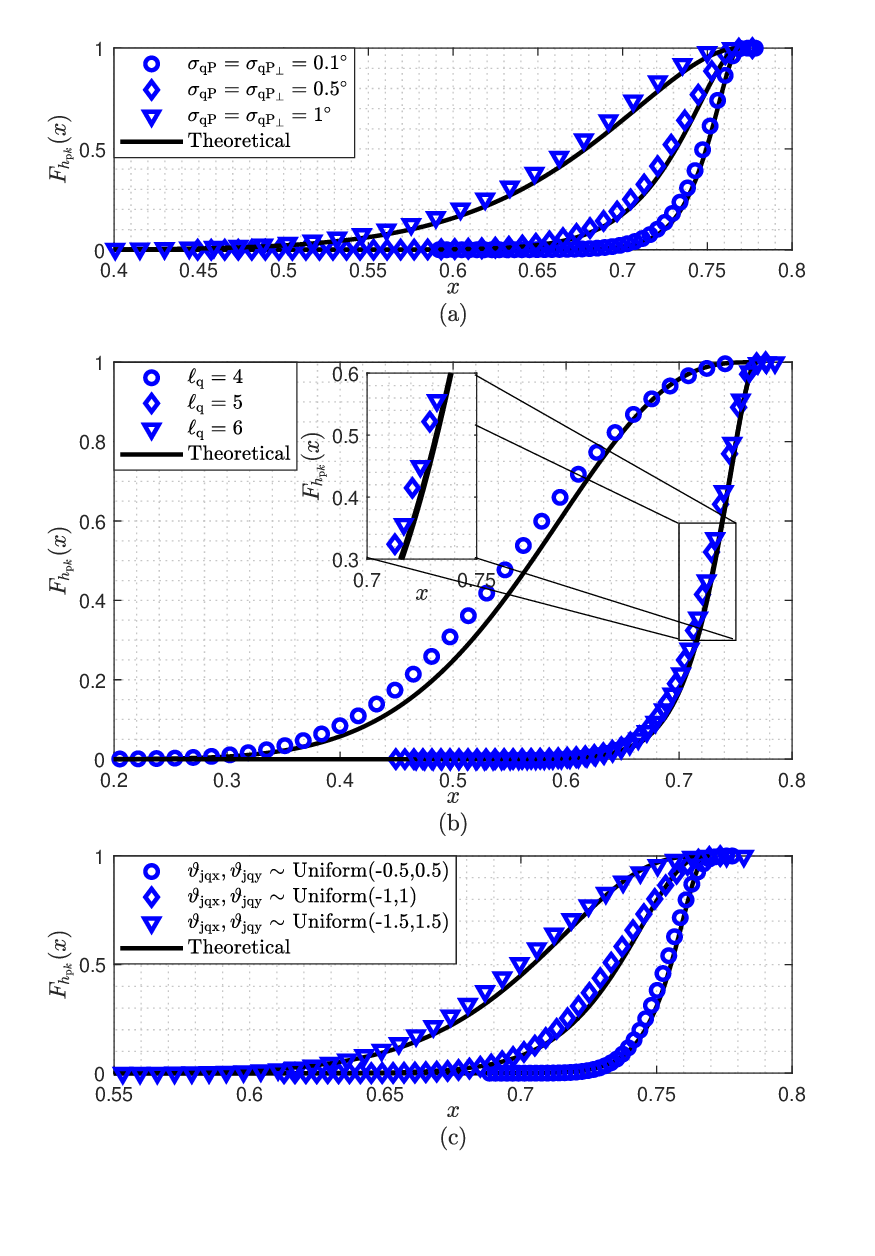}
    \vspace{-45pt}
    \caption{The CDF of the pointing error coefficient $h_{{\rm{p}}k}$ for (a) different angle estimation error variance, (b) different number of quantization bits, and (c) uniform distributed mechanical jitter and angle estimation errors.}
    \label{fig002}
\end{figure}

In Fig. \ref{fig002}, the impacts of different levels of disturbances and different distributions on the pointing error characteristics at the center frequency are analyzed for $16 \times 16$ arrays at transmitter and receiver with analog beamforming. Here, the radiation direction is chosen as $\theta_{\rm{q}} = 20^\circ$ and $\phi_{\rm{q}} = 0^\circ$, whereas the maximum steering range is set to $\theta_{\rm{q,max}} = 45^\circ$. In Fig. \ref{fig002}a, the standard deviation of the AoD/AoA estimation errors and mechanical jitters are taken as $\sigma_{\rm{qP}} = \sigma_{\rm{qP}_\perp} = 0.1^\circ, 0.5^\circ, 1^\circ$ and $\sigma_{\rm{jqx}} = \sigma_{\rm{jqy}} = 0.5^\circ$. Also, $\ell_{\rm{q}} = 5$ number of quantization bits is considered. It can be seen that as the standard deviation of the AoD/AoA estimation errors increases, the CDF becomes wider as both transmitter and receiver are more likely to steer their beams to incorrect directions, leading to a more severe pointing error characteristics. The CDF of the pointing error is illustrated in Fig. \ref{fig002}b for $\ell_{\rm{q}} = 4,5,6$, $\sigma_{\rm{qP}} = \sigma_{\rm{qP}_\perp} = 0.5^\circ$, and $\sigma_{\rm{jqx}} = \sigma_{\rm{jqy}} = 0.5^\circ$. It is observed from the figure that $\ell_{\rm{q}} = 4$ results in a more pronounced pointing error impact. Increasing the number of quantization bits to $\ell_{\rm{q}} = 5$ improves the robustness, however, further increment does not bring any advantage. This can be explained by the sufficient angular resolution achieved by $\ell_{\rm{q}} = 5$ for the considered steering range in this scenario. 
Moreover, in Fig. \ref{fig002}c, the uniform distribution is assumed for AoD/AoA estimation errors, i.e. $\vartheta_{\rm{qP}}, \vartheta_{\rm{qP}_\perp} \sim \text{Uniform}(-0.5,0.5)$, and mechanical jitters with different distribution intervals. Here, $\ell_{\rm{q}} = 5$ is assumed, and $E[Z_{{\rm{q}}k}]$ and $\text{Var}(Z_{{\rm{q}}k})$ are obtained by the simulated data. It can be deduced from the figure that as the distribution interval increases, the CDF of the pointing error gets wider. All curves in Fig. \ref{fig002}a-c illustrate the flexibility of the proposed NLSG model under different distributions and severity levels of mechanical jitters, AoD/AoA estimation errors, and phase shift quantization errors.

\begin{figure}[!t]
    \centering
    \vspace{-4.25pt}
    \includegraphics[width=1\linewidth]{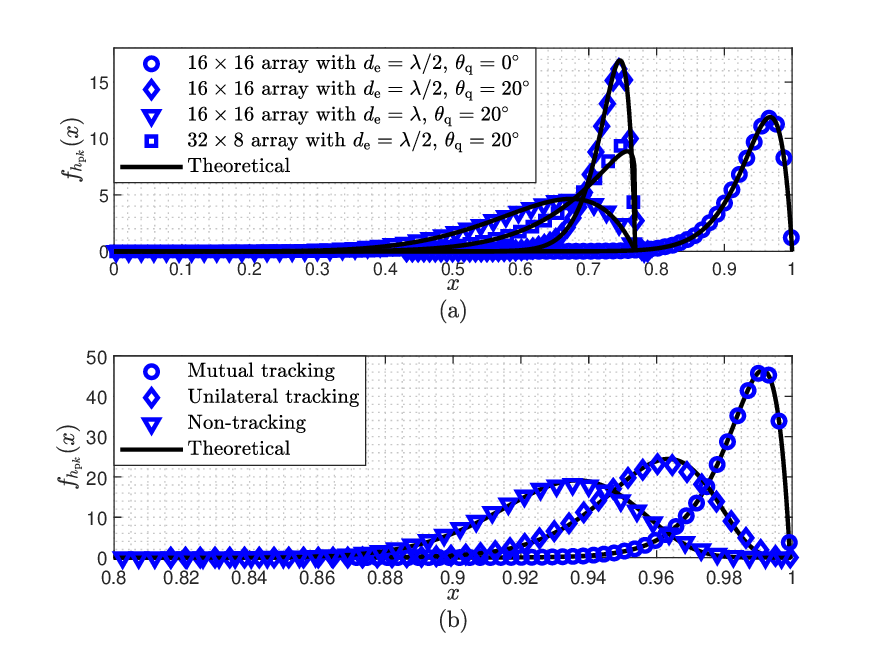}
    \vspace{-20pt}
    \caption{The PDF of the pointing error coefficient $h_{{\rm{p}}k}$ (a) at the center frequency for different array configurations and steering angles (b) in different tracking modes.}
    \label{fig003}
\end{figure}

In Fig. \ref{fig003}a, the PDF of the pointing error coefficient at the center frequency is shown for different array configurations and steering angles ($\theta_{\rm{q}} = 0^\circ, 20^\circ$ and $\phi_{\rm{q}} = 0^\circ$) with $\sigma_{\rm{jqx}} = \sigma_{\rm{jqy}} = 0.5^\circ$, $\sigma_{\rm{qP}} = \sigma_{\rm{qP}_\perp} = 0.5^\circ$, and $\ell_{\rm{q}} = 5$. Here, digital beamforming is considered at both sides of the communication system. It can be observed from the figure that as the steering angle $\theta_{\rm{q}}$ increases, the maximum value of the pointing error coefficient decreases due to scanning loss. In other words, it is more likely that the less power is received. This can be explained by the nominal gain drop to the element pattern in the case of steering. This phenomenon is fully characterized by the relationship between the steering angle and the location parameter of the NLSG model. Moreover, it can be deduced that larger element spacing results in a wider PDF which indicates a more severe pointing error characteristics. This is because larger element spacing narrows the main beam and makes the system more susceptible to misalignment. Additionally, comparing the diamond ($\bDiamond$) and square ($\square$) markers, the asymmetry in the array configuration leads to a wider PDF, which originates from the main beam being narrower in the $\rm{x}$-$\rm{z}$ plane compared to $\rm{y}$-$\rm{z}$ plane for $32 \times 8$ array as the array is larger along the $\rm{x}$-axis. Hence, it can be said that the NLSG model successfully captures the relationship between the pointing error characteristics and the element pattern, array configurations, and steering angles. 
In Fig. \ref{fig003}b, the pointing error characteristics at the center frequency are compared in different tracking modes, namely, mutual tracking, unilateral tracking, and non-tracking modes. Here, the disturbances are characterized by $\sigma_{\rm{jqx}} = \sigma_{\rm{jqy}} = 0.25^\circ$, $\sigma_{\rm{qP}} = \sigma_{\rm{qP}_\perp} = 0.25^\circ$, and $\ell_{\rm{q}} = 5$, and maximum steering range is taken as $\theta_{\rm{q,max}} = 20^\circ$. Additionally, hybrid beamforming architecture is employed at both transmitter and receiver with the subarray size $4 \times 4$ and superarray size $4 \times 4$. In the considered setup, the transmitter and receiver are located at $\theta_{\rm{q}} = 1^\circ$ and $\phi_{\rm{q}} = 0^\circ$ relative to each other. As can be seen, mutual tracking has the most favorable PDF since both sides of the communication link are directing their main beams towards each other to obtain higher array gain. Unilateral tracking provides an intermediate characteristics between the mutual tracking mode and non-tracking mode as at least one of the nodes (transmitter or receiver) tracks the other to achieve higher gain, whereas in the non-tracking case neither node attains its maximum achievable gain, degrading the pointing error characteristics.

\begin{figure}[!t]
    \centering
    \vspace{-4.25pt}
    \includegraphics[width=1\linewidth]{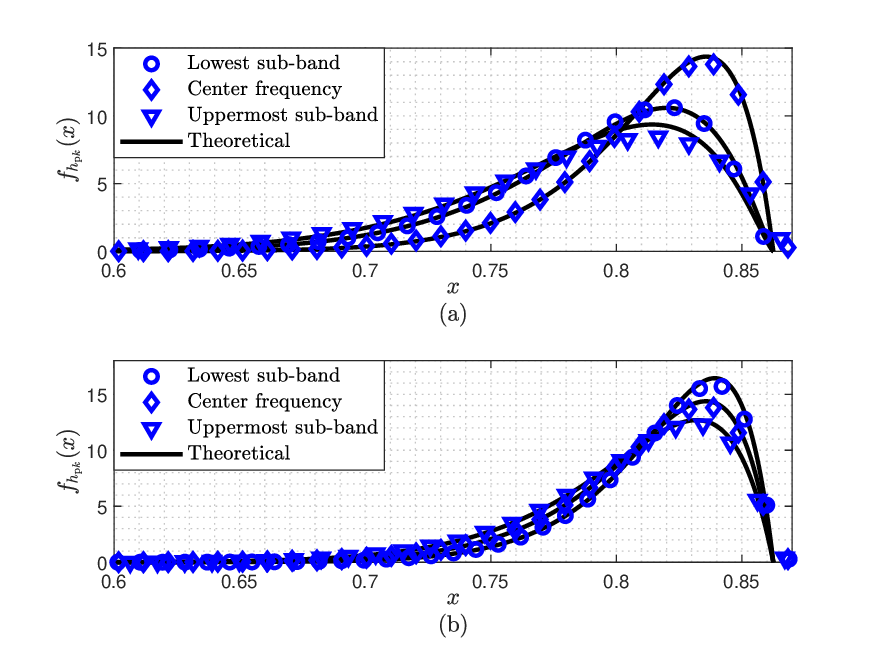}
    \vspace{-20pt}
    \caption{The theoretical and simulation PDFs of the pointing error coefficient $h_{{\rm{p}}k}$ at the lowest sub-band, center frequency, and uppermost sub-band for analog beamforming ($16 \times 16$ array) (a) without squint compensation and (b) with squint compensation.} 
    \label{fig004}
\end{figure}

The PDF of the pointing error coefficient in the lowest sub-band ($k = 1$), center frequency, and the uppermost sub-band ($k = K$) is illustrated in the presence of beam squint in Fig. \ref{fig004}a and in the case of compensated beam squint in Fig. \ref{fig004}b for analog beamforming with $16 \times 16$ arrays at both transmitter and receiver. The mechanical jitters and angle estimation errors are characterized by $\sigma_{\rm{jqx}} = \sigma_{\rm{jqy}} = 0.5^\circ$ and $\sigma_{\rm{qP}} = \sigma_{\rm{qP}_\perp} = 0.5^\circ$, respectively, and the number of quantization bits is $\ell_{\rm{q}} = 5$. Here, the steering angles are $\theta_{\rm{q}} = 15^\circ$ and $\phi_{\rm{q}} = 0^\circ$. In Fig. \ref{fig004}a, it can be seen that the lowest and uppermost sub-bands are more prone to pointing errors in uncompensated beam squint case. This stems from the nominal misalignment caused by the beam squint effect. Despite being influenced by the same level of mechanical jitter, AoD/AoA estimation errors, and quantization errors, the lowest and uppermost sub-bands are slightly misaligned due to beam squint, which reduces the observed gains at both transmitter and receiver. On the other hand, in the compensated squint case in Fig. \ref{fig004}b, the lowest sub-band is more robust to misalignment compared to the center frequency and the uppermost sub-band. The reason for this is that the main beam is slightly wider in the lowest sub-band as the operating wavelength in this sub-band ($\lambda_{k}$ for $k = 1$) is larger compared to the uppermost sub-band ($\lambda_{k}$ for $k = K$) as the array is designed according to the wavelength which corresponds to the center frequency ($\lambda_{\rm{c}}$). Therefore, it can be deduced that the NLSG model accurately depicts the impact of the beam squint effect in wideband systems in both compensated and uncompensated beam squint scenarios.

\subsection{Validation of the Special Cases}

In this subsection, the PDFs and CDFs of the pointing error coefficient are presented for the special cases introduced in Section VII-B. Similar to the previous subsection, $10^{6}$ samples are randomly generated for simulations by using the array pattern in \eqref{NLSG-Eq03}, element pattern described in \ref{Sec-2}-A, array factors in the Dirichlet kernel form in \eqref{NLSG-Eq05}, and pointing error coefficient in \eqref{NLSG-Eq25.5}, whereas the theoretical curves are obtained by using the equations corresponding to each special case. Moreover, the operating frequency is chosen as 300 GHz and the system bandwidth is assumed as $B = 1$ GHz.

\begin{figure}[!t]
    \centering
    \includegraphics[width=1\linewidth]{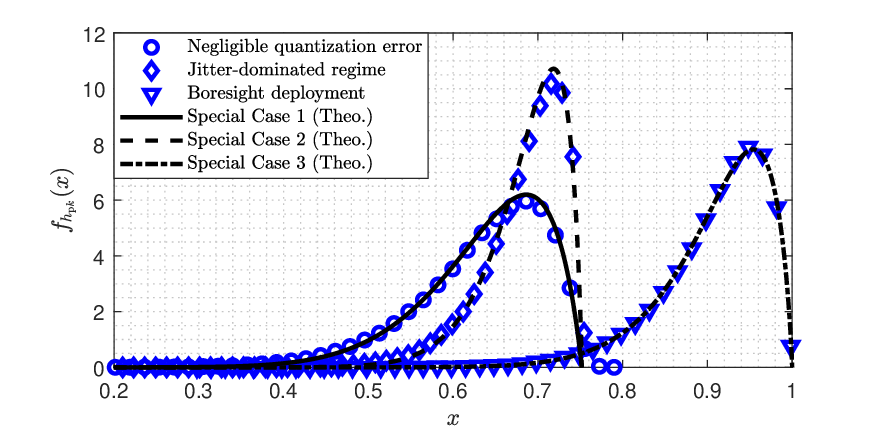}
    \vspace{-20pt}
    \caption{Simulation and theoretical PDFs of the pointing error coefficient $h_{{\rm{p}}k}$ for Special Cases 1, 2, and 3.}
    \label{fig005}
\end{figure}

In Fig. \ref{fig005}, the simulation and theoretical PDFs are illustrated for the first three special cases. Here, cuboid antennas (characterized by $G^{{\rm{q}}k}_{\rm{e,max}} = 14.5$ dBi, $\theta_{\rm{ex,3dB}} = 31.2^\circ$, and $\theta_{\rm{ey,3dB}} = 30.2^\circ$ \cite{9998671}) are used for the antenna array with the subarray size $2 \times 3$ and superarray size $4\times 3$ with the element spacing $d_{\rm{qx}} = d_{\rm{qy}} = d_{\rm{e}} = 1.5\lambda_{\rm{c}}$ due to physical size of the antenna element. Also, the maximum steering range is $\theta_{\rm{q,max}} = 30^\circ$, while the radiation direction is chosen as $\theta_{\rm{q}} = 10^\circ$ and $\phi_{\rm{q}} = 0^\circ$. For jitter and angle estimation error standard deviations in Special Case 1 (negligible quantization error), the stable link setup in Table \ref{table_params_fig_9} is used. In this case, the theoretical curves are obtained by using \eqref{NLSG-Eq29-SC-1} and \eqref{NLSG-Eq45-SC-1} instead of \eqref{NLSG-Eq29} and \eqref{NLSG-Eq45}, respectively, which provides slightly easier calculation of the distribution parameters. Since the number of quantization bits is $\ell_{\rm{q}} = 7$ in this setup, the impact of quantization error diminishes, and the theoretical curve perfectly matches with the simulation one. Moreover, for Special Case 2 (jitter-dominated regime), the angle estimation error standard deviations in stable link setup are reduced by a factor of 1000 to characterize a highly accurate tracking mechanism. In this case, \eqref{NLSG-Eq29-SC-2} and \eqref{NLSG-Eq45-SC-2} are substituted in \eqref{NLSG-Eq28.6} and \eqref{NLSG-Eq28.7} to obtain the distribution parameters, providing even simpler calculation. As can be seen, with the reduced angle estimation errors, the PDF becomes narrower, indicating a more favorable pointing error statistics. 
Furthermore, for the Special Case 3 (boresight deployment in jitter-dominated regime), the radiation direction is set to $\theta_{\rm{q}} = 0$ and $\phi_{\rm{q}} = 0$, simplifying the expressions further. In this scenario, \eqref{NLSG-Eq29-SC-4} and \eqref{NLSG-Eq45-SC-4} are used for the shape and scale parameters, whereas the location parameter is nullified $\rho_{k} = 0$, yielding the NLG distribution in \eqref{NLSG-Eq49-SC-3}. It is observed from the figure that with the boresight deployment, the PDF leans towards the right, reflecting lower probabilities of observing low values of the pointing error coefficient. It is inferred from the figure that the simplified expressions accurately capture the pointing error statistics.   

\begin{figure}[!t]
    \centering
    \vspace{-4.25pt}
    \includegraphics[width=1\linewidth]{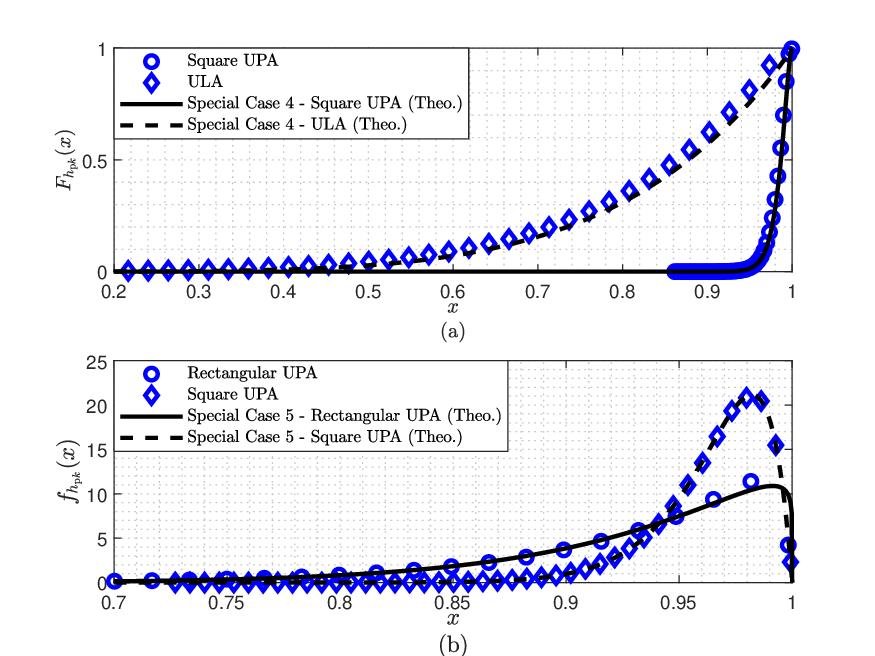}
    \vspace{-20pt}
    \caption{(a) The CDF of the pointing error coefficient $h_{{\rm{p}}k}$ for Special Case 4 with square UPA and ULA of isotropic antennas. (b) The PDF of the pointing error coefficient $h_{{\rm{p}}k}$ for Special Case 5 with rectangular and square UPAs of isotropic antennas under symmetrical jitter characteristics.}
    \label{fig006}
\end{figure}

In Fig. \ref{fig006}, the pointing error characteristics are shown in the case of isotropic antenna arrays in jitter-dominated regime. In Fig. \ref{fig006}a, the CDF of the pointing error coefficient in square UPA and ULA scenarios, as in Special Case 4 (standard UPAs and ULAs of isotropic antennas), are illustrated. For the square UPA scenario $5 \times 5$ array is used, whereas $25 \times 1$ array is used for the ULA scenario. For square UPA scenario \eqref{NLSG-Eq29-SC-4.6} and \eqref{NLSG-Eq45-SC-4.6} are used, whereas \eqref{NLSG-Eq29-SC-4.7} and \eqref{NLSG-Eq45-SC-4.7} are used for the ULA scenario. Here, the mechanical jitters are characterized by the unstable link scenario in Table \ref{table_params_fig_9}. It can be deduced from the figure that a ULA equipped with the same number of elements as UPA is severely affected by the mechanical jitters. This can be explained by significantly narrow beam on the x-z plane in ULA case compared to square UPA, making ULA configuration much more sensitive to mechanical jitters. In Fig. \ref{fig006}b, rectangular and square UPA scenarios are considered with symmetrical jitters with $\sigma_{\rm{j}} = 0.7^\circ$. $24 \times 6$ and $12 \times 12$ arrays are used for the rectangular and square UPA cases, respectively. It can be observed from the figure that the pattern asymmetry introduced by the rectangular array results in less desirable pointing error characteristics. Additionally, it is obvious from both Fig. \ref{fig006}a and Fig. \ref{fig006}b that the theoretical curves perfectly match with the simulation results for all special cases.

\subsection{Modeling Accuracy}

\begin{table*}[!t]
\centering
{
\caption{Distribution Parameters and Error Metrics for the Statistics in Various Scenarios Shown in Fig. \ref{fig001}-\ref{fig006}}
\label{table_fitness_test}
\resizebox{0.99\linewidth}{!}{
\begin{tabular}{|l|c|c|c|c|c|c|c|c|}
\hline
Figure & Marker & Distribution parameters ($\varsigma_{k},\nu_{k},\rho_{k}$) & $\text{NME}_{1}$ [\%] & $\text{NME}_{2}$ [\%] & $\text{NME}_{3}$ [\%] & $\text{NME}_{4}$ [\%] & JS divergence & KS statistics \\ \hline \hline 
\multirow{3}{3em}{Fig. \ref{fig001}}             & $\circ$     & (1.86924, 0.06582, 0.06574) & 0.21689 & 0.37303 & 0.48694 & 0.57070 & 0.00154 & 0.00788 \\ 
                                                 & $\bDiamond$ & (1.19524, 0.06453, 0.06574) & 0.13518 & 0.23349 & 0.31048 & 0.37608 & 0.01343 & 0.03658 \\ 
                                                 & $\nabla$    & (1.94063, 0.00981, 0.06574) & 0.00371 & 0.00358 & 0.00027 & 0.00779 & 0.00778 & 0.04547 \\ \hline \hline 
\multirow{3}{3em}{Fig. \ref{fig002}a}            & $\circ$     & (2.00145, 0.01522, 0.26298) & 0.00982 & 0.01058 & 0.00265 & 0.01362 & 0.01428 & 0.04449 \\ 
                                                 & $\bDiamond$ & (1.99840, 0.02969, 0.26298) & 0.05040 & 0.08662 & 0.10960 & 0.12021 & 0.00662 & 0.02691 \\ 
                                                 & $\nabla$    & (1.99508, 0.07497, 0.26298) & 0.31795 & 0.54312 & 0.69950 & 0.80348 & 0.00057 & 0.01205 \\ \hline 
\multirow{2}{4em}{Fig. \ref{fig002}b${}^{*}$}    & $\circ$     & (3.37534, 0.09817, 0.26298) & 1.13897 & 1.79569 & 2.00311 & 1.80856 & 0.00393 & 0.04455 \\ 
                                                 & $\nabla$    & (1.99794, 0.02917, 0.26298) & 0.04768 & 0.08154 & 0.10251 & 0.11142 & 0.00812 & 0.02256 \\ \hline 
\multirow{3}{3em}{Fig. \ref{fig002}c}            & $\circ$     & (2.30122, 0.00892, 0.26298) & 0.00002 & 0.00019 & 0.00065 & 0.00152 & 0.01611 & 0.04465 \\ 
                                                 & $\bDiamond$ & (2.78753, 0.01757, 0.26298) & 0.00020 & 0.00160 & 0.00528 & 0.01221 & 0.00955 & 0.03456 \\ 
                                                 & $\nabla$    & (3.41033, 0.02835, 0.26298) & 0.00103 & 0.00793 & 0.02569 & 0.05846 & 0.00650 & 0.03043 \\ \hline \hline 
\multirow{4}{3em}{Fig. \ref{fig003}a}            & $\circ$     & (2      , 0.03194, 0      ) & 0.05185 & 0.09715 & 0.13677 & 0.17145 & 0.00003 & 0.00403 \\ 
                                                 & $\bDiamond$ & (1.99794, 0.02917, 0.26298) & 0.03981 & 0.06617 & 0.08006 & 0.08229 & 0.00860 & 0.02399 \\ 
                                                 & $\nabla$    & (1.99794, 0.11668, 0.26298) & 0.70310 & 1.12102 & 1.36792 & 1.50641 & 0.00062 & 0.01550 \\ 
                                                 & $\square$   & (1.13076, 0.10664, 0.26298) & 0.33907 & 0.54976 & 0.69131 & 0.79623 & 0.00574 & 0.06376 \\ \hline 
\multirow{3}{4em}{Fig. \ref{fig003}b}            & $\circ$     & (2.00346, 0.00797, 0.00065) & 0.00220 & 0.00428 & 0.00624 & 0.00810 & 0.00007 & 0.00241 \\ 
                                                 & $\bDiamond$ & (5.76451, 0.00761, 0.00065) & 0.02904 & 0.05656 & 0.08252 & 0.10687 & 0.00108 & 0.01601 \\ 
                                                 & $\nabla$    & (9.53103, 0.00754, 0.00065) & 0.05524 & 0.10767 & 0.15717 & 0.20365 & 0.00085 & 0.01918 \\ \hline \hline 
\multirow{3}{3em}{Fig. \ref{fig004}a}            & $\circ$     & (2.26483, 0.03823, 0.14792) & 0.11551 & 0.23568 & 0.35991 & 0.48767 & 0.00040 & 0.01351 \\ 
                                                 & $\bDiamond$ & (1.99941, 0.03061, 0.14792) & 0.05033 & 0.09003 & 0.12006 & 0.14127 & 0.00106 & 0.00725 \\ 
                                                 & $\nabla$    & (2.26462, 0.04356, 0.14792) & 0.13205 & 0.20325 & 0.21926 & 0.18506 & 0.00534 & 0.02511 \\ \hline 
\multirow{2}{4em}{Fig. \ref{fig004}b${}^{**}$}   & $\circ$     & (1.99941, 0.02666, 0.14792) & 0.03768 & 0.06672 & 0.08773 & 0.10126 & 0.00113 & 0.00873 \\ 
                                                 & $\nabla$    & (1.99941, 0.03483, 0.14792) & 0.06574 & 0.11811 & 0.15860 & 0.18845 & 0.00157 & 0.00791 \\ \hline \hline 
\multirow{3}{3em}{Fig. \ref{fig005}}             & $\circ$     & (1.94063, 0.08836, 0.28535) & 0.37387 & 0.56703 & 0.63333 & 0.60581 & 0.00487 & 0.01369 \\ 
                                                 & $\bDiamond$ & (1.83546, 0.05144, 0.28535) & 0.11119 & 0.13851 & 0.09954 & 0.00742 & 0.00645 & 0.02262 \\ 
                                                 & $\nabla$    & (1.83546, 0.05303, 0)       & 0.15229 & 0.27599 & 0.38074 & 0.47313 & 0.00046 & 0.00887 \\ \hline \hline 
\multirow{2}{3.5em}{Fig. \ref{fig006}a}            & $\circ$     & (1.92878, 0.00834, 0)       & 0.00425 & 0.00834 & 0.01228 & 0.01610 & 0.00011 & 0.00558 \\ 
                                                 & $\bDiamond$ & (0.96872, 0.19756, 0)       & 0.93749 & 1.33607 & 1.54030 & 1.66147 & 0.00149 & 0.01705 \\ \hline 
\multirow{2}{3.5em}{Fig. \ref{fig006}b}            & $\circ$     & (1.12133, 0.06735, 0)       & 0.14443 & 0.25475 & 0.34355 & 0.41925 & 0.00481 & 0.06356 \\ 
                                                 & $\bDiamond$ & (2,       0.01755, 0)       & 0.01804 & 0.03473 & 0.05019 & 0.06451 & 0.00003 & 0.00275 \\ 

\hline
\multicolumn{9}{l}{${}^{*\textcolor{white}{*}}$ The values for the diamond markers ($\bDiamond$) in Fig. \ref{fig002}b are as the same as the values for the diamond markers ($\bDiamond$) in Fig. \ref{fig002}a.} \\
\multicolumn{9}{l}{${}^{**}$ The values for the diamond markers ($\bDiamond$) in Fig. \ref{fig004}b are as the same as the values for the diamond markers ($\bDiamond$) in Fig. \ref{fig004}a.}
\end{tabular}}
}
\end{table*}

Up to this point, impacts of various communication scenarios and design parameters on the pointing error characteristics are shown via the PDF and CDF of the pointing error coefficient. In Table \ref{table_fitness_test}, the distribution parameters evaluated in Fig. \ref{fig001}-\ref{fig006} are listed along with the NMEs for the first four moments, JS divergence values, and KS statistics. By examining the third column in Table \ref{table_fitness_test} and the corresponding figures, a clear relationship between the distribution parameters and the pointing error severity can be observed: As the value of either distribution parameter increases, the PDF/CDF leans towards left, meaning that the pointing error coefficient is more likely to have lower values which indicates a more severe pointing error characteristics. Moreover, all statistical error metrics show that the proposed NLSG model accurately characterizes the pointing error in a wide range of scenarios discussed above with NMEs mostly below 1\%, JS divergence values all below 0.02, and KS statistics mostly below 0.05. These results illustrate exceptional flexibility and accuracy,  which stems from the precise modeling of the array pattern. Since the defining factor in pointing error characteristics in highly directional communication systems is the shape of the pattern, modeling the main lobe with high accuracy leads to almost perfect statistical fit across various communication scenarios. In short, the proposed NLSG model is proven to be a powerful tool while designing a highly directional communication system to analyze the impacts of design parameters on pointing error statistics.

\section{Performance Analysis and Engineering Insights}

In a highly directional beamforming system, the performance is significantly influenced by the pointing errors, as it leads to a reduction in the received signal power. In this section, the ergodic capacity is analyzed to explore the impact of design parameters, such as array size, considered beamforming architecture, number of quantization bits, etc., on the system performance through pointing error statistics derived in Section VI.

\subsection{Ergodic Capacity}

The ergodic capacity is defined as the average achievable data rate in a wireless channel. In wideband systems, the ergodic capacity can be formulated as the sum of the average achievable data rates in all sub-bands as
\begin{equation}
\begin{split}
C_{\rm erg} = \sum_{k = 1}^{K} E\left[ \frac{B}{K} \log_{2} \left(1 + \gamma_{k}\right) \right], 
\label{NLSG-Eq53}
\end{split}
\end{equation}
where $\frac{B}{K}$ and $\gamma_{k} = \frac{P_{k} h_{{\rm{f}}k}^{2} h_{{\rm{p}}k}^{2}}{L_{k} \sigma_{w}^{2}}$ are the bandwidth of each sub-band and the instantaneous {signal-to-noise ratio (SNR)} of the $k$-th sub-band, respectively. By using the Jensen's inequality, the ergodic capacity can be upper bounded as
\begin{equation}
\begin{split}
C_{\rm erg} \leq \sum_{k = 1}^{K}  \frac{B}{K} \log_{2} \left(1 + E\left[\gamma_{k} \right]\right). 
\label{NLSG-Eq54}
\end{split}
\end{equation}
Here, the average value of the SNR in the $k$-th sub-band can be obtained as
\begin{equation}
\begin{split}
E\left[\gamma_{k} \right] = \frac{P_{k} E[h_{{\rm{f}}k}^{2}] E[h_{{\rm{p}}k}^{2}]}{L_{k} \sigma_{w}^{2}}.
\label{NLSG-Eq55}
\end{split}
\end{equation}
By using \cite[Eq. (106)]{8610080} and \eqref{NLSG-Eq52}, the above-equation can be rewritten as
\begin{equation}
\begin{split}
E\left[\gamma_{k} \right] = \frac{P_{k}}{L_{k} \sigma_{w}^{2}} \frac{(\hat{h}_{{\rm{f}}k})^2}{\mu^{\frac{2}{\alpha}}}\frac{\Gamma\left(\frac{2}{\alpha}+\mu\right)}{\Gamma(\mu)} (1 + 2 \nu_{k})^{-\varsigma_{k}} e^{-2\rho_{k}}.
\label{NLSG-Eq56}
\end{split}
\end{equation}
Plugging the above-equation in \eqref{NLSG-Eq54}, the upper bound for the ergodic capacity can be found as
\begin{equation}
{\footnotesize
\begin{split}
C_{\rm erg} &\leq \sum_{k = 1}^{K}  \frac{B}{K} \log_{2} \Bigg(1 + 
\frac{P_{k} (\hat{h}_{{\rm{f}}k})^2 \Gamma\left(\frac{2}{\alpha}+\mu\right)}{L_{k} \sigma_{w}^{2} \mu^{\frac{2}{\alpha}} \Gamma(\mu)} 
(1 + 2 \nu_{k})^{-\varsigma_{k}} e^{-2\rho_{k}}\Bigg).
\label{NLSG-Eq57}
\end{split}}
\end{equation}
Notice here that the shape parameter $\varsigma_{k}$ and scale parameter $\nu_{k}$ are related to the array design parameters, such as beamforming architectures, number of quantization bits as well as the jitter variance, AoD/AoA estimation errors, etc. Owing to the proposed model in this paper, the impact of the design parameters on the ergodic capacity can be analyzed. This emphasizes the capability of the NLSG model to guide the system design through pointing error characteristics in various practical scenarios.


\subsection{Engineering Insights}

In the following, ergodic capacity of a highly directional communication system is analyzed under various conditions. Unlike the previous discussions, the impact of antenna/array design, beamforming architecture, jitter/angle estimation error characteristics, and narrowband/wideband characteristics are directly examined based on the achievable data rate in the system to gain performance insights and to guide the system design. In the subsequent discussions, the operating frequency is selected as 300 GHz, and the total bandwidth $B = 40$ GHz with each sub-band having a width of 1 GHz ($K = 40$). Additionally, throughout the simulations, $\frac{P_{k}}{L_{k} \sigma_{w}^{2}} = 20$ dB is assumed. Unless otherwise stated, the antenna element is characterized by $G^{{\rm{q}}k}_{\rm{e,max}} = 8$ dBi and $\theta_{\rm{ex,3dB}} = \theta_{\rm{ey,3dB}} = 65^\circ$, \textcolor{black}{and the element spacing is taken as $d_{\rm{e}} = \lambda_{\rm{c}}/2$.}


\begin{figure}[!t]
    \centering
    \includegraphics[width=1\linewidth]{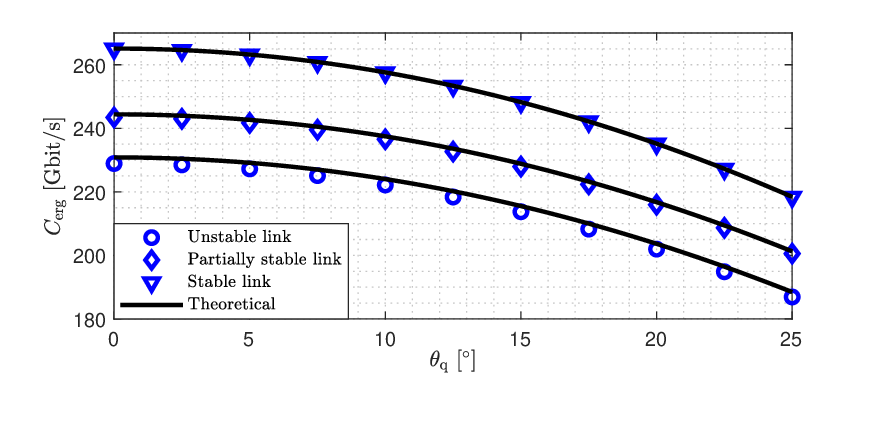}
    \vspace{-35pt}
    \caption{The theoretical and simulation ergodic capacity curves for unstable, partially stable, and stable links.}
    \label{fig09}
\end{figure}

In Fig. \ref{fig09}, the ergodic capacity curves are shown for unstable, partially stable, and stable links considering hybrid beamforming architecture with the subarray size $4 \times 4$ and superarray size $4 \times 4$. Here, $\phi_{\rm{q}} = 0^\circ$ and $\theta_{\rm{q,max}} = 25^\circ$ are assumed. For the unstable link, both transmitter and receiver are subject to high jitter variance and angle estimation error variance. The standard deviations of mechanical jitters and angle estimation errors at the transmitter and receiver are given in Table \ref{table_params_fig_9} together with the number of quantization bits. It can be seen that the stability of the link in terms of the mechanical jitter level and angle estimation errors has significant influence on the capacity. The system has the highest ergodic capacity if both of the nodes are subject to low jitter and angle estimation errors, whereas it has the lowest capacity if disturbances are high at both nodes. It can be inferred from the figure that the impact of asymmetric mechanical jitters and angle estimation errors on the ergodic capacity performance is fully captured by the proposed model. This feature of the analytical model proposed in this paper offers a distinct advantage in analyzing and designing practical systems, such as inter-satellite or satellite-ground links, where the jitters and angle estimation errors are asymmetric.

\begin{figure}[!t]
    \centering
    \vspace{-4.25pt}
    \includegraphics[width=1\linewidth]{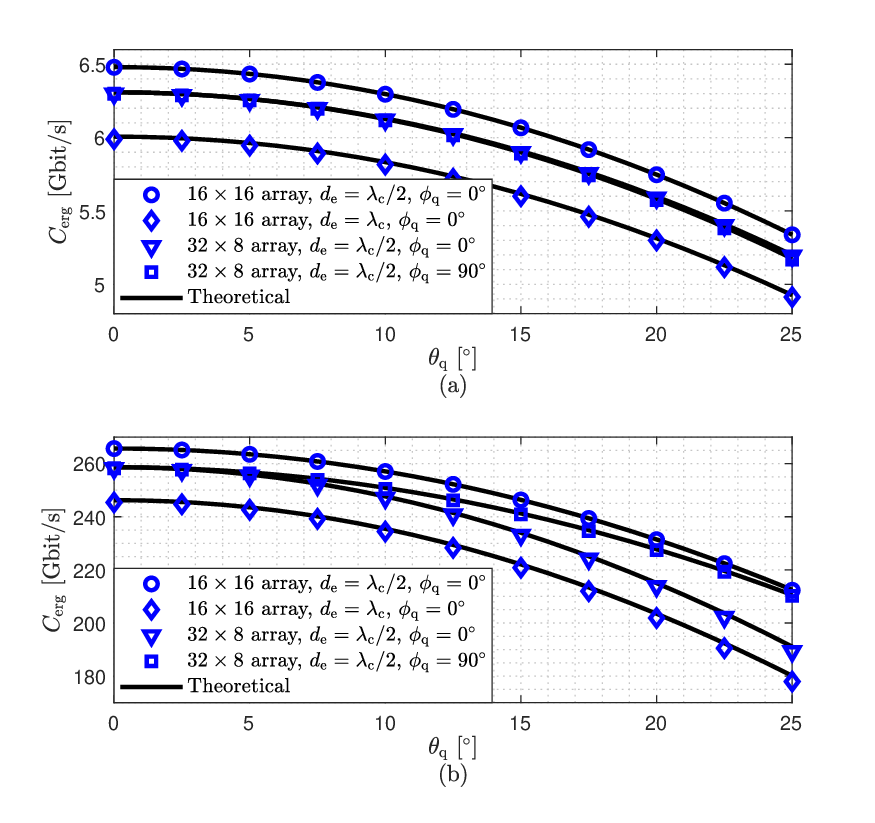}
    \vspace{-25pt}
    \caption{The theoretical and simulation ergodic capacity curves for different array configurations with analog beamforming in (a) narrowband systems, (b) wideband systems.}
    \label{fig05}
\end{figure}

In Fig. \ref{fig05}, ergodic capacity curves are shown for different array configurations and steering angles for narrowband systems (Fig. \ref{fig05}a) and wideband systems (Fig. \ref{fig05}b). \textcolor{black}{In this figure, the total bandwidth in the narrowband case is taken as $1$ GHz.} Here, analog beamforming architecture is considered without beam squint compensation. Also, the mechanical jitters and angle estimation errors are characterized by $\sigma_{\rm{jqx}} = \sigma_{\rm{jqy}} = 0.5^\circ$ and $\sigma_{\rm{qP}} = \sigma_{\rm{qP}_\perp} = 0.5^\circ$, respectively. Moreover, $\ell_{\rm{q}} = 5$ number of quantization bits are used, whereas the maximum steering angle is chosen as $\theta_{\rm{q,max}} = 25^\circ$. It can be deduced that the increment in the element spacing results in a significant reduction in the ergodic capacity as the width of the main lobe gets narrower, leading to more severe pointing error impact. Additionally, the asymmetry in the pattern introduced by a rectangular array configuration causes an ergodic capacity loss. This is because the main lobe is narrower along the longer side of the array and is wider along the shorter side of the array. Along the longer and shorter sides of the array, the system is slightly more susceptible to and more robust against the jitters and angle estimation errors, respectively. However, on average, the asymmetric pattern has poorer capacity performance compared to the symmetric pattern. Also, it is observed that the ergodic capacity performance remains almost the same under the considered setup for $\phi_{\rm{q}} = 0^\circ$ and $\phi_{\rm{q}} = 90^\circ$. In Fig. \ref{fig05}b, the same scenarios are discussed for wideband systems. In addition to the similar relationships between the array configuration and the achieved ergodic capacity performance, a significant capacity loss is observed in $32 \times 8$ array when steered towards $\phi_{\rm{q}} = 0^\circ$ and $\phi_{\rm{q}} = 90^\circ$ especially at high $\theta_{\rm{q}}$. This disparity stems from the beam squint effect in wideband analog beamforming systems since higher steering angles in the elevation direction lead the different sub-bands to be steered towards slightly off the desired direction, which results in a more severe pointing error characteristics in the corresponding sub-bands (as discussed in Fig. \ref{fig004}), therefore reducing the ergodic capacity.

\begin{figure}[!t]
    \centering
    \includegraphics[width=1\linewidth]{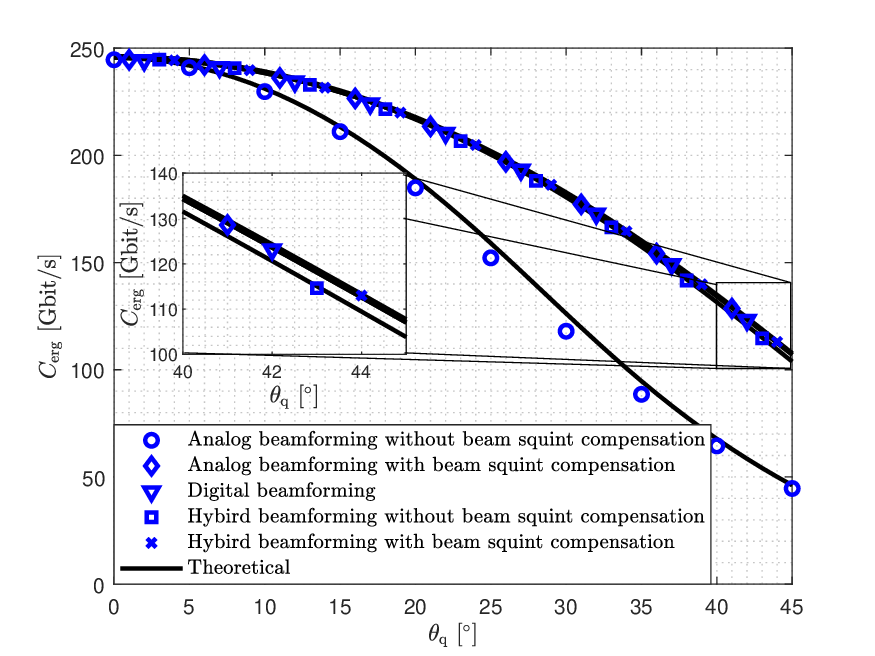}
    \vspace{-20pt}
    \caption{The theoretical and simulation ergodic capacity curves for analog ($64 \times 64$ array), digital ($64 \times 64$ array), and hybrid beamforming ($8 \times 8$ subarray and $8 \times 8$ superarray) for wideband systems.}
    \label{fig06}
\end{figure}

The ergodic capacity performance of analog, digital, and hybrid beamforming architectures are presented for wideband systems in Fig. \ref{fig06} for $\phi_{\rm{q}} = 0^\circ$. Here, $64 \times 64$ arrays are considered for analog and digital beamforming systems, whereas $8 \times 8$ subarray and $8 \times 8$ superarray configuration is considered in the hybrid beamforming counterpart to maintain the same maximum boresight array gain (with the same total number of antennas). Furthermore, $\ell_{\rm{q}} = 5$ number of quantization bits is used in the analog beamforming stages with the maximum steering range $\theta_{\rm{q,max}} = 45^\circ$. Also, jitter and angle estimation errors are characterized by $\sigma_{\rm{jqx}} = \sigma_{\rm{jqy}} = 0.25^\circ$ and $\sigma_{\rm{qP}} = \sigma_{\rm{qP}_\perp} = 0.25^\circ$, respectively. It can be inferred from the figure that the beamforming systems perform almost the same if the beam squint effect is compensated. However, a capacity loss is observed if the beam squint effect is not compensated (circle ($\circ$) and square ($\square$) markers). Here, the beam squint effect has more impact on the analog beamforming compared to the hybrid case since only the subarrays are prone to beam squint in the hybrid beamforming architecture, making it inherently more robust against the beam squint. As $\theta_{\rm{q}}$ increases, the sub-bands get further and further away from the desired steering angle due to beam squint in analog beamforming, which explains more rapid capacity reduction compared to the rest. After a certain steering angle around $\theta_{\rm{q}} = 40^\circ$, the sub-bands are already out of use due to beam squint, hence, only the center frequency (and sub-bands very close to it) are in use, which are less affected by this phenomenon. Therefore, the capacity reduction rate decreases, which explains the slight bend observed in the analog beamforming without beam squint compensation case around $\theta_{\rm{q}} = 40^\circ$.

\begin{figure}[!t]
    \centering
    \includegraphics[width=1\linewidth]{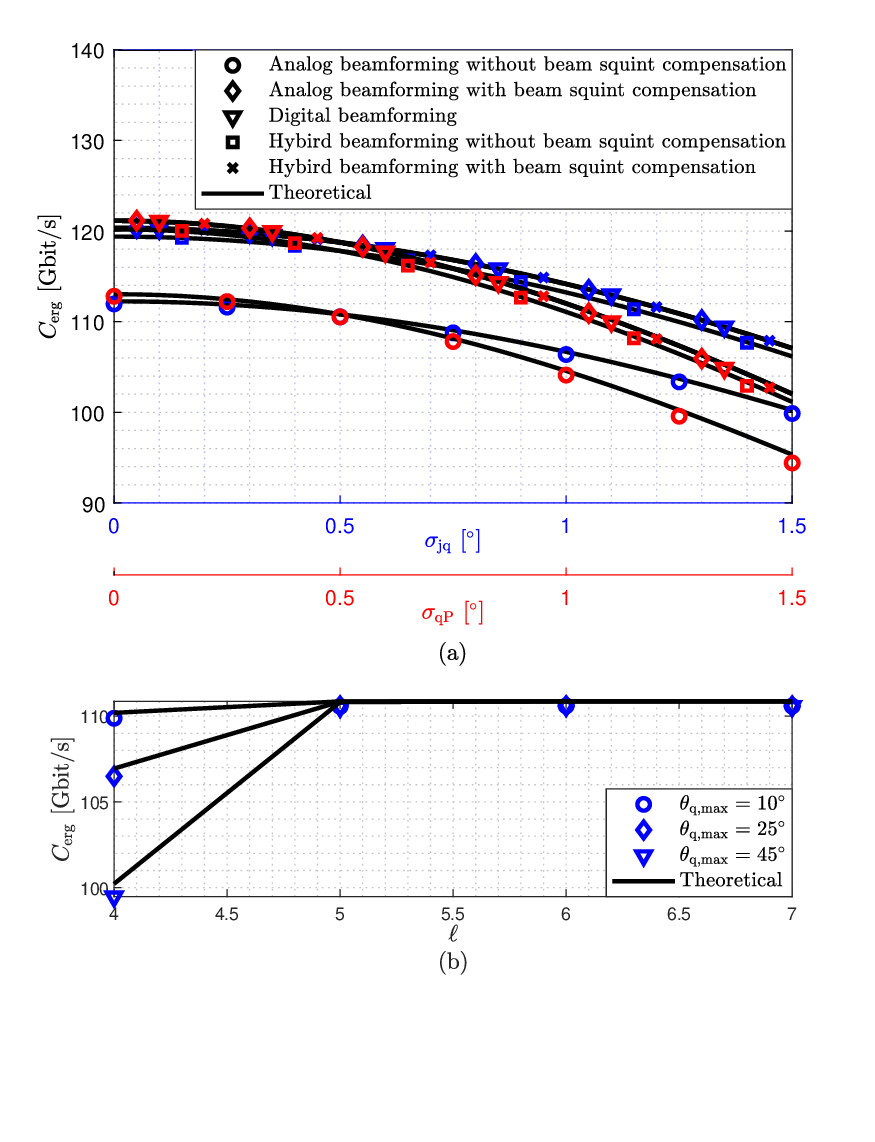}
    \vspace{-65pt}
    \caption{The impact of (a) mechanical jitters and angle estimation error (b) the number of quantization bits on the ergodic capacity performance in wideband systems with different beamforming architectures.}
    \label{fig07}
\end{figure}

The influence of the disturbance levels are shown in Fig. \ref{fig07} for $\theta_{\rm{q}} = 45^\circ$, $\phi_{\rm{q}} = 0^\circ$, and $\theta_{\rm{q,max}} = 45^\circ$. In this setup, $16 \times 16$ arrays are used for analog and digital beamforming systems, whereas $4 \times 4$ subarray and $4 \times 4$ superarray configuration is considered for the hybrid beamforming architecture. The impact of mechanical jitters and angle estimation error on the ergodic capacity performance in wideband systems is illustrated in Fig. \ref{fig07}a. Here, the blue markers show the change in the ergodic capacity with respect to the standard deviation of the jitters, whereas the red markers presents that with respect to the standard deviation of the angle estimation errors. It can be observed that the influence of the angle estimation errors is slightly more critical than the jitters as the red  markers have steeper slope compared to blue ones. Moreover, the figure shows that digital and hybrid beamforming architectures are more robust against both jitters and angle estimation errors, while analog and hybrid beamforming achieve lower capacities due to beam squint. In Fig. \ref{fig07}b, the influence of the number of quantization bits are shown for different steering ranges. Although the improvements for $\ell_{\rm{q}} \geq 5$ are barely visible, increasing the number of quantization bits results in a more favorable pointing error characteristics, and thus higher capacities. Lower values of $\ell_{\rm{q}}$ can cause capacity reduction especially in larger steering range cases due to lower steering resolution. Hence, it can be said that $\ell_{\rm{q}} \geq 5$ can be considered in general, but $\ell_{\rm{q}} = 4$ can also be used to reduce the hardware cost if the required steering range is narrow.

\begin{figure}[!t]
    \centering
    \includegraphics[width=1\linewidth]{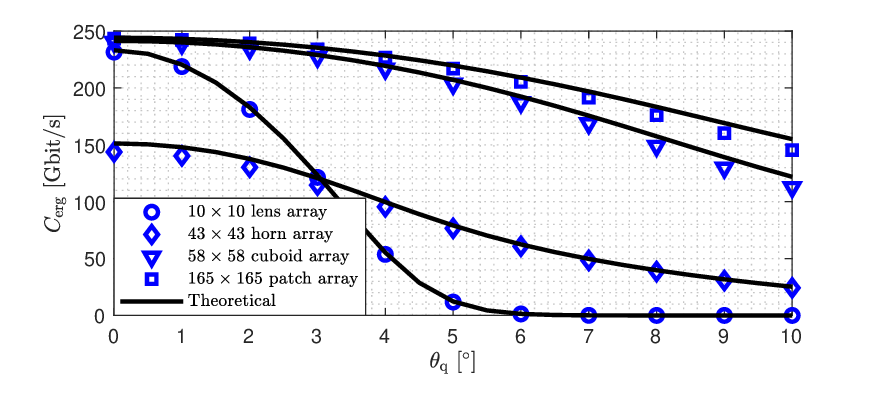}
    \vspace{-30pt}
    \caption{The impact of the antenna type on the ergodic capacity performance with respect to elevation angle under pointing errors.}
    \label{fig08}
\end{figure}

\begin{table}[!t]
\centering
{
\caption{Maximum Gain and 3 dB Beamwidth of the Antenna Elements with Element Spacing Used in Fig. \ref{fig08}}
\label{table_params_fig_13}
\begin{tabular}{|l|c|c|c|c|}
\hline
Antenna type                  & $G^{{\rm{q}}k}_{\rm{e,max}}$ & $\theta_{\rm{ex,3dB}}$ & $\theta_{\rm{ey,3dB}}$ & $d_{\rm{e}}$ \\ \hline
Lens antenna \cite{10287222}  & $29.63$ dBi                & $5.31^\circ$          & $5.31^\circ$          & $10 \lambda_{\rm{c}}$ \\ \hline
Horn antenna \cite{8608702}   & $17$ dBi                   & $22^\circ$            & $22^\circ$            & $5 \lambda_{\rm{c}}$ \\ \hline
Cuboid antenna \cite{9998671} & $14.5$ dBi                 & $31.2^\circ$          & $30.2^\circ$          & $1.5 \lambda_{\rm{c}}$ \\ \hline
Patch antenna \cite{9998628}  & $5.3$ dBi                  & $60^\circ$            & $60^\circ$            & $0.5 \lambda_{\rm{c}}$ \\ \hline
\end{tabular}
}
\end{table}

In Fig. \ref{fig08}, the ergodic capacity curves are shown with respect to elevation angle for different types of antenna element. Here, $10 \times 10$ lens antenna array, $43 \times 43$ horn antenna array, $58 \times 58$ cuboid antenna array, and $165 \times 165$ patch antenna array are considered with analog beamforming without beam squint compensation. All four configurations provide approximately $49.6$ dBi array gain in the boresight direction. Even though the total gain achieved by the transmitter and receiver is the same at $\theta_{\rm{q}} = 0^\circ$, there is a significant capacity reduction in horn antenna array. This originates from the width of the main lobe in horn antenna array configuration being narrower compared to the other three configurations, making it more susceptible to pointing errors. Furthermore, steering the beams at higher elevation angles reduces the ergodic capacity, which originates from decrement in the achieved gain because of the antenna pattern, i.e. scanning loss. Moreover, the ergodic capacity is almost zero in the lens antenna array configuration for $\theta_{\rm{q}} \geq 6^\circ$ since the element pattern is so much narrower compared to horn and cuboid antennas that it provides almost zero power intensity in that direction. The patch antenna array provides the best ergodic capacity performance, however, it comes at the cost of hardware complexity. Hence, it can be said that not only the achieved maximum gain but also how that maximum gain is achieved is important while designing a highly directional system.

\subsection{Design Guidelines}

As thoroughly examined in the previous subsections, numerous design parameters have non-negligible impact on the pointing error statistics, thus on the system performance. It is crucial to account for these considerations, relationships, and tradeoffs while designing a highly directional wideband communication systems. The most important points in the design of such a system are listed as the guidelines as follows:
\begin{itemize}
    \item Based on the system of interest, required minimum total array gain (including the gains at the transmitter and receiver) can be found to mitigate the total loss factor. Depending on the physical constraints, such as size or allowed mass, the total gain can be partitioned between the transmitter and receiver, i.e. {$G^{{\rm{t}}k}_{\rm{max}}$ and $G^{{\rm{r}}k}_{\rm{max}}$}.

    \item Considering the mobility characteristics or the steering range in the system, an appropriate antenna type can be chosen. If the system requires high steering ranges, antenna elements with wider beamwidths can be used to avoid the scanning loss. According to the selected antenna type, the sizes of the arrays at both transmitter and receiver can be obtained by $G^{{\rm{q}}k}_{\rm{max}} = N_{\rm{q}} G^{{\rm{q}}k}_{\rm{e,max}}$. To minimize the impact of the beam squint effect, the size of the antenna arrays can be determined, e.g. square array ($N_{\rm{qx}} = N_{\rm{qy}}$) diminishes the beam squint most efficiently.

    \item To reduce cost and hardware complexity, analog, digital, or hybrid beamforming architectures can be preferred. Here, the beam squint effect should be taken into account: analog beamforming is the most influenced by beam squint, whereas digital beamforming is the least.

    \item In case of analog or hybrid beamforming, the number of quantization bits can be chosen as $\ell_{\rm{q}} \geq 5$ to mitigate the pointing error introduced by quantization. For systems with lower steering ranges, $\ell_{\rm{q}} = 4$ can also be used. Additionally, for higher steering ranges, beam squint compensation can be employed either by digital signal processing or by using TTD elements.

    \item Based on the design parameters, the pointing error statistics can be calculated. If high values of distribution parameters are obtained, which indicates severe pointing error characteristics, the design choices can be refined to reduce the impact of misalignment.
\end{itemize}

\begin{table}[!t]
\centering
{
\caption{NLSG Distribution Parameters for Link Stability Regimes}
\label{look_up_table}
\resizebox{1\linewidth}{!}{
\begin{tabular}{|l|c|l|}
\hline
Link stability regime                        & Distribution parameters ($\varsigma_{k},\nu_{k},\rho_{k}$)     & Practical communication scenarios \\ \hline
\multirow{3}{2.5cm}{Unstable links ($0.1 < \varsigma_{k} \nu_{k}$)}         & (0.96872, 0.19756, 0)       & \multirow{3}{4cm}{Inter-satellite, UAV-to-UAV, high-speed vehicular, dynamic maritime}                                  \\ 
                                             & (1.86924, 0.06582, 0.06574) &                                   \\ 
                                             & (3.37534, 0.09817, 0.26298) &                                   \\ \hline
\multirow{3}{2.5cm}{Partially stable links ($0.05 < \varsigma_{k} \nu_{k} \leq 0.1$)} & (1.12133, 0.06735, 0)       & \multirow{3}{4cm}{Satellite-to-ground, HAPS-to-ground, UAV-to-ground, vehicle-to-infrastructure}                                  \\ 
                                             & (1.19524, 0.06453, 0.06574) &                                   \\ 
                                             & (9.53103, 0.00754, 0.00065) &                                   \\ \hline
\multirow{3}{2.5cm}{Stable links ($\varsigma_{k} \nu_{k} \leq 0.05$)}           & (2, 0.01755, 0)             & \multirow{3}{4.25cm}{Fixed wireless access, point-to-point terrestrial backhaul/fronthaul, stationary indoor ISAC} \\ 
                                             & (1.94063, 0.00981, 0.06574) &                                   \\ 
                                             & (2.00346, 0.00797, 0.00065) &                                   \\ \hline
\end{tabular}}
}
\end{table}

By following such a design approach, the impact of pointing error in a highly directional wideband communication system can be minimized, enhancing the overall system performance. To provide an immediate reference for such engineering evaluations, Table \ref{look_up_table} categorizes representative NLSG distribution parameters mapped to practical communication scenarios and stability regimes.\footnote{Here, the stability categorization is based on the product $\varsigma_{k} \nu_{k}$ to focus the impacts of jitters, angle estimation errors, and quantization errors. The influence of the location parameter $\rho_{k}$ can be taken into consideration in the link budget analysis as scanning loss during system design.}

\section{Conclusion}

In this paper, a generalized pointing error model is proposed for highly directional wideband beamforming systems by accounting for various design parameters and disturbances. To achieve this, first, the element pattern is modeled for both high- and low-gain antennas, and the array factor is approximated as a Gaussian beam for the subarrays and superarrays to accommodate analog, digital, and hybrid beamforming architectures. After obtaining the main lobe of the array pattern, the mechanical jitters are modeled as the 3-dimensional rotations of the array. Then, impact of inaccurate AoD/AoA estimations are incorporated through the erroneous phase excitations of the antenna array. Moreover, for the analog beamforming stage, the phase shifters' quantization is also included in the model as the quantization error. Afterwards, the statistics of the pointing error coefficient are derived. It is shown that the pointing error follows a novel distribution, namely, the NLSG distribution, which is described by three parameters, i.e. the shape, scale, and location parameters. Having simple closed-form expressions for the CDF, PDF, and the moments, this new distribution is applicable to a wide range of scenarios, accommodating narrowband and wideband communications with analog, digital, or hybrid beamforming architectures in different tracking modes. Furthermore, the model incorporates the beam squint effect in wideband systems as well as the cases where it is compensated. The NLSG model is able to capture the pointing error characteristics even if the jitters and angle estimation errors are symmetrical or asymmetrical and Gaussian or non-Gaussian. The impact of pattern asymmetry, caused either by the antenna element or the rectangular arrays, is also included in the NLSG model. Extensive verification is performed for all scenarios mentioned above, with goodness-of-fit tests by using NME, JS divergence, and KS statistics. The results have shown that the model is exceptionally accurate with NMEs mostly below 1\%, JS divergence values all below $0.02$, and KS statistics mostly below $0.05$. By utilizing the proposed model, the ergodic capacity analysis is performed under various scenarios in the presence of pointing errors and multipath fading in order to examine the influence of all physical disturbances and design parameters, providing important engineering insights about the system design. The results have demonstrated the importance of the design choices, such as the type of the antenna, element spacing, shape of the array (square or rectangular), beamforming architecture, and number of quantization bits. Finally, some important guidelines are provided to design a highly directional wideband beamforming system.

\section*{Acknowledgment}

This research was supported by the Istanbul Technical University Research Fund under grant number 2026-49083.

\begin{appendices}

\section*{Appendix A\\Circular Pattern Radiating Towards the z-Axis}

The pattern function for low-gain antenna pointed towards the $x$-axis is given as \cite[Eq. (8.1.18)]{vaughan2003channels}
\begin{equation}
\begin{split}
g^{{\rm{q}}k}(\vartheta,\varphi) = \sin^{n_{\vartheta}} \vartheta \cos^{n_{\varphi}}\left(\frac{\varphi}{2}\right),
\end{split}
\label{NLSG-EqAp-01}
\end{equation}
where $n_{\vartheta}$ and $n_{\varphi}$ are the beamwidth parameters which are related to the 3 dB beamwidths on the $x$-$z$ plane and $x$-$y$ plane as $\theta_{{\rm{e}}xz{\rm ,3dB}} = 2\sin^{-1}\left(2^{-1/n_{\vartheta}}\right)$ and $\theta_{{\rm{e}}xy{\rm ,3dB}} = 4\cos^{-1}\left(2^{-1/n_{\varphi}}\right)$. To align this pattern with the intended radiation direction, the intermediate coordinate system ($x$, $y$, $z$) is defined by rotating the local frame (${\rm{x}}'$, ${\rm{y}}'$, ${\rm{z}}'$) by $-\frac{\pi}{2}$ rad about the ${\rm{y}}'$-axis. Consequently, the pattern's principal beam, inherently directed along the $x$-axis, perfectly coincides with the ${\rm{z}}'$-direction of the (${\rm{x}}'$, ${\rm{y}}'$, ${\rm{z}}'$) frame. Based on this transformation, the cartesian coordinates of the intermediate frame $(x, y, z)$ map to the local frame $({\rm{x}}', {\rm{y}}', {\rm{z}}')$ as $x = {\rm{z}}'$, $y = {\rm{y}}'$, and $z = -{\rm{x}}'$. Moreover, $\theta_{{\rm{e}}xz{\rm ,3dB}}$ and $\theta_{{\rm{e}}xy{\rm ,3dB}}$ are replaced by the 3 dB beamwidths on the ${\rm{x}}'$-${\rm{z}}'$ plane and ${\rm{y}}'$-${\rm{z}}'$ plane as $\theta_{\rm{ex,3dB}}$ and $\theta_{\rm{ey,3dB}}$, respectively. 

By applying standard spherical-to-cartesian conversions (${\rm{x}}' = \sin\theta'\cos\phi'$, ${\rm{y}}' = \sin\theta'\sin\phi'$, ${\rm{z}}' = \cos\theta'$), the following relationships can be derived:
\begin{equation}
\begin{split}
\sin\vartheta &= \sqrt{1 - \sin^2\theta'\cos^2\phi'}, \\
\cos\varphi &= \frac{\cos\theta'}{\sqrt{1 - \sin^2\theta'\cos^2\phi'}}.
\end{split}
\label{NLSG-EqAp-02}
\end{equation}
By using \eqref{NLSG-EqAp-02} with the help of the half-angle identity, i.e. $\cos(\varphi/2) = \sqrt{(1 + \cos\varphi)/2}$, the pattern function can be written in the local frame ($\theta'$,$\phi'$) as in \eqref{NLSG-Eq03.1}.

\section*{Appendix B\\Proof of Lemma 1}

Denoting the continuous phase adjustments as $b_{\rm{qx}} = \zeta_{\rm{qx}}\vartheta_{\rm{qP}} + \eta_{\rm{qx}}\vartheta_{\rm{qP}_\perp} + \mu_{b_{\rm{qx}}}$ and $b_{\rm{qy}} = \zeta_{\rm{qy}}\vartheta_{\rm{qP}} + \eta_{\rm{qy}}\vartheta_{\rm{qP}_\perp} + \mu_{b_{\rm{qy}}}$ for brevity, where $\vartheta_{\rm{qP}} \sim \mathcal{N}(0,\sigma_{\rm{qP}}^{2})$ and $\vartheta_{{\rm{qP}_\perp}} \sim \mathcal{N}(0,\sigma_{{\rm{qP}_\perp}}^{2})$ are two independent Gaussian random variables, $b_{\rm{qx}}$ and $b_{\rm{qy}}$ are Gaussian distributed as $b_{\rm{qx}} \sim \mathcal{N}(\mu_{b_{\rm{qx}}},\sigma_{b_{\rm{qx}}}^{2})$ and $b_{\rm{qy}} \sim \mathcal{N}(\mu_{b_{\rm{qy}}},\sigma_{b_{\rm{qy}}}^{2})$, respectively, with means $\mu_{b_{\rm{qx}}}$ and $\mu_{b_{\rm{qy}}}$ and variances $\sigma_{b_{\rm{qx}}}^{2} = \zeta_{\rm{qx}}^{2} \sigma_{\rm{qP}}^{2} + \eta_{\rm{qx}}^{2} \sigma_{\rm{qP}_\perp}^{2}$ and $\sigma_{b_{\rm{qy}}}^{2} = \zeta_{\rm{qy}}^{2} \sigma_{\rm{qP}}^{2} + \eta_{\rm{qy}}^{2} \sigma_{\rm{qP}_\perp}^{2}$. Then, $\text{Var}(b_{\rm{qx}}^{2})$ can be found as
\begin{equation}
\begin{split}
\text{Var}(b_{\rm{qx}}^{2}) = E[b_{\rm{qx}}^{4}] - E^{2}[b_{\rm{qx}}^{2}].
\end{split}
\label{NLSG-EqA-01}
\end{equation}
By using the second and fourth non-central moments of the Gaussian random variable $b_{\rm{qx}}$, i.e. $E[b_{\rm{qx}}^{2}] = \mu_{b_{\rm{qx}}}^{2} + \sigma_{b_{\rm{qx}}}^{2}$ and $E[b_{\rm{qx}}^{4}] = \mu_{b_{\rm{qx}}}^{4} + 6 \mu_{b_{\rm{qx}}}^{2} \sigma_{b_{\rm{qx}}}^{2} + 3 \sigma_{b_{\rm{qx}}}^{4}$, the above-equation can be rewritten as
\begin{equation}
\begin{split}
\text{Var}(b_{\rm{qx}}^{2}) = 4 \mu_{b_{\rm{qx}}}^{2} \sigma_{b_{\rm{qx}}}^{2} + 2 \sigma_{b_{\rm{qx}}}^{4}.
\end{split}
\label{NLSG-EqA-02}
\end{equation}
By following the similar manipulations, $\text{Var}(b_{\rm{qy}}^{2})$ can be obtained as
\begin{equation}
\begin{split}
\text{Var}(b_{\rm{qy}}^{2}) = 4 \mu_{b_{\rm{qy}}}^{2} \sigma_{b_{\rm{qy}}}^{2} + 2 \sigma_{b_{\rm{qy}}}^{4}.
\end{split}
\label{NLSG-EqA-03}
\end{equation}

The covariance between $b_{\rm{qx}}^{2}$ and $b_{\rm{qx}}$ can be evaluated as
\begin{equation}
\begin{split}
\text{Cov}(b_{\rm{qx}}^{2},b_{\rm{qx}}) &= E[b_{\rm{qx}}^{3}] - E[b_{\rm{qx}}^{2}] E[b_{\rm{qx}}],
\end{split}
\label{NLSG-EqA-04}
\end{equation}
where the first and third non-central moments of $b_{\rm{qx}}$ are $E[b_{\rm{qx}}] = \mu_{b_{\rm{qx}}}$ and $E[b_{\rm{qx}}^{3}] = \mu_{b_{\rm{qx}}}^{3} + 3 \mu_{b_{\rm{qx}}} \sigma_{b_{\rm{qx}}}^{2}$, respectively. Hence, $\text{Cov}(b_{\rm{qx}}^{2},b_{\rm{qx}})$ yields
\begin{equation}
\begin{split}
\text{Cov}(b_{\rm{qx}}^{2},b_{\rm{qx}}) &= 2\mu_{b_{\rm{qx}}}\sigma_{b_{\rm{qx}}}^{2}.
\end{split}
\label{NLSG-EqA-04}
\end{equation}
By the similar approach, $\text{Cov}(b_{\rm{qy}}^{2},b_{\rm{qy}})$ can be derived as
\begin{equation}
\begin{split}
\text{Cov}(b_{\rm{qy}}^{2},b_{\rm{qy}}) &= 2\mu_{b_{\rm{qy}}}\sigma_{b_{\rm{qy}}}^{2}.
\end{split}
\label{NLSG-EqA-05}
\end{equation}

The covariance between the continuous phase adjustments $b_{\rm{qx}}$ and $b_{\rm{qy}}$ can be written as $\text{Cov}(b_{\rm{qx}},b_{\rm{qy}}) = E[b_{\rm{qx}} b_{\rm{qy}}] - E[b_{\rm{qx}}] E[b_{\rm{qy}}]$
By substituting the definitions of $b_{\rm{qx}}$ and $b_{\rm{qy}}$ in the above-equation, $\text{Cov}(b_{\rm{qx}},b_{\rm{qy}})$ can be rewritten after some manipulations as
\begin{equation}
{\small
\begin{split}
\text{Cov}(b_{\rm{qx}},b_{\rm{qy}}) &= 
  \zeta_{\rm{qx}}\zeta_{\rm{qy}}E[\vartheta_{\rm{qP}}^{2}] + \eta_{\rm{qx}}\eta_{\rm{qy}}E[\vartheta_{\rm{qP}_\perp}^{2}] + \mu_{b_{\rm{qx}}}\mu_{b_{\rm{qy}}} \\
&+(\zeta_{\rm{qx}}\eta_{\rm{qy}} + \eta_{\rm{qx}}\zeta_{\rm{qy}})E[\vartheta_{\rm{qP}}]E[\vartheta_{\rm{qP}_\perp}] \\
&+(\zeta_{\rm{qx}}\mu_{b_{\rm{qy}}} + \mu_{b_{\rm{qx}}}\zeta_{\rm{qy}})E[\vartheta_{\rm{qP}}] \\
&+(\eta_{\rm{qx}}\mu_{b_{\rm{qy}}} + \mu_{b_{\rm{qx}}}\eta_{\rm{qy}})E[\vartheta_{\rm{qP}_\perp}] - \mu_{b_{\rm{qx}}} \mu_{b_{\rm{qy}}}.
\end{split}}
\label{NLSG-EqA-07}
\end{equation}
Since $\vartheta_{\rm{qP}}$ and $\vartheta_{\rm{qP}_\perp}$ are zero-mean Gaussian random variables, their $r$-th moment is zero for odd $r$. Therefore, $\text{Cov}(b_{\rm{qx}},b_{\rm{qy}})$ can be found by using $E[\vartheta_{\rm{qP}}^{2}] = \sigma_{\rm{qP}}^{2}$, $E[\vartheta_{\rm{qP}_\perp}^{2}] = \sigma_{\rm{qP}_\perp}^{2}$ as 
\begin{equation}
{
\begin{split}
\text{Cov}(b_{\rm{qx}},b_{\rm{qy}}) &= 
  \zeta_{\rm{qx}}\zeta_{\rm{qy}} \sigma_{\rm{qP}}^{2} + \eta_{\rm{qx}}\eta_{\rm{qy}} \sigma_{\rm{qP}_\perp}^{2}.
\end{split}}
\label{NLSG-EqA-07}
\end{equation}

Similarly, $\text{Cov}(b_{\rm{qx}}^{2},b_{\rm{qy}})$ can be evaluated as
\begin{equation}
{\small
\begin{split}
\text{Cov}(b_{\rm{qx}}^{2},b_{\rm{qy}}) 
&= (\zeta_{\rm{qx}}^{2}\mu_{b_{\rm{qy}}} + 2\zeta_{\rm{qx}}\mu_{b_{\rm{qx}}}\zeta_{\rm{qy}})E[\vartheta_{\rm{qP}}^{2}] \\
&\hspace{-10pt}+ (\eta_{\rm{qx}}^{2}\mu_{b_{\rm{qy}}} + 2\eta_{\rm{qx}}\mu_{b_{\rm{qx}}}\eta_{\rm{qy}})E[\vartheta_{\rm{qP}_\perp}^{2}] \\
&\hspace{-10pt}+ \mu_{b_{\rm{qx}}}^{2}\mu_{b_{\rm{qy}}} - (\mu_{b_{\rm{qx}}}^{2} + \zeta_{\rm{qx}}^{2} \sigma_{\rm{qP}}^{2} + \eta_{\rm{qx}}^{2} \sigma_{\rm{qP}_\perp}^{2}\hspace{-2pt})\mu_{b_{\rm{qy}}},
\end{split}}
\label{NLSG-EqA-08}
\end{equation}
by omitting the odd moments of $\vartheta_{\rm{qP}}$ and $\vartheta_{\rm{qP}_\perp}$ as they are zero-mean, and by using $E[b_{\rm{qx}}^{2}] = \mu_{b_{\rm{qx}}}^{2} + \sigma_{b_{\rm{qx}}}^{2}$ and $E[b_{\rm{qy}}] = \mu_{b_{\rm{qy}}}$. Afterwards, substituting the second moments of $\vartheta_{\rm{qP}}$ and $\vartheta_{\rm{qP}_\perp}$ yields 
\begin{equation}
\begin{split}
\text{Cov}(b_{\rm{qx}}^{2},b_{\rm{qy}}) &= 2 \zeta_{\rm{qx}} \mu_{b_{\rm{qx}}} \zeta_{\rm{qy}} \sigma_{\rm{qP}}^{2} + 2 \eta_{\rm{qx}} \mu_{b_{\rm{qx}}} \eta_{\rm{qy}} \sigma_{\rm{qP}_\perp}^{2}.
\end{split}
\label{NLSG-EqA-09}
\end{equation}
By following the similar manipulations, \textcolor{black}{$\text{Cov}(b_{\rm{qx}},b_{\rm{qy}}^{2})$} can be obtained as
\begin{equation}
\begin{split}
\text{Cov}(b_{\rm{qx}},b_{\rm{qy}}^{2}) &= 2 \zeta_{\rm{qy}} \mu_{b_{\rm{qy}}} \zeta_{\rm{qx}} \sigma_{\rm{qP}}^{2} + 2 \eta_{\rm{qy}} \mu_{b_{\rm{qy}}} \eta_{\rm{qx}} \sigma_{\rm{qP}_\perp}^{2}.
\end{split}
\label{NLSG-EqA-10}
\end{equation}

Finally, $\text{Cov}(b_{\rm{qx}}^{2},b_{\rm{qy}}^{2})$ can be written as
\begin{equation}
{\footnotesize
\begin{split}
\text{Cov}(b_{\rm{qx}}^{2},b_{\rm{qy}}^{2}) 
&= \zeta_{\rm{qx}}^{2}\zeta_{\rm{qy}}^{2}E[\vartheta_{\rm{qP}}^{4}] + \eta_{\rm{qx}}^{2}\eta_{\rm{qy}}^{2}E[\vartheta_{\rm{qP}_\perp}^{4}] +  \mu_{b_{\rm{qx}}}^{2}\mu_{b_{\rm{qy}}}^{2}\\
&\hspace{-45pt}+ (\zeta_{\rm{qx}}^{2}\eta_{\rm{qy}}^{2} + \eta_{\rm{qx}}^{2}\zeta_{\rm{qy}}^{2} + 4\zeta_{\rm{qx}}\eta_{\rm{qx}}\zeta_{\rm{qy}}\eta_{\rm{qy}})E[\vartheta_{\rm{qP}}^{2}]E[\vartheta_{\rm{qP}_\perp}^{2}] \\
&\hspace{-45pt}+(\zeta_{\rm{qx}}^{2}\mu_{b_{\rm{qy}}}^{2} + \mu_{b_{\rm{qx}}}^{2}\zeta_{\rm{qy}}^{2} + 4\zeta_{\rm{qx}}\mu_{b_{\rm{qx}}}\zeta_{\rm{qy}}\mu_{b_{\rm{qy}}})E[\vartheta_{\rm{qP}}^{2}] \\
&\hspace{-45pt}+ (\eta_{\rm{qx}}^{2}\mu_{b_{\rm{qy}}}^{2} + \mu_{b_{\rm{qx}}}^{2}\eta_{\rm{qy}}^{2} + 4\eta_{\rm{qx}}\mu_{b_{\rm{qx}}}\eta_{\rm{qy}}\mu_{b_{\rm{qy}}})E[\vartheta_{\rm{qP}_\perp}^{2}]\\
&\hspace{-45pt}- (\mu_{b_{\rm{qx}}}^{2} + \zeta_{\rm{qx}}^{2} \sigma_{\rm{qP}}^{2} + \eta_{\rm{qx}}^{2} \sigma_{\rm{qP}_\perp}^{2}\hspace{-2pt})(\mu_{b_{\rm{qy}}}^{2} + \zeta_{\rm{qy}}^{2} \sigma_{\rm{qP}}^{2} + \eta_{\rm{qy}}^{2} \sigma_{\rm{qP}_\perp}^{2}\hspace{-2pt}),
\end{split}}
\label{NLSG-EqA-11}
\end{equation}
by omitting the odd moments of $\vartheta_{\rm{qP}}$ and $\vartheta_{\rm{qP}_\perp}$ as they are zero-mean, and by using $E[b_{\rm{qx}}^{2}] = \mu_{b_{\rm{qx}}}^{2} + \sigma_{b_{\rm{qx}}}^{2}$ and $E[b_{\rm{qy}}^{2}] = \mu_{b_{\rm{qy}}}^{2} + \sigma_{b_{\rm{qy}}}^{2}$. By using the second and fourth moments of $\vartheta_{\rm{qP}}$ and $\vartheta_{\rm{qP}_\perp}$, i.e. $E[\vartheta_{\rm{qP}}^{4}] = 3\sigma_{\rm{qP}}^{4}$, $E[\vartheta_{\rm{qP}_\perp}^{4}] = 3\sigma_{\rm{qP}_\perp}^{4}$, $E[\vartheta_{\rm{qP}}^{2}] = \sigma_{\rm{qP}}^{2}$, and $E[\vartheta_{\rm{qP}_\perp}^{2}] = \sigma_{\rm{qP}_\perp}^{2}$, the covariance between $b_{\rm{qx}}^{2}$ and $b_{\rm{qy}}^{2}$ can be obtained as
\begin{equation}
{\small
\begin{split}
\text{Cov}(b_{\rm{qx}}^{2},b_{\rm{qy}}^{2}) &= 2 \zeta_{\rm{qx}}^{2} \zeta_{\rm{qy}}^{2} \sigma_{\rm{qP}}^{4} + 2 \eta_{\rm{qx}}^{2} \eta_{\rm{qy}}^{2} \sigma_{\rm{qP}_\perp}^{4} \\
&\hspace{-10pt}+ 4 \zeta_{\rm{qx}} \mu_{b_{\rm{qx}}} \zeta_{\rm{qy}} \mu_{b_{\rm{qy}}} \sigma_{\rm{qP}}^{2} + 4 \eta_{\rm{qx}} \mu_{b_{\rm{qx}}} \eta_{\rm{qy}} \mu_{b_{\rm{qy}}} \sigma_{\rm{qP}_\perp}^{2}\\
&\hspace{-10pt}+ \textcolor{black}{4 \zeta_{\rm{qx}} \eta_{\rm{qx}} \zeta_{\rm{qy}} \eta_{\rm{qy}} \sigma_{\rm{qP}}^{2} \sigma_{\rm{qP}_\perp}^{2}},
\end{split}}
\label{NLSG-EqA-11}
\end{equation}
which concludes the proof.

\end{appendices}


\bibliographystyle{IEEEtran}
\bibliography{main.bib}

\end{document}